\documentclass{ws-ijmpa}
\usepackage[super,compress]{cite}
\usepackage{graphicx}
\usepackage{hyperref}
\usepackage{color}

\usepackage{placeins}
\usepackage{epsfig}

\newcommand{\ipb}{pb$^{-1}$}
\newcommand{\ifb}{fb$^{-1}$}

\newcommand{\Sleuth}{{\sc sleuth}}

\newcommand{\met}       {\mbox{$\not\!\!E_T$}}

\newcommand{\wpwz}{$W'\rightarrow WZ$}

\newcommand{\eeggMet}{$ee\gamma\gamma\met$}
\newcommand{\eeggmet}{ee\gamma\gamma\met}
\newcommand{\Grav}{\tilde{G}}
\newcommand{\chioneO}{\tilde{\chi}_1^0}
\newcommand{\tanBeta}{$\tan\beta$}
\newcommand{\sqdec}{\tilde{q}\to q\chioneO}
\newcommand{\gldec}{\tilde{g}\to q\tilde{q}\chioneO}

\newcommand{\sctop}{\tilde{t}}

\def\citeall{\cite{PhysRevLett.63.1447,
PhysRevLett.62.1825,
PhysRevLett.64.147,
PhysRevLett.67.2609,
PhysRevD.46.R1889,
PhysRevLett.69.3439,
PhysRevLett.68.1463,
PhysRevD.48.R3939,
PhysRevLett.71.2542,
PhysRevLett.72.3004,
PhysRevLett.74.3538,
PhysRevD.51.R949,
PhysRevLett.75.1012,
PhysRevLett.75.613,
PhysRevLett.74.2900,
PhysRevLett.76.2006,
PhysRevLett.76.4307,
PhysRevLett.78.2906,
PhysRevD.55.R5263,
PhysRevLett.80.5275,
PhysRevD.58.051102,
PhysRevLett.81.1791,
PhysRevLett.81.4806,
PhysRevLett.82.2038,
PhysRevD.59.092002,
PhysRevLett.82.3206,
PhysRevLett.83.2133,
PhysRevLett.83.3124,
PhysRevLett.84.5704,
PhysRevLett.84.1110,
PhysRevLett.84.835,
PhysRevLett.84.5716,
PhysRevLett.84.5273,
PhysRevLett.85.2062,
PhysRevLett.85.1378,
PhysRevLett.85.2056,
PhysRevD.63.091101,
PhysRevD.64.092002,
PhysRevLett.87.231803,
PhysRevLett.88.041801,
PhysRevLett.87.251803,
PhysRevD.65.052006,
PhysRevLett.88.071806,
PhysRevD.66.012004,
PhysRevLett.89.041802,
PhysRevLett.89.281801,
PhysRevLett.90.131801,
PhysRevLett.90.251801,
PhysRevLett.91.171602,
PhysRevLett.92.051803,
PhysRevLett.92.121802,
PhysRevLett.93.061802,
PhysRevLett.75.618,
Abachi1995405,
PhysRevLett.76.2228,
PhysRevLett.76.2222,
PhysRevLett.76.3271,
Abachi1996471,
PhysRevLett.78.2070,
PhysRevLett.78.3818,
PhysRevLett.80.442,
PhysRevLett.80.1591,
PhysRevD.57.589,
PhysRevLett.80.666,
PhysRevLett.81.524,
PhysRevLett.82.29,
PhysRevLett.83.4937,
PhysRevD.60.031101,
PhysRevLett.83.4476,
PhysRevLett.82.2457,
PhysRevLett.82.4769,
PhysRevD.62.071701,
PhysRevD.62.031101,
PhysRevD.62.092004,
PhysRevD.63.091102,
PhysRevLett.86.1156,
PhysRevLett.87.061802,
PhysRevD.64.012004,
PhysRevLett.87.231801,
PhysRevLett.88.171802,
PhysRevLett.89.171801,
PhysRevLett.89.261801,
PhysRevD.66.112001,
PhysRevLett.90.251802,
Abazov2004147,
PhysRevLett.93.011801,
PhysRevLett.92.221801,
PhysRevD.69.111101
}}

\begin{document}
\markboth{D. Toback, L. \v{Z}ivkovi\'{c}}
{Searches for New Particles and Interactions}

\catchline{}{}{}{}{}
%
%%%%%%%%%%%%%%%%%%%%%%%%%%%%%%%%%%%%%%%%%%%%%%%%%%%%%%%%%%%%%%%%%%%%

\title{REVIEW OF PHYSICS RESULTS FROM THE TEVATRON: SEARCHES FOR NEW PARTICLES AND INTERACTIONS}

\author{DAVID TOBACK}

\address{Mitchell Institute for Fundamental Physics and Astronomy\\
Department of Physics and Astronomy, Texas A$\&$M University\\
College Station, TX 77843--4242,
USA\\
toback@tamu.edu}

\author{LIDIJA \v{Z}IVKOVI\'{C}}

\address{
Laboratory for High Energy Physics,
Institute of Physics Belgrade\\
Pregrevica 118,
11080 Zemun,
Serbia\\
lidiaz@fnal.gov}

\maketitle

\begin{history}
\received{Day Month Year}
\revised{Day Month Year}
\end{history}

\begin{abstract}
We present a summary of results for searches for new particles and interactions at the Fermilab Tevatron collider 
by the CDF and the D0 experiments. 
These include results from Run~I as well as Run~II for the time period up to July
2014. 
We focus on searches for
supersymmetry, as well as other models of new physics such as
new fermions and bosons,
various models of excited fermions, leptoquarks, technicolor, hidden--valley model particles,
long--lived particles,
extra dimensions, dark matter particles,  and signature--based searches.

\end{abstract}

\ccode{PACS numbers: 14.65.Jk, 14.70.Pw, 14.70.Kv, 14.80.Nb, 14.80.Sv, 14.80.Tt, 14.80.Ly, 14.80.Rt, 13.85.Rm}
\clearpage

\tableofcontents

\newpage

\section{\textbf{Introduction}}	
The standard model (SM) of particle physics has had great success describing the known 
particles, their properties and the interactions between them, up to energies 
of about 1~TeV\cite{Baak:2012kk}. The recent discovery of the Higgs 
boson\cite{Aad:2012tfa,Chatrchyan:2012ufa} completed the model, 
but there are still many unanswered questions as well as many unexplained phenomena that remain.

In this review we present a summary of results for searches for new particles and interactions at the Fermilab Tevatron collider.
These include results from Run~I as well as Run~II
 which produced about 10~\ifb\ of $p{\bar p}$ collisions at $\sqrt{s}=1.96$~TeV 
recorded by each 
experiment.
We focus on searches for supersymmetry (SUSY)\footnote{Note 
that for simplicity, when we say we are looking for a model, like SUSY, we are looking for evidence of new particles and/or interactions.},
new fermions and
bosons, excited fermions, leptoquarks, technicolor particles,  hidden--valley model particles, long--lived particles,
 extra dimensions, 
 dark matter,  and signature--based searches.
While we will not discuss the full set of searches, the references contain a fairly complete set of results. 
Many other searches for new particles and interactions that are not presented here (e.g. non--SM Higgs boson searches,
$B_s\to\mu\mu$)
are presented in the different chapters of this review.

We begin with a quick overview of some of the theoretical motivations
that influenced the set of searches that were ultimately done by the experiments.
In section~\ref{sec_run1}, we provide a historical review of some of the Run~I results that had a large impact 
on the world--wide searches, including the \eeggMet\ candidate event, follow--ups on 
the leptoquark hints from the DESY $ep$ collider (HERA),
signature--based 
searches like \Sleuth\ and other searches that kept the Fermilab Tevatron collider experiments at the frontier.
 In section~\ref{sec_susy} we discuss the Run~II SUSY results and in section~\ref{sec_nonsusy} 
 we discuss the various other beyond the standard model (BSM) 
searches results from Run~II. In section~\ref{sec_sum} we conclude.

\section{\textbf{Theoretical Motivation}}\label{sec_theory}

There are many reasons to search for new particles and 
interactions beyond the 
SM, and different theoretical viewpoints can guide the ways in which we search. 
On one end of the search strategy spectrum is the fact that we have many 
compelling and well specified models of BSM physics which predict new particles and how to look for them. 
On the opposite end of the spectrum, it is possible that we have not guessed the new physics,
 but the 
 Tevatron collider
has the ability to produce these new particles. 
Searching must also be done thoughtfully and carefully
 in more model--independent ways
  to be ready for 
  surprises.
In this section we provide an overview of both types of motivations, 
with others in between the two extremes, 
 with an eye towards searches. 
 We will point the reader to more details on theoretical issues as they are 
 discussed extensively in the literature; 
 our references here are not intended to be complete but rather a guide for the reader to get started. 
  Phenomenological issues, like production mechanisms, decay products,
 final states and relevant models parameters 
 are discussed in sections~\ref{sec_susy} and~\ref{sec_nonsusy}. 
  
 \subsection{\textbf{\textit{Supersymmetry}}}

The motivations for SUSY are well known and 
documented\cite{Wess:1974tw,Fayet:1976cr,Nilles:1983ge,Haber:1984rc}
 and include 
its ability to potentially solve hierarchy problem for the Higgs boson mass,
provide a dark matter candidate, and satisfy consistency requirements of 
modern models of string theory.
Inherent in the theory is that for every fermion observed in the SM 
there is a supersymmetric boson partner that has not yet been observed;
the same is true for the known bosons, including the observed Higgs boson, and the hypothetical graviton (the particle 
mediator of gravity).
The non–-observation
of low--mass sparticles with equal masses to their SM counterparts 
has focused 
efforts on SUSY models with broken symmetry.

Since there are many new particles to be searched for, and a 
128 free parameters
 in the most general models, other ``clues'' and possible tie–-ins have been used by model builders
  to focus on weak-scale SUSY~\cite{Martin:1997ns}.
The hallmark of these SUSY models is their ability to provide a 
 dark matter candidate, and not contradict other observations\cite{Beringer:1900zz}. 
 Since experimental results from the proton and electron lifetime measurements imply
conservation of baryon number and lepton number, it is not unreasonable that SUSY has an additional 
 symmetry, known as $R$--parity\footnote{$R=(-1)^{3B+L+2S}$, 
 where $B$ is baryon
number, $S$ is spin and $L$ is lepton number.}. If $R$--parity is conserved this has the consequence that 
the lightest SUSY particle (LSP) must be stable, potentially making it a dark matter candidate. We note 
for now that for many SUSY models the LSP couple to normal matter with a tiny strength 
and when produced in a collision, it would leave the detector without a trace,
yielding significant missing transverse energy, \met,
giving a signature for SUSY that is searched for in many models.

With this in mind we quickly mention the models focused on at the Fermilab Tevatron collider
 which are typically selected for simplicity and general features. 
These include: (i) gravity--mediated SUSY breaking (minimal SuperGravity or mSUGRA) 
type models, where the lightest neutralino is the LSP, has a mass at the 
electroweak scale and becomes a natural cold--dark matter candidate (discussed in section~\ref{sec_susy1}), (ii) 
gauge--mediated SUSY breaking models (GMSB) which have a $\sim$keV mass gravitino 
as the LSP, and often have a photon and \met\ in the final state (discussed in section~\ref{sec_susy2}), and
(iii) $R$--parity violating (RPV) searches which release the desire to solve the 
dark matter problem with SUSY,
 but must be considered in the most general SUSY frameworks (discussed in~\ref{sec_susy3}). 
Other models that contain SUSY, like hidden--valley models, 
models that include
charged massive stable particles  
(CHAMPS), etc, are discussed more in section~\ref{sec_theo_ll} for their theory, and section~\ref{sec_LL} for results. 

\subsection[\textbf{\textit{Resonances}}]{\textbf{\textit{Resonances: new fermions and bosons, excited fermions, leptoquarks,
technicolor and other new particles}}}
  
 There are many other models of new physics  which also predict new particles.
 For example, models which extend the gauge structure of the SM 
generically predict new gauge bosons and possibly new scalars and fermions.
These new patterns may be produced on-- (or nearly on--) shell 
and decay into SM particles yielding a tell--tale bump in an invariant mass spectrum.
We search  
 for resonances in a general way, but optimize and report 
 our sensitivity to a small number of specific model types. 
These include models that contain new fermions and/or bosons, excited fermions, leptoquarks, 
as well as particles from technicolor and other models.
We next describe some of the models that garnered the most attention during Runs~I and II. 
Note that typically the mass of any new particle is the most relevant 
parameter of the theory, from a phenomenological standpoint, but when there are other parameters of importance we note them 
in section~\ref{sec_nonsusy}.

New fermions are predicted in many BSM models. While there are significant experimental constraints from 
LEP\cite{Collaborations:2000aa}, 
and many models of extra dimensions would give an unobserved large enhancement of the Higgs boson cross section
 if there are extra fermions~\cite{PhysRevD.76.075016}, 
there is currently
no compelling theoretical reason for there to be three and only
 three fermion generations
in the SM. Thus, it becomes natural to look for fourth generation chiral quarks and 
leptons\cite{Frampton:1999xi}, and  
vector--like quarks\cite{Atre:2008iu}
(which have right--handed and left--handed components that transform in the same way under $SU(3)\times SU(2)\times U(1)$).

Similarly, additional bosons are predicted in many new models. 
For example, 
new gauge bosons are predicted in
the minimal extensions of the SM that restore   
 left--right symmetry\cite{Mohapatra:1974hk,Mohapatra:1974gc,Senjanovic:1975rk}
 with the gauge group
$SU(2)_L\times SU(2)_R$. 
In these theories additional $W$ and $Z$ bosons, usually denoted as $W'$ and $Z'$,
will couple to the right--handed fermions with weak coupling strength. In addition, grand unified theories and other theories also predict 
the existence of new heavy bosons, where often the gauge group can be broken to the SM gauge group
or have
 additional $U(1)$'s which could yield multiple $Z'$ bosons.
For a review of neutral heavy bosons, see Ref.~\citen{Langacker:2008yv}.

The simple organization of the SM particles into a table that resembles the
periodic table of elements is suggestive that the known ``fundamental'' particles
may actually be composite or otherwise have substructure\cite{Baur:1989kv}. 
This idea is inherent in string theory\cite{Schwarz:2000ew}  or models of technicolor (more below).
If the
known particles were composite, 
excited versions of each SM particle could be produced (like the excited states of atoms or hadrons);
signatures of excited leptons
could involve the production and decay  
 $\ell^*\to\ell\gamma$, or, for excited quarks, $q^*\to q\gamma$.
 
 Many grand unified
theory models have unification of the quarks and leptons  at the highest energies,
suggesting the possibility of leptoquarks ($LQ$) in nature\cite{Buchmuller:1986zs}.
 These new particles
are color--triplet bosons,
 carry both quark and lepton quantum numbers, and have fractional electric charge, but their
spin can be 0 (scalar $LQ$) or 1 (vector $LQ$).
They could produce resonant signatures in the $\ell q$ or $\nu q$ final state.

Theories of strong dynamics, such as technicolor\cite{Weinberg:1979bn,Susskind:1978ms,Hill:2002ap,Lane:2002sm}, 
predict a host of new particles known as technifermions. In many ways this 
model posits that there are no fundamental bosons, and that the vector bosons
and the Higgs boson are composite objects made of technifermions. An advantage of this 
model is that it removed the need of the only fundamental scalar in the SM, the Higgs boson 
and/or explain why it had not been observed in Run~I or at LEP. With the discovery of the Higgs boson 
with SM properties, these models have fallen out of favor. 
One of the up sides of these models, is that they did 
 provide natural search strategies for a number of resonances which were followed. 
 
 Another resonance search is for the production of light axigluons 
 which can produce an anomalous 
top--quark forward--backward asymmetry\cite{Aaltonen:2008hc,Abazov:2007ab,Aaltonen:2011kc}
 $A_{fb}$\footnote{More details  can be found in the top
quark chapter of this review\cite{TopReview}}.
Alternative axigluon decay modes include low mass, strongly interacting
particles which will further decay to pairs of jets, yielding resonances in the  four--jet final states.
These final states are also predicted by various theories where no intermediate resonance is necessary.

The searches for resonances from new fermions and bosons are presented in section~\ref{sec_reson1}.
Similarly, searches for excited fermions, leptoquarks and technicolor are presented in sections~\ref{sec_reson2}--\ref{sec_reson4}.
Other searches for other resonances,
such as $Z\to\gamma\gamma$ and the $W+\rm{dijet}$
 search from CDF in Run~II\cite{PhysRevLett.106.171801},
 are presented in section~\ref{sec_reson5}.

\subsection[\textbf{\textit{Hidden--valley models, CHAMPS and long-lived particles}}]{\textbf{\textit{Hidden--valley models, 
CHAMPS and other long-lived particles}}}\label{sec_theo_ll}

During Run~II hidden--valley (HV) models were constructed 
that predict a new, confining gauge group 
that is weakly coupled to the standard model, leading to the production of new particles.
These low mass particles could help explain potential hints in astrophysical and 
dark matter searches.
These models are often incorporated into SUSY
models with sparticles known as ``dark particles'', and a hallmark of their production 
is the decay of long--lived particles with unusual signatures in the detectors\cite{Han:2007ae,ArkaniHamed:2008qn,Baumgart:2009tn}. 

Other models
of new long-–lived particles include 
charged massive particles (CHAMPS) which are predicted in many models of new physics, especially in 
 SUSY models\cite{PhysRevD.66.075007}. Dirac monopoles 
 have also been predicted for many years in GUT models~\cite{Beringer:1900zz}, 
and models that symmetrize electromagnetism. 
Finally, a new class of models predict new particles known as quirks which
arise when there is a new, unbroken $SU(N)$ gauge group added beyond the SM\cite{Kang:2008ea}.

\subsection{\textbf{\textit{Extra dimensions and dark matter}}}

Many versions of string theory posit (and in most cases require) the 
existence of other dimensions in addition to our three spatial + one time dimensions. 
 There are a number of different types of models which have received the most attention.  
The first are  
 large extra dimension (LED) models~\cite{ArkaniHamed:1998rs} 
 which postulate the existence of two or more extra dimensions in which only gravity can propagate.
The weakness of gravity can thus be explained by a propagation through higher--dimension space.
In universal extra dimensions (UED) models\cite{PhysRevD.64.035002} extra spatial dimensions are accessible to all SM fields.
Consequently, the difference between UED and LED is that the spatial dimensions in UED are compactified 
resulting in KK excitations, which are ``towers'' of the SM fields.
A third model is warped extra dimensions~\cite{PhysRevLett.83.3370} in which the existence of the
fifth dimension, with a warped spacetime metric, is bounded by two three--dimensional branes. 
The SM fields and gravity are on different branes with a small overlap,
causing gravity to appear weak at the TeV scale.

Dark matter has been inferred from the dynamics of
galaxies and clusters of galaxies for 
decades, and the evidence that 
 it is due to a new kind of elementary particle
 has steadily mounted\cite{Bertone:2004pz,Feng:DM}. 
There have been many models of dark matter put forward by the 
community, but from the perspective of searches at 
the Fermilab Tevatron collider there have been just a 
few types that can be 
searched for:
(i)~production and decay of SUSY particles into dark matter particles (typically the LSP),
(ii)~axions\cite{Duffy:2009ig}, 
and (iii)~theory--independent models of production.  
While SUSY has already been mentioned, we quickly note that axion production is
 out of reach of colliders. Recently, more model--independent searches have been done with few 
assumptions about the new physics  and focus
on the direct production 
of dark matter.

\subsection{\textbf{\textit{Signature--based searches and model--independent searches}}}

While model--based searches have always been favored by the theory community, 
signature--based searches became an important part of the search 
program as a more model--independent way to search 
for physics beyond the SM. Perhaps by observing the unexpected we could find 
explanations of 
various unexplained phenomena (dark matter, electroweak symmetry breaking 
etc.). Indeed, looking at history, many of the major discoveries in particle 
physics have been made in unexpected ways (the muon is a prime example).
For these reasons, 
powerful and systematic ways of searching for new physics were  
developed to help ensure that
the unexpected was not missed. New methodologies focused on the idea 
of just looking at the final state particles to see if there was any indication of an unexpected 
resonance, or anomalous number of events or kinematic distribution when considering  
 combinations of high transverse momentum ($p_T$) particles. This method of searching 
with a set of 
 final state particles is known as a ``signature--based'' search;  because it is not 
 looking for any particular new model, it is known as model--independent. 
Many different variations on these themes were 
created and executed, starting strongly in Run~I (especially after the 
unexpected observation of an event with two electrons, two photons and \met)
 and continuing throughout the search program.

\section{\textbf{Run~I Results}}\label{sec_run1}

The Run~I dataset consists of $\sim$100~\ipb\ of collision data at $\sqrt{s}=1.8$~TeV.
 Most searches~\citeall focused on the simplest resonance models, mSUGRA searches and RPV models.
  With the world's then
  highest energy collisions, these were the cutting edge,
  bracketed on each side by complementary searches from LEP. 
  We learned that the SM worked up to a higher energy scale since there was no
  evidence for new physics other than the discovery of the top quark. 
  However, there were some results that were exciting enough to have significant impact on the field. We mention three 
  before proceeding  to the Run~II results. 
 
\subsection{\textbf{\textit{The \eeggMet\ candidate event and its influence}}}\label{sec_eeggmet}

During Run~I, the CDF experiment observed a very unusual event which created 
significant interest\cite{PhysRevD.64.092002,PhysRevLett.81.1791}.
This event had two high energy electron candidates, two high 
energy photons and large \met\ (see Fig.~\ref{f1}(a)). Of particular note was 
that the \met\ was 55 GeV
 and that the event could not be readily explained as a $W\to e\nu$, $Z\to ee$
 or radiative versions of any combination of the above. 
There were no searches for 
this type of event at the time, and while the large \met\ was suggestive  of SUSY, there were no 
models that were in favor that had photons in the final state. 

\begin{figure}[htb]
\begin{center}
\includegraphics[height=5.5cm]{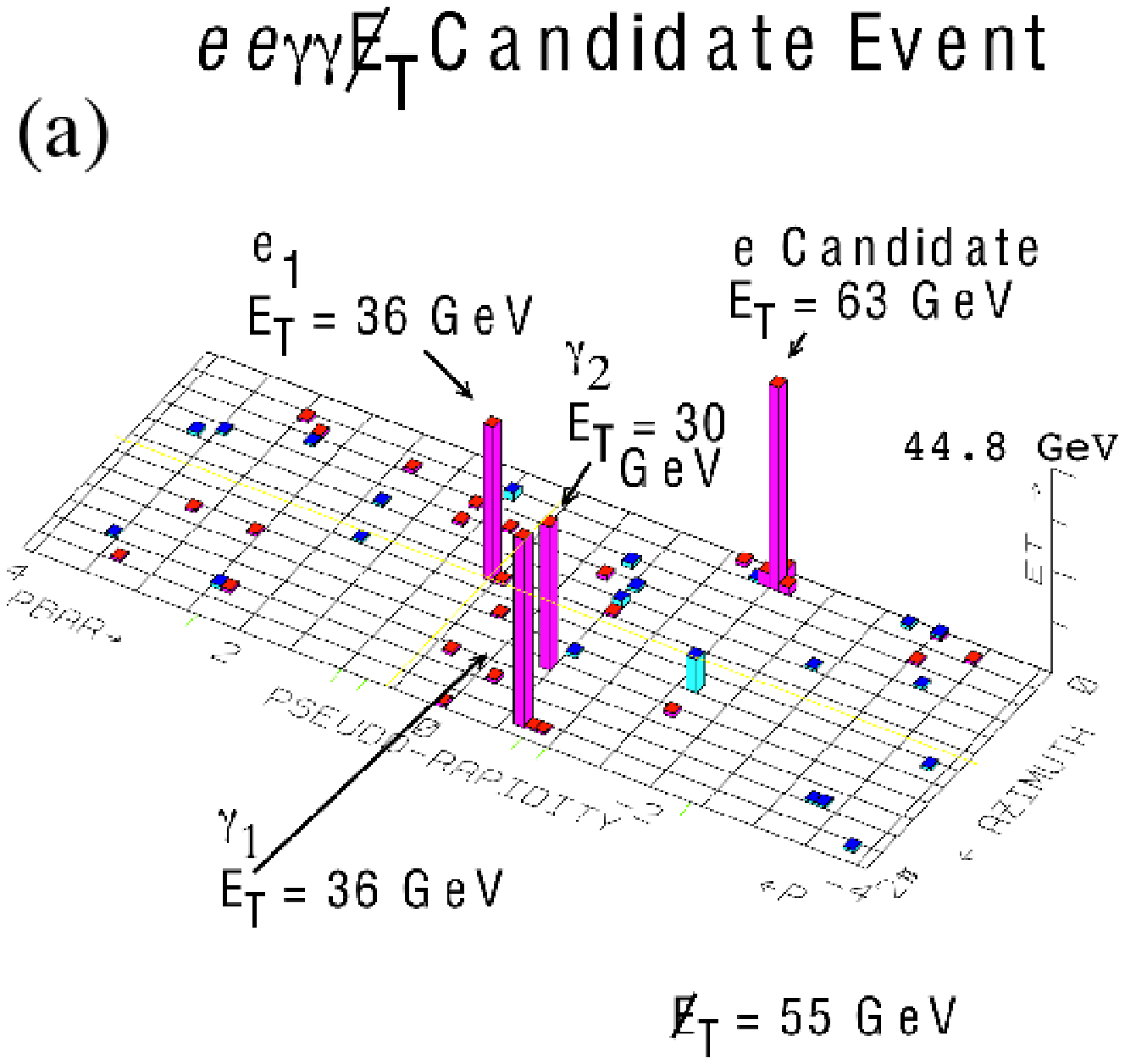}\hfill
\includegraphics[height=5.5cm]{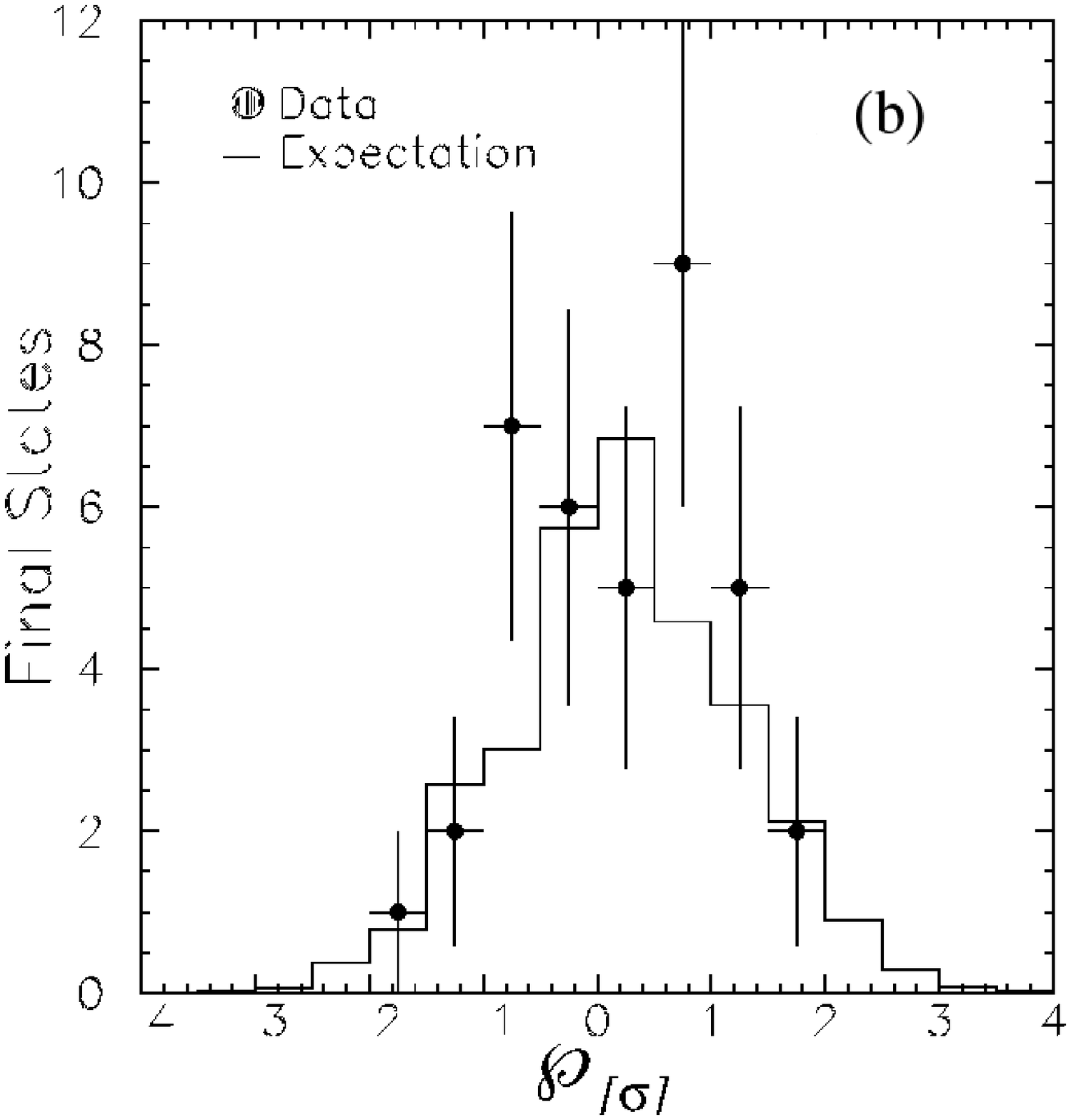}
\end{center}
\caption{(color online) (a) An event display of the CDF
\eeggMet\ candidate event observed in Run~I. (b) The
significance ${\cal P}$ of the excess, in units of standard deviations, 
obtained using \Sleuth\ at the D0 experiment from Run~I. \label{f1}}
\end{figure}

While a detailed description of the set of models which were proposed to explain it is 
beyond the scope of this review, a long lasting impact was the rise of interest in  
GMSB SUSY. 
Example production and decay chains include 
${\tilde e}{\tilde e} \rightarrow (e\chioneO)(e\chioneO) \rightarrow e(\gamma\Grav)e(\gamma\Grav) \rightarrow ee\gamma\gamma\met$, 
or similar 
ones from chargino pair production and decay with virtual $W$ bosons\cite{Martin:1997ns}.
GMSB 
has been a popular hunting ground ever since
although no other hint for GMSB or other versions of SUSY were found in 
Run~I~\cite{PhysRevD.64.092002,PhysRevLett.81.1791,PhysRevLett.78.2070,PhysRevLett.80.442}. 
Run~II searches for GMSB are described in section~\ref{sec_susy2}.

The second major thing that came out of this observation was the 
clear need to be on the lookout for hints of new particles using more model--independent methods; 
if this event was an example of a new particle decay,
then it becomes natural to speculate about what kind of 
particles 
produced it and search for other events ``like it'' in the hopes of providing evidence one way or the other.  
Unbiased follow up was 
difficult because, since there was no \textit{a priori} search for this event, \textit{a posteriori}
methods had to be determined.
The simplest quasi--model--independent search method 
used the idea that this event could have been produced by anomalous 
$WW\gamma\gamma$ production and decay 
(SM $WW\gamma\gamma\to e\nu e\nu\gamma\gamma\to\eeggmet$ 
production
 and decay was the dominant background to this event type with $10^{-6}$ events expected). 
The signature--based way to look for 
this type of production is to 
consider all $\gamma\gamma$ events and search each for evidence of associated $WW$ production and decay,
 for example in the 
 $WW\gamma\gamma\to(jj)(jj)\gamma\gamma$ final state. 
No excess 
in this or other $\gamma\gamma$ or $\ell+\gamma+\met$
searches~\cite{PhysRevD.64.092002,PhysRevLett.81.1791,PhysRevLett.78.2070} turned up any further indications of new particles. 
Other, more model--dependent, but still signature--based searches\cite{PhysRevD.66.012004,PhysRevLett.89.041802,PhysRevD.65.052006}
 also found no evidence of new physics in Run~I or Run~II.
 Ultimately, it was recognized that new, {\it a priori} methods of 
 finding and following up on  interesting events needed to found, and developed in ways that avoid potential biases. 
Model--independent and signature--based searches, 
in particular \Sleuth, which is discussed bellow, arose at the D0 experiment in Run~I
 for these reasons
 (see sections~\ref{sec_sleuth} and~\ref{sec_sbmi}). 
 Ultimately, it is not clear what was the source
of the event (perhaps it was a very unusual example of whatever it was), but the
legacy of this event is still with us.

\subsection{\textbf{\textit{Follow up on the leptoquark hints from DESY}}}

In 1997 the H1\cite{Adloff:1997fg} and the ZEUS\cite{Breitweg:1997ff}  experiments at the DESY $ep$ collider (HERA)
reported an excess of events at high momentum transfer $Q^2$ with a potential explanation being the  
production of a single first generation
scalar $LQ$ with a mass of around 200~GeV. By that time, the D0 and the CDF experiments 
 had already excluded scalar $LQ$ masses of up to 
130~GeV\cite{PhysRevD.48.R3939,PhysRevLett.72.965,PhysRevLett.75.1012,PhysRevLett.75.3618}.
Both experiments quickly followed up on these hints in the same final state, but with $LQ$ pair production and decay
and they were able to 
exclude first generation scalar 
leptoquarks in the simplest models above 200 GeV.
They were then expanded to second and third generation searches,
and from $LQ\to\ell q$ to include $LQ\to\nu q'$ modes. 
 Limits were set in both scalar and vector resonances. 
Ultimately, all the results were found to be consistent with the 
SM\cite{PhysRevLett.79.4327, PhysRevLett.79.4321, PhysRevLett.80.2051, PhysRevD.64.092004,
PhysRevLett.81.4806,PhysRevLett.85.2056, PhysRevLett.83.2896, PhysRevLett.84.2088,  PhysRevLett.78.2906, 
PhysRevLett.82.3206,PhysRevLett.81.38,PhysRevLett.81.5742,PhysRevLett.88.191801}. 
These searches were extended in 
Run~II, again with null results  (see section~\ref{sec_reson3}). 

\subsection{\textbf{\textit{Signature--based searches and SLEUTH}}}\label{sec_sleuth}

Signature--based searches emerged at the end of Run~I. In each the analysis selection 
criteria are established before doing the search using systematic ways to 
 separate any data event into a unique group based on its final 
state signature; specifically based on the 
 set of final state particle objects. 
 For example, those objects
passing  standardized 
lepton, photon, \met, jet, $b$--tagging ID requirements, and above various $p_T$ thresholds
are selected.
With a clear definition of all event requirements 
this allows for
 definite predictions of the
rates and kinematic properties of 
 events from SM background processes.
 Note that there is no prediction of what new physics might arise, just a comparison to the SM--only hypothesis,
  and, consequently, there is nothing that can be optimized for sensitivity. As previously 
  mentioned, many searches were done in Run~I and Run~II which followed this methodology. 

The major leap forward in this area was the development of the 
quasi--model--independent \Sleuth\
formalism at the D0 experiment\cite{PhysRevD.62.092004}.  
\Sleuth\ 
traded the ability to optimize for a particular model of new 
physics, for breadth in covering previously unsearched territory.
By looking for excesses on the tails of distributions 
(with a bias towards the large $Q^2$ interactions as it is more likely that new physics has a large scale or
 mass as the lower scales and masses are already well probed) \Sleuth\ looked for regions in the data that were not well 
 described by the 
SM--only 
 background predictions. It made a novel use of pseudoexperiments 
 (and was a powerful early user of these methods) to quantify how unusual the largest 
 observed deviation was. As a test,
\Sleuth\ was able to show that it could discover $WW$ and top--quark production in the dilepton final state
 at many standard deviations (s.d.) in 
the case that neither were included in the background modeling. It was the same with 
leptoquarks, at a certain production level and mass, in lepton+jets. 
 Ultimately, \Sleuth\ was 
run on $\sim\!\!50$ final states at the D0 experiment and compared the fluctuations to 
expectations\cite{PhysRevD.64.012004,Abbott:2001ke} (see Fig.~\ref{f1}(b)). 
The distribution of the fluctuations were consistent with statistical
expectations.
This methodology was eventually adopted by other experiments,
 for example at the HERA experiments~\cite{Aktas:2004pz}, 
and the CDF experiment in Run~II  where it was extended for the other types of systematic, model--independent search strategies
 (described in section~\ref{sec_sbmi}). 
This methodology is currently less used 
 for a number of reasons, primarily that 
 it is not always clear how to quantify
 the search sensitivity to new physics.

\section{\textbf{Supersymmetry}}\label{sec_susy}

In this section we focus on the searches for SUSY during Run~II. We note that
there are good reasons to expect to
 find sparticles at Tevatron energies, in particular 
since the Higgs
mass would potentially diverge if there are no sparticles 
(most importantly the top squark, or stop for short) with a mass at or below the TeV scale\cite{Martin:1997ns}.
Until the observation of the \eeggMet\ candidate event in Run~I, most analyses
focused on mSUGRA--type models
with a hierarchy of heavy colored states and a light LSP
as a dark matter candidate
 and its
smoking--gun signature of large \met.
 We will focus on mSUGRA in section~\ref{sec_susy1}.
In section~\ref{sec_susy2} we will discuss GMSB SUSY searches with
their smoking gun final states of photons and \met\ from light gravitinos.
Finally, in section \ref{sec_susy3} we will discuss $R$--parity violating (RPV) scenarios.
Ultimately, most of these results are from the first half of Run~II data
as the turn on of the LHC, with its larger energy and production cross
sections and comparable luminosities, eclipsed the Fermilab Tevatron collider sensitivity and obviated the need
to get results using the full dataset. Other searches with 
SUSY interpretations, such as hidden--valley models, and other long--lived particles like CHAMPS, are described in section~\ref{sec_nonsusy}.
 Other searches, like $B_s \to \mu\mu$, which have important SUSY interpretations are 
found in the heavy flavor chapter of this review\cite{BPhysReview}.

\subsection{\textbf{\textit{mSUGRA/Heavy LSP models}}}\label{sec_susy1}

While mSUGRA models have many 
theoretical advantages, from an experimental standpoint they are valued because
they simplify the 128 parameter model down to four 
parameters\footnote{The four mSUGRA  parameters are: 
(i) $m_0$ -- common mass parameter of scalars (squarks, sleptons, Higgs bosons) at the GUT scale;
(ii) $m_{1/2}$ -- common mass of gauginos and higgsinos  at the GUT scale;
(iii) $A_0$ -- common trilinear coupling; and
(iv) \tanBeta\ -- ratio of Higgs boson vacuum expectation values.} 
 and a sign\footnote{$\rm{sign}(\mu)=\pm1$ -- sign of $\mu$ SUSY conserving higgsino mass parameter.} 
which specify the
sparticle masses and decay products. 
Equally valuable is that 
large chunks of the regions turn out to be 
qualitatively similar and much of the 
 non--excluded parameter space
has the lightest neutralino, $\chioneO$, as the LSP, which  
  provides a cold--dark matter 
candidate. 
An important difference is between low  and high \tanBeta; at
low \tanBeta\ the sparticles from all three generations
are degenerate (or nearly degenerate) in mass, while in models with 
high \tanBeta, the third generation sparticles  (stops, sbottoms and staus) can be
much lighter than all other generations, leading to final states enriched with $\tau$--leptons and/or $b$--quarks.

While the couplings of the sparticles to their SM counterparts is important, perhaps the most important issue in 
the production of sparticles at the Fermilab Tevatron collider is their masses. If the 
colored objects (squarks and gluinos) are light enough, their production cross sections will dominate; 
if they are too heavy, they cannot be produced in significant enough quantities to be seen. 
Typically the gauginos are much lighter in mass so an important region of parameter space is the case where the 
colored objects are out of reach and 
gaugino pair 
production dominates the overall sparticle production. 
These two cases are again separated by high and low values of \tanBeta.

\subsubsection{\textit{Light flavor squarks and gluinos}}

If the masses of the gluinos or the first and second generation squarks 
are favourable, then the large production rate of these sparticles provides
a golden channel for the search for SUSY.
Squarks and gluinos are expected to be  produced in pairs, $\tilde{g}\tilde{g}$, $\tilde{g}\tilde{q}$ and $\tilde{q}\tilde{q}$,
and then to decay via 
 $\sqdec$ and $\gldec$. Each will lead to final states with jets and large 
 \met, with the number of jets depending on the 
whether the squark or gluino is  
 heavier. Alternatively, 
 leptonic decay modes of squarks from gluino pair production
(e.g.  $\tilde{g}\to q\tilde{q}\to qq\tilde{\chi}^{\pm}_1\to qq\ell\nu\chioneO$) 
  can lead to final states with 
 same--sign leptons, jets  and large \met \cite{PhysRevLett.87.251803}.
 Both the CDF\cite{PhysRevLett.102.121801} and the D0\cite{Abazov:2006bj,Abazov:2007aa} experiments 
 searched for evidence of these particles, with limits 
in different scenarios shown in Fig.~\ref{f2}(a,b)
interpreted as limits on squarks and gluinos on the one hand and in the
$m_0$ vs. $m_{1/2}$ mSUGRA parameter plane on the other. 
For many years these results were the 
most sensitive, with no substantive competition from LEP. 
To be complementary, a new set of high \tanBeta\ searches emerged 
with squarks decaying to 
 jets, $\tau$--leptons and large \met\ in the final state\cite{Abazov:2009rj}. 
 No evidence was observed as shown in Fig.~\ref{f2}(c).
These were, at the time, the world's most sensitive searches. 

\begin{figure}[htb]
\begin{center}
\includegraphics[height=5.cm]{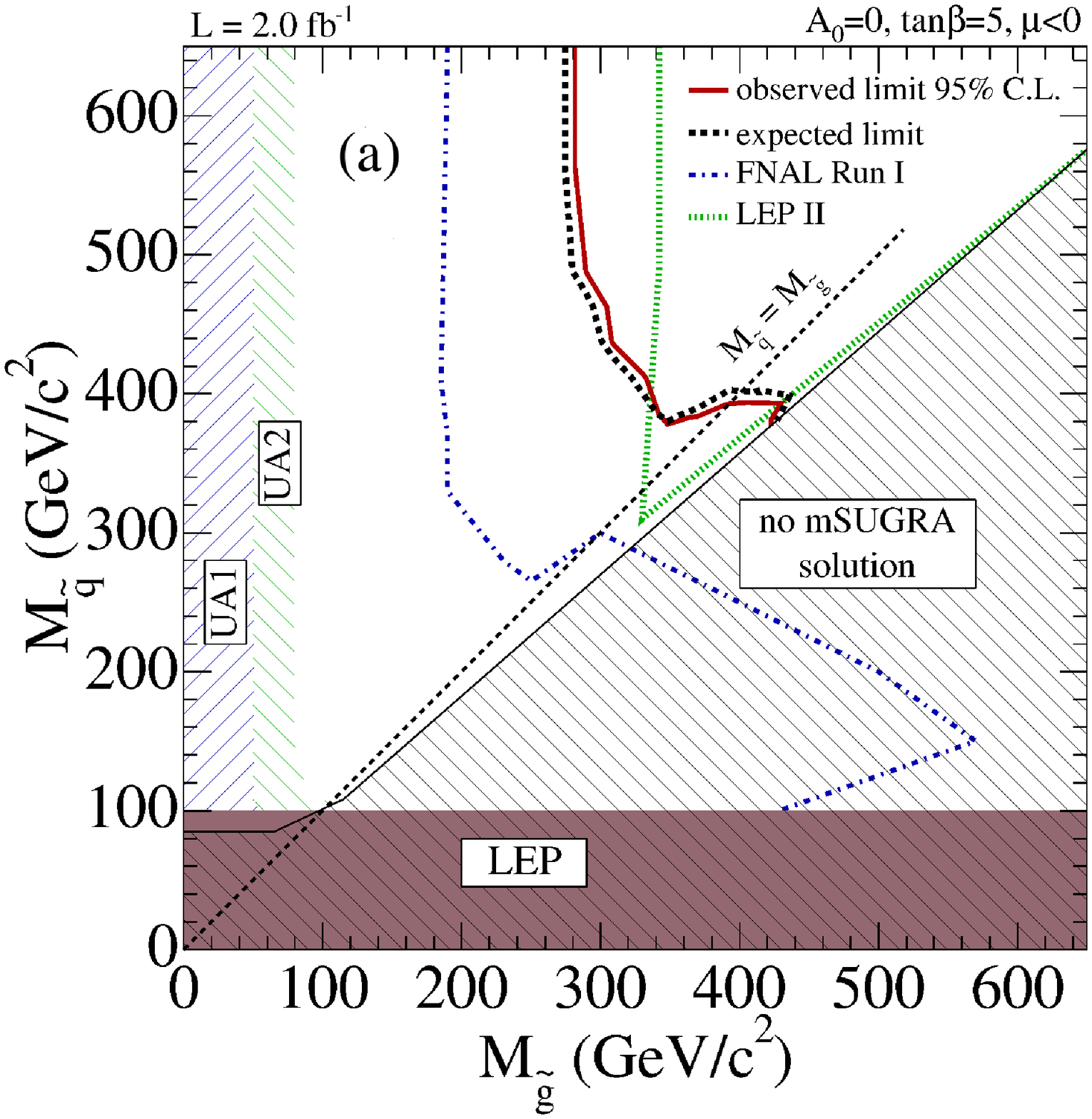}
\includegraphics[height=5.cm]{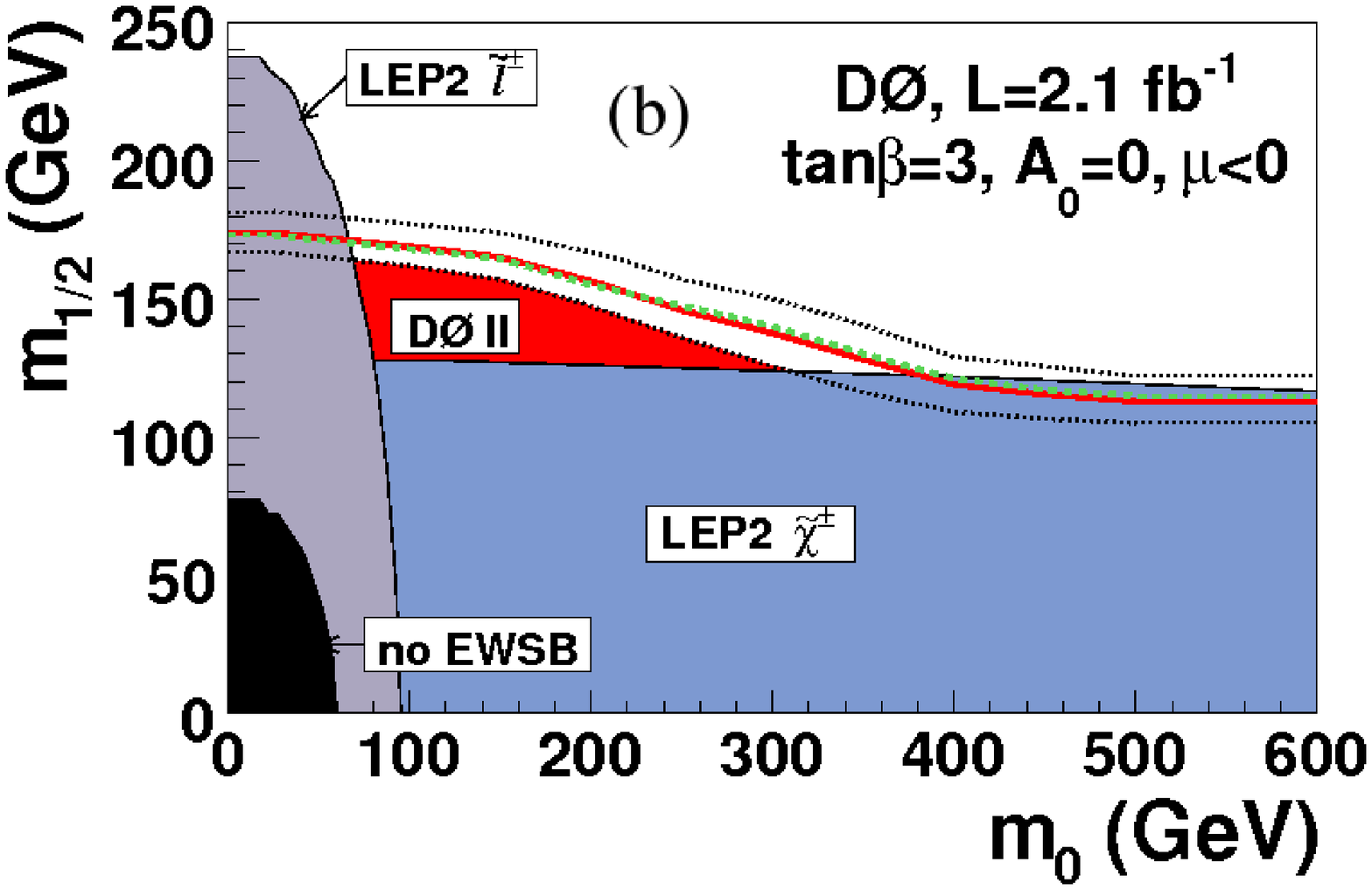}
\includegraphics[height=5.2cm]{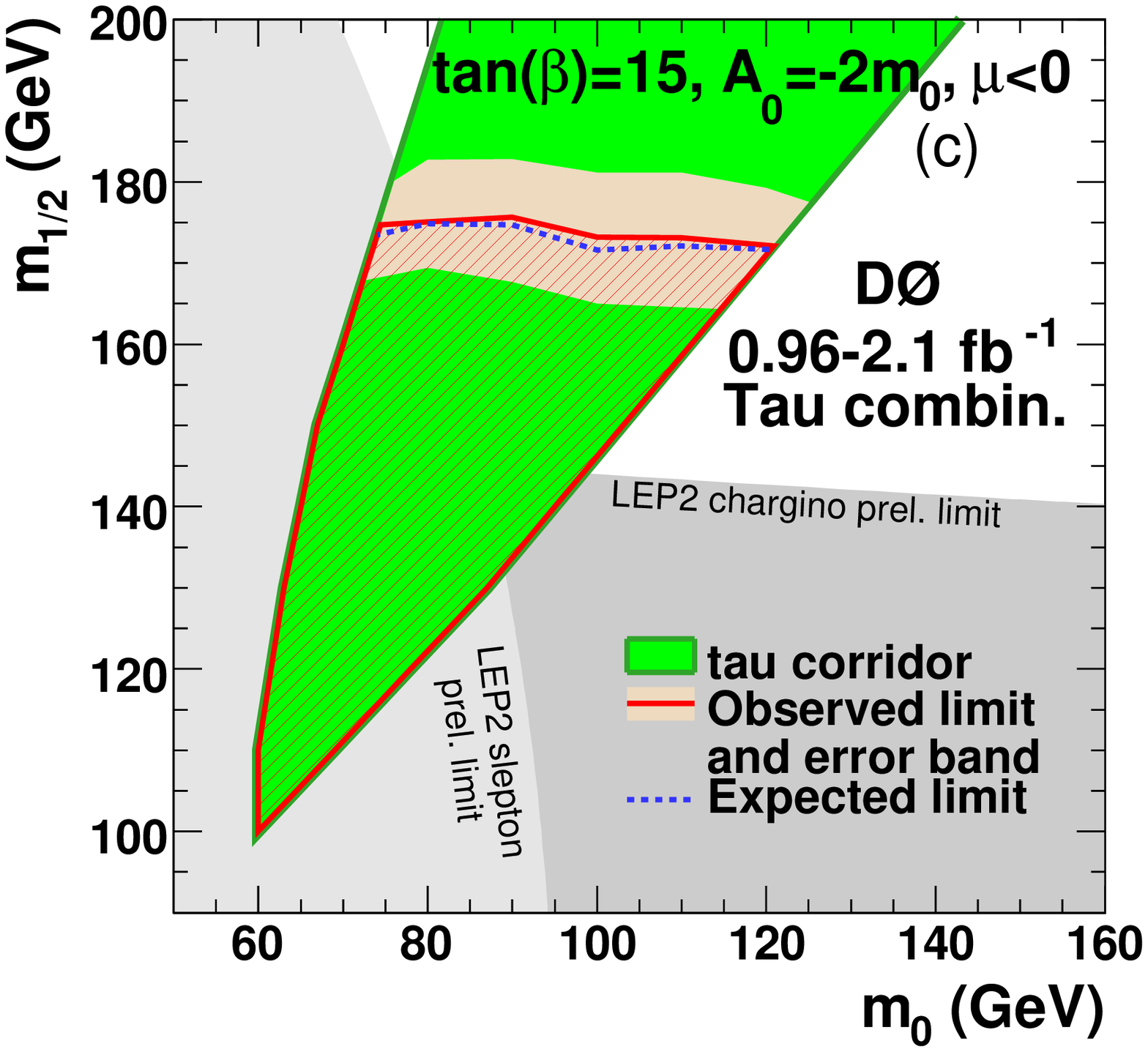}
\end{center}
\caption{(color online) Limits from the search for squarks and gluino 
production in the $\met$+jets final state, including jets for hadronically decaying
$\tau$--lepton. Shown are the 
excluded region (a) in the $m_{\tilde{q}}$ vs. $m_{\tilde{g}}$ plane from the CDF experiment,
and (b) in the $m_{1/2}$ vs. $m_0$ plane from the D0 experiment.
(c) Limits in the $m_{1/2}$ vs. $m_0$ plane from the large \tanBeta\ search with hadronic $\tau$--lepton  combined
with the jets+\met\ search from the D0 experiment.
  \label{f2}}
\end{figure}

\subsubsection{\textit{Bottom and top squarks}}

The large mass difference between the third generation particles in the SM and their lighter counterparts in the first and second 
generations suggests that perhaps the third generation is special. This specialness can manifest itself 
 in SUSY models with high \tanBeta\ which predict that third generation squarks (and/or sleptons) 
are lighter than their first and second generation counterparts
and decay differently, often to third generation particles like $\tau$--leptons and $b$--quarks. 
One strategy to search for bottom squarks
is to use similar analysis techniques for 
 light flavor squarks and gluinos searches in the jets+\met\ final state,
 but with the additional
requirement of $b$--tagging of one or more of the jets.
The simplest is direct sbottom pair production with 
$\tilde{b}\tilde{b}\to b\chioneO b\chioneO\to bb+\met$\cite{PhysRevLett.105.081802, PhysRevLett.97.171806,Abazov:2010wq}.
 Complementary searches can be done where the sbottoms
 are produced as the decay products of a gluino. Specifically, 
 $\tilde{g}\tilde{g}\to b\tilde{b} b\tilde{b}\to bb\chioneO bb\chioneO\to 4b+\met$\cite{PhysRevLett.96.171802,PhysRevLett.102.221801}.
The results are shown in Fig.~\ref{f3}. 

\begin{figure}[htb]
\begin{center}
\includegraphics[height=4.9cm]{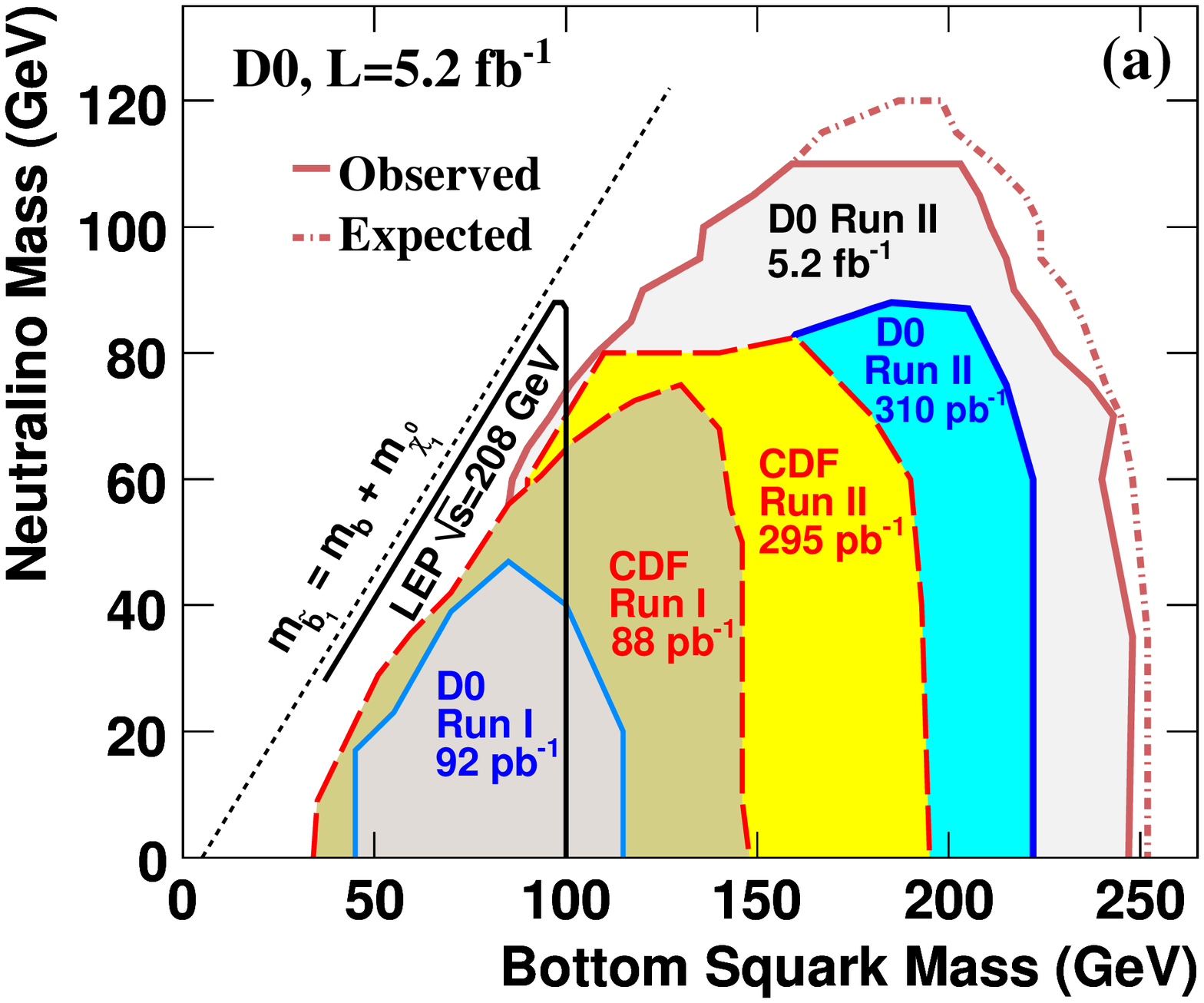}\hfill
\includegraphics[height=5.5cm]{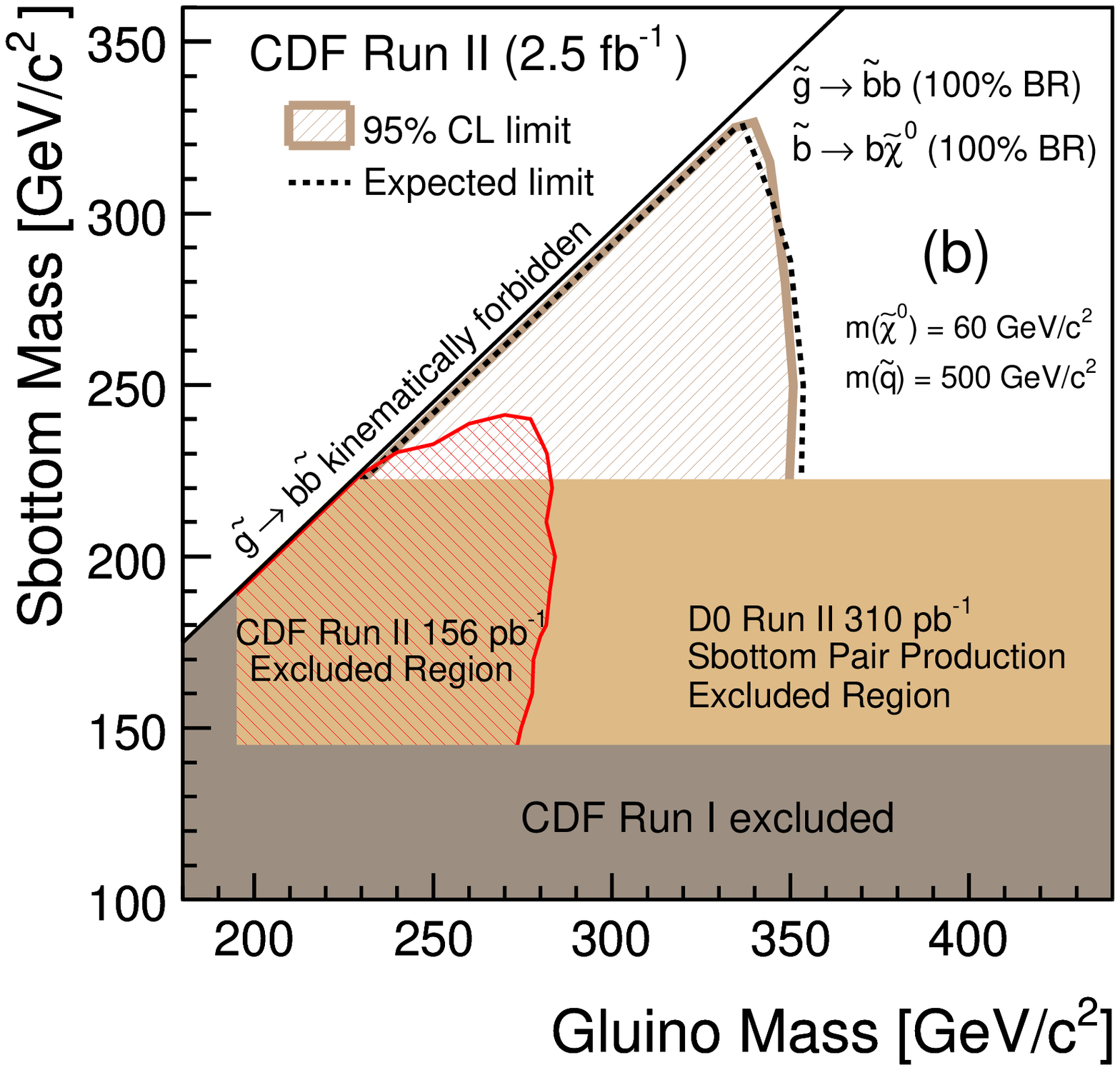}
\end{center}
\caption{(color online) (a) The 
excluded region
 in the $m_{\chioneO}$ vs. $m_{\tilde{b}}$ from the D0 experiment from 
the search for  
 direct sbottom production and decay.
 (b) The exclusions in the 
  $m_{\tilde{b}}$ vs. $m_{\tilde{g}}$ plane from the CDF experiment 
    in the search for sbottoms
 from gluino pair production and decay. \label{f3}}
\end{figure}

In recent years, the case for light
 stops has been motivated by the need
to regularize the Higgs boson mass\cite{Martin:1997ns}.
Many more versions of stop searches
are required because of
 the many possible 
decay modes of the stop. For example, 
the stop can decay via 
charged or neutral modes. In the charged modes, 
 the stop decays via 
$\sctop\to b\tilde{\chi}^{\pm}_1$ 
 and the chargino decays via 
$\nu\tilde{\ell}$,
$\ell\tilde{\nu}$ or
$ bW\chioneO$,
where the $W$ boson is
 either real or virtual depending on the mass differences.
  In all cases, we get $\sctop\to b\ell+\met$, 
but with different kinematics depending on the masses. There are multiple searches in this final 
state~\cite{PhysRevD.82.092001,Abazov2008500,Abazov:2008kz,Abazov:2010xm,Abazov:2012cz,PhysRevLett.104.251801,Abazov:2009ps} 
with results shown in Fig.~\ref{fig_stop}(a,b). In the 
neutral modes, 
$\sctop\to t\chioneO$ or $\sctop\to c+\chioneO$\cite{Aaltonen:2012tq,PhysRevD.76.072010,Abazov2007119,Abazov:2008rc},
there are also a number of searches, again with null results and limits shown in Fig.~\ref{fig_stop}(c).

\begin{figure}[htb]
\begin{center}
\includegraphics[height=6.5cm]{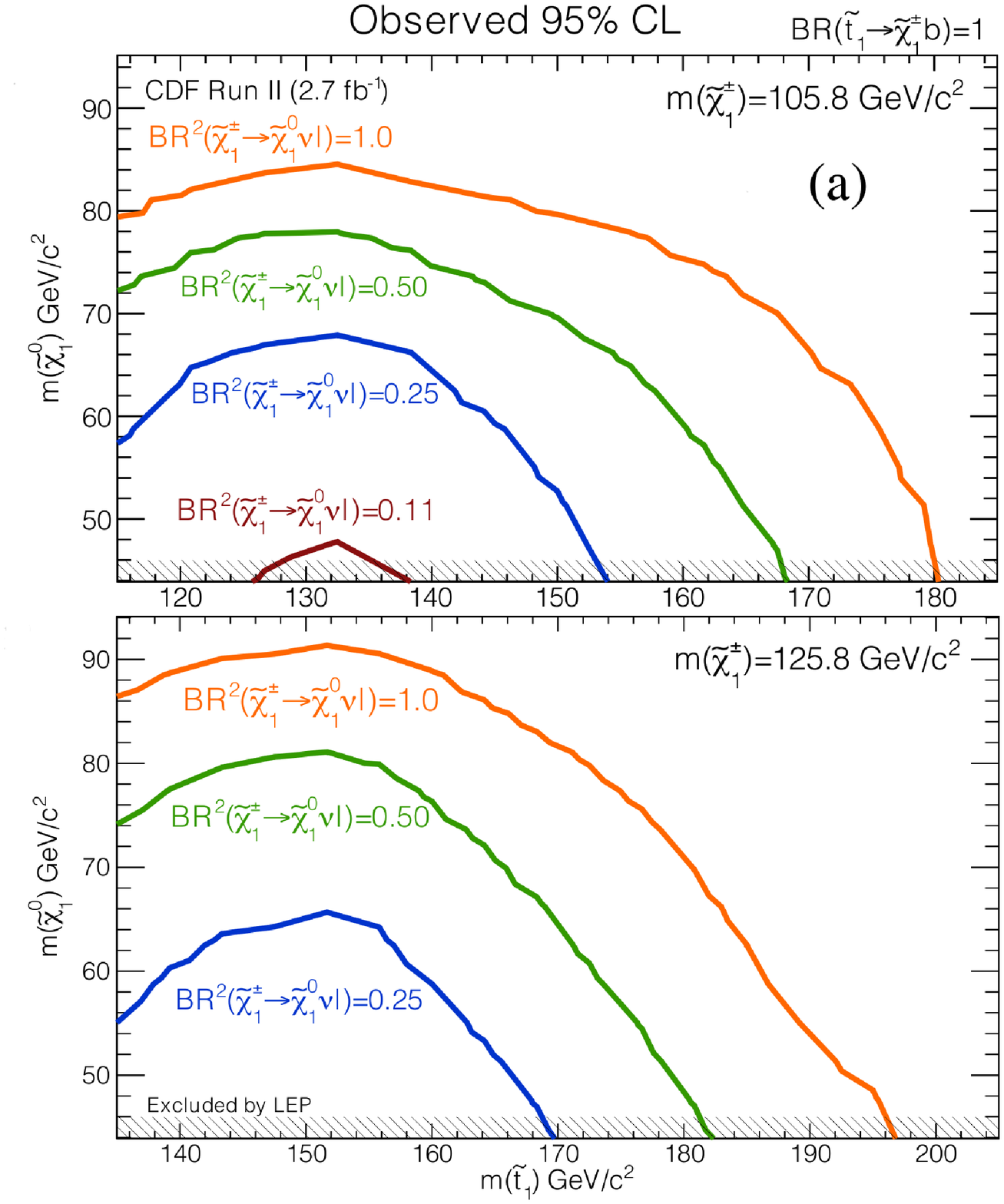}\\
\includegraphics[height=5cm]{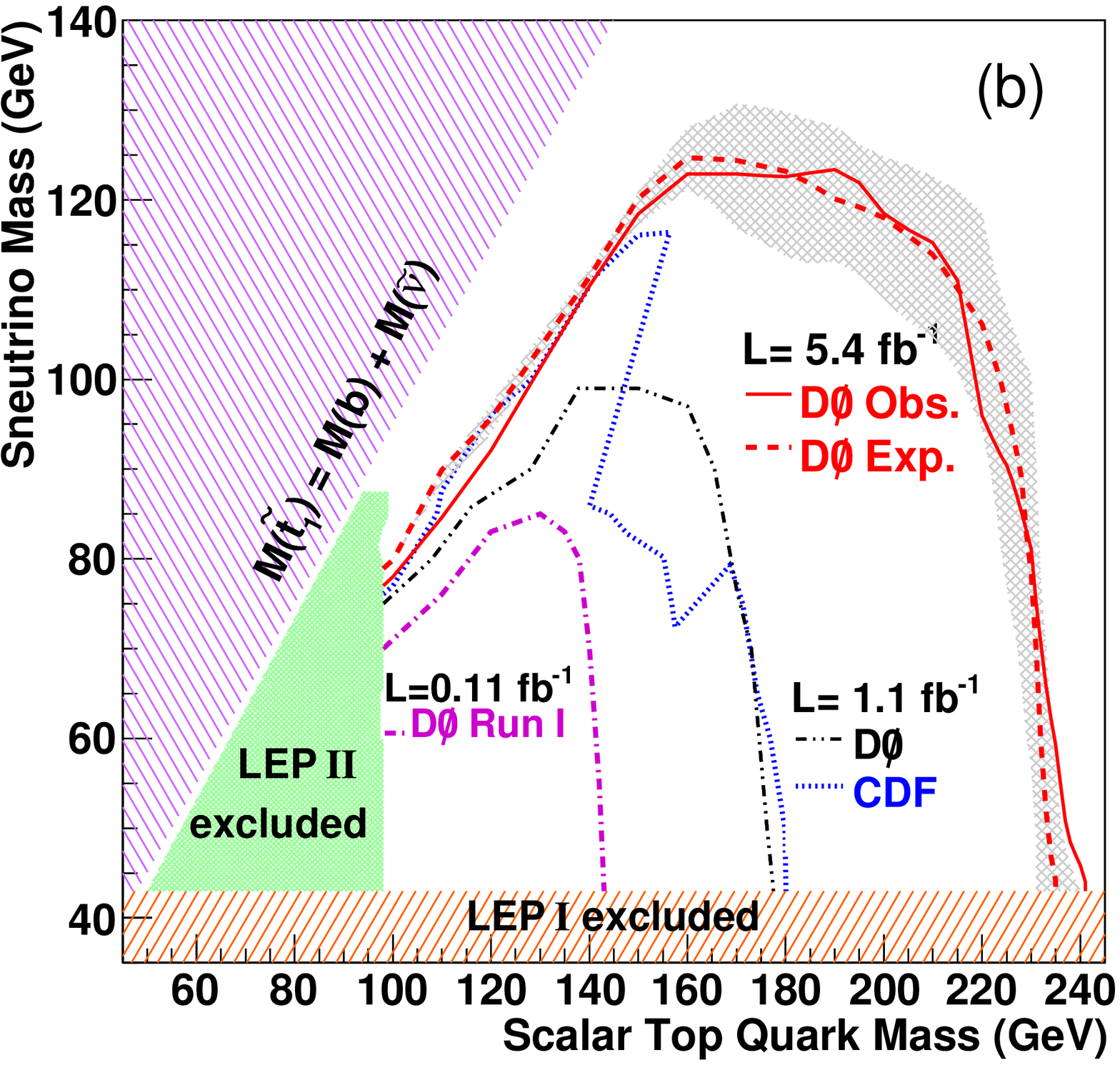}\hspace{1cm}
\includegraphics[height=5.5cm]{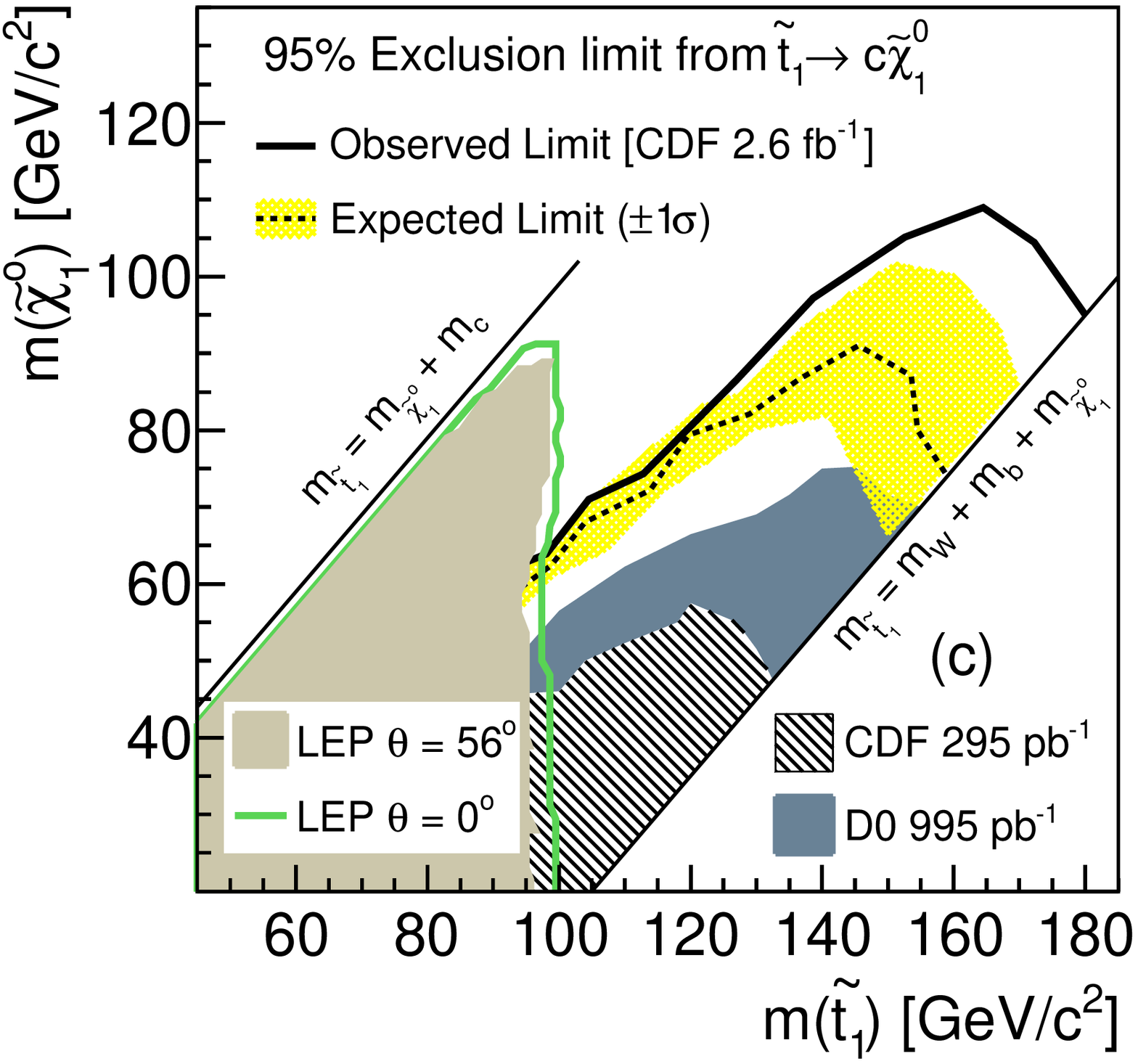}
\end{center}
\caption{(color online) Exclusion regions in the searches for stops. (a) The results 
in the $m_{\chioneO}$ vs. $m_{\sctop}$ plane
 for several values of ${\cal BR}(\tilde{\chi}^{\pm}_1\to\chioneO\ell\nu)$
  and two different chargino masses form the CDF experiment. (b) The results 
in the $m_{\tilde{\nu}}$ vs. $m_{\sctop}$ plane 
 from the  search for the $\sctop\to \tilde{\chi}^{\pm}_1 b$
  through sleptons and sneutrinos
  form the D0 experiment,
and  (c) the results 
in the $m_{\chioneO}$ vs. $m_{\sctop}$ plane
from the search for the $\sctop\to \chioneO c$ 
 from the CDF experiment.\label{fig_stop}}
\end{figure}

\subsubsection{\textit{Gauginos}}

While the squarks and gluinos have the largest production cross section 
(at given mass) it is very possible (and favored in some scenarios) 
that their masses are out of reach of the Fermilab Tevatron collider. Thus, a full set of searches for the lighter sparticles, in particular the lightest
chargino and the next--to--lightest neutralino are crucial. 
The golden final state modes for gaugino pair production and decay is 
 $\ell\ell\ell+\met$
(the trilepton final state)\cite{PhysRevLett.101.251801,Aaltonen:2013vca,Abazov:2009zi,PhysRevLett.95.151805}
or two same--sign leptons + \met\cite{PhysRevLett.98.221803,PhysRevLett.110.201802}
 as there are very few SM backgrounds for each. 
Late in Run~II many of the searches included $\tau$--lepton final states to extend 
the searches to include  higher \tanBeta~\cite{Aaltonen:2013vca}.  
 No evidence of these sparticles have been observed at the CDF or the D0 experiments, with limits shown in 
 Fig.~\ref{fig_gaugino}.
 
 \begin{figure}[htb]
\begin{center}
\includegraphics[height=3.97cm]{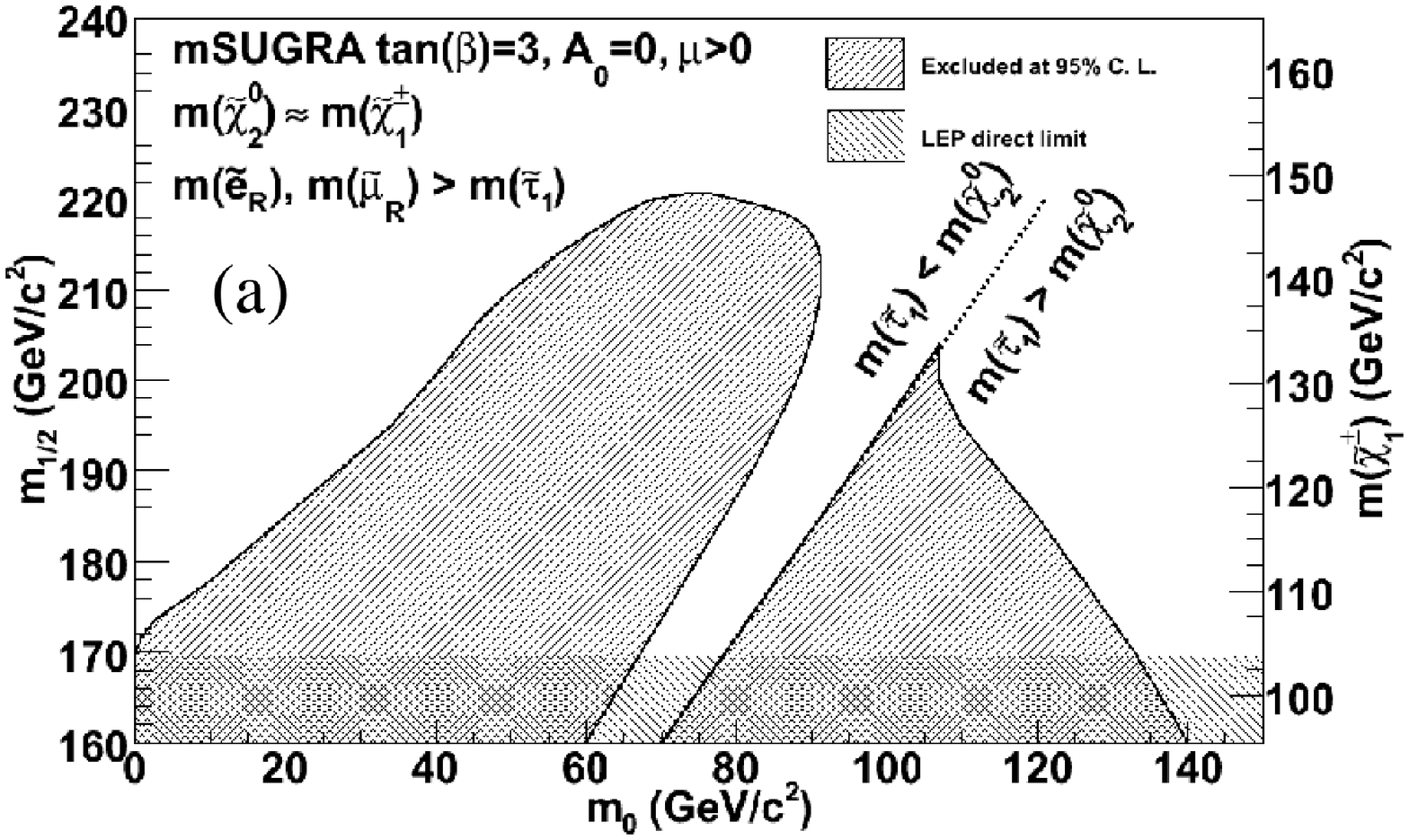}
\includegraphics[height=4cm]{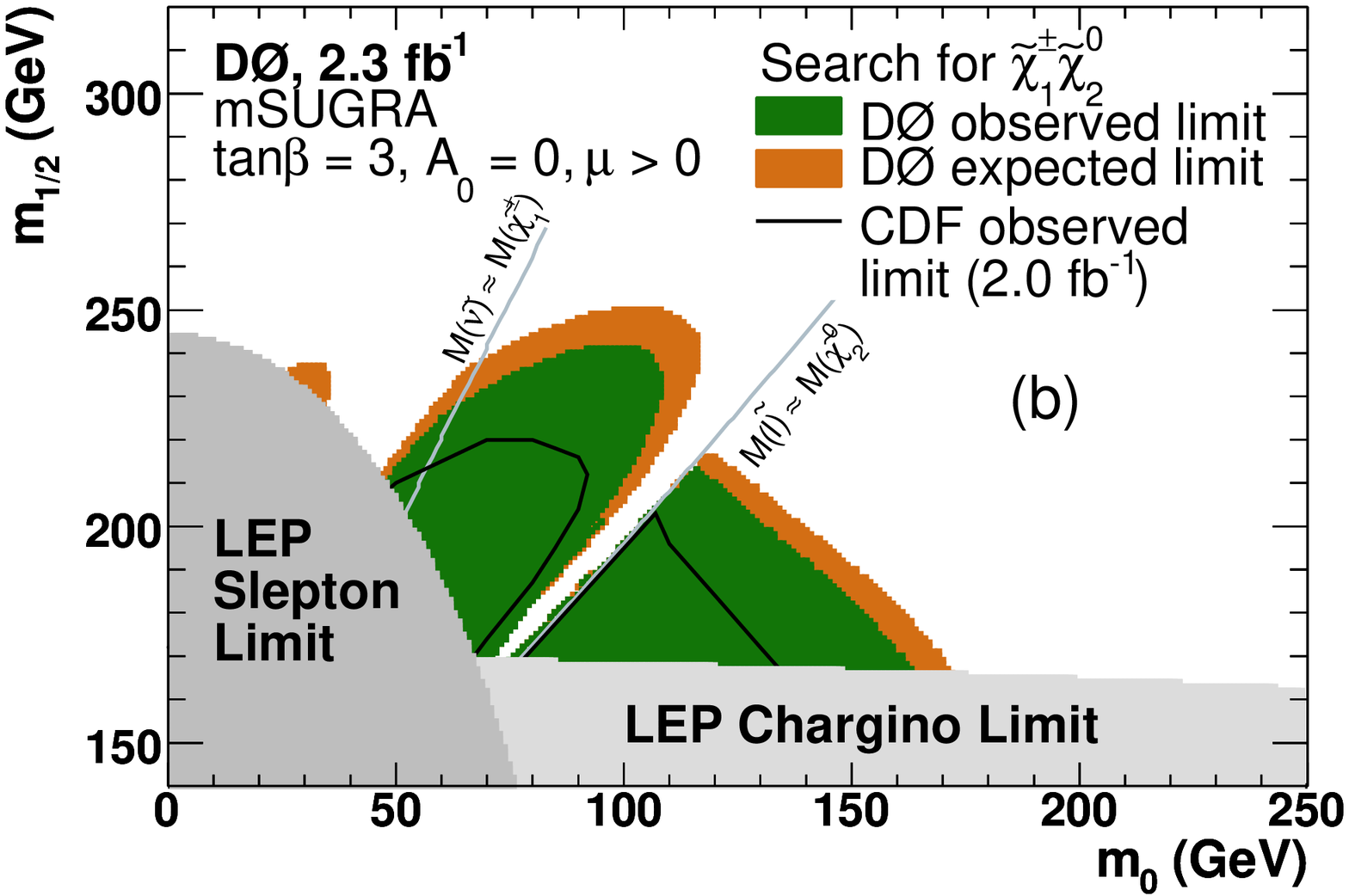}
\end{center}
\caption{(color online) Excluded region in the $m_{1/2}$ vs. $m_0$ plane in the search for charginos and 
neutralinos in the final state with three leptons from (a) the CDF experiment and (b) the D0 experiment. \label{fig_gaugino}}
\end{figure}

\subsection{\textbf{\textit{Gauge--mediated supersymmetry breaking (GMSB) models}}}\label{sec_susy2}

The hallmark of gauge mediated supersymmetry breaking 
models is the light gravitino, $\Grav$, and searches for GMSB are usually (but not always) done 
 in the context of  minimal models, typically using 
the SPS--8 relations\cite{Allanach:2002nj}. This essentially leaves two 
free theory parameters: 
the masses (which typically assume
fixed mass relations) 
and the lifetime ($\tau_{\chioneO}$) of the lightest neutralino which is the next--to--lighest sparticle (NLSP). 
Since the parameter space where
squarks and gluinos are accessible 
 at Tevatron collider energies 
is easily ruled out
(although not done explicitly), searches focus on 
the pair production
of the lightest charginos and neutralinos. 
Through cascade decays, each chargino and/or neutralino typically decays to a $\chioneO$ or $\tilde{\tau}$
accompanied by other high $p_T$ SM 
light particles.
The $\chioneO$ typically decays  to $\gamma\tilde{G}$
(if the mass is low) or to $Z\tilde{G}$ if the mass is large; the $\tilde{\tau}$  decays via 
$\tilde{\tau}\to\tau\tilde{G}$.
In all cases, the 
$\tilde{G}$, like the $\chioneO$
in mSUGRA models, leaves the detector and gives significant \met. The lifetime and
masses of the sparticles dictate the different final states. Specifically, 
in the $\chioneO\to\gamma\tilde{G}$
scenario with $\tau_{\chioneO}\leqslant 1$~ns, both 
$\chioneO$ decay in the detector,  giving a final state of 
$\gamma\gamma+\met+X$. 
For intermediate lifetimes, $1\leqslant \tau_{\chioneO}\leqslant 50$~ns,
frequently one $\chioneO$ 
travels a significant distance in the detector
 before decaying
 and the other 
 leaves the detector without decaying or interacting. 
If this occurs, the 
event will be reconstructed as a $\gamma+\met$ event, where  
the time--of--arrival of the photon at the calorimeter
will be slightly later than ``expected''; these photons are known as 
``delayed  photon'' $\gamma_{\rm{delayed}}$\cite{PhysRevD.70.114032}.
The different lifetime scenarios are considered separately\footnote{For large lifetimes,
both neutralinos can leave the detector and are indistinguishable from mSUGRA
scenarios.}. For the scenario where 
the $\chioneO$ can decay via 
$\chioneO\to Z\tilde{G}$
we can have both $ZZ+\met$ and $Z\gamma+\met$ final states. 
For decays with $\tilde{\tau}$ sleptons as the intermediate sparticle, we can have 
multiple $\tau$--leptons and \met\ in the final state. 

In Run~II, the CDF and the D0 experiments did a full suite of searches for GMSB.
The short--lifetime searches were done in the 
$\gamma\gamma+\met$ final 
state~\cite{PhysRevD.71.031104,PhysRevLett.104.011801,Abazov:2007ag,PhysRevLett.94.041801,PhysRevLett.105.221802} and 
were a natural follow up to the searches done in Run~I  for the \eeggMet\ candidate event.
In the scenarios with the intermediate $\tau_{\chioneO}$, 
the CDF experiment used 
 the electromagnetic calorimeter timing readout system 
 installed in Run~II~\cite{2006NIMPA.565..543G}, 
 and searches were done in the $\gamma_{\rm{delayed}}+\rm{jet}+\met$ final state~\cite{PhysRevD.78.032015,PhysRevLett.99.121801}. 
The D0 experiment
did the first search in $Z\gamma+\met$\cite{PhysRevD.86.071701}
 and the CDF experiment did a search with same--sign $\tau$--leptons+\met\cite{PhysRevLett.110.201802}. 
 No evidence was observed 
and limits are shown in Fig.~\ref{f5}. 
Recently, scenarios with a light neutralino and gravitino (with all other sparticles out of the reach of colliders) have 
been proposed\cite{Mason2011377}, and searches for this final state (without limits) have been done at the CDF experiment, in the exclusive 
$\gamma_{\rm{delayed}}+\met$ with no evidence for new 
physics\cite{PhysRevD.88.031103}. 

\begin{figure}[hbt]
\begin{center}
\includegraphics[height=4.5cm]{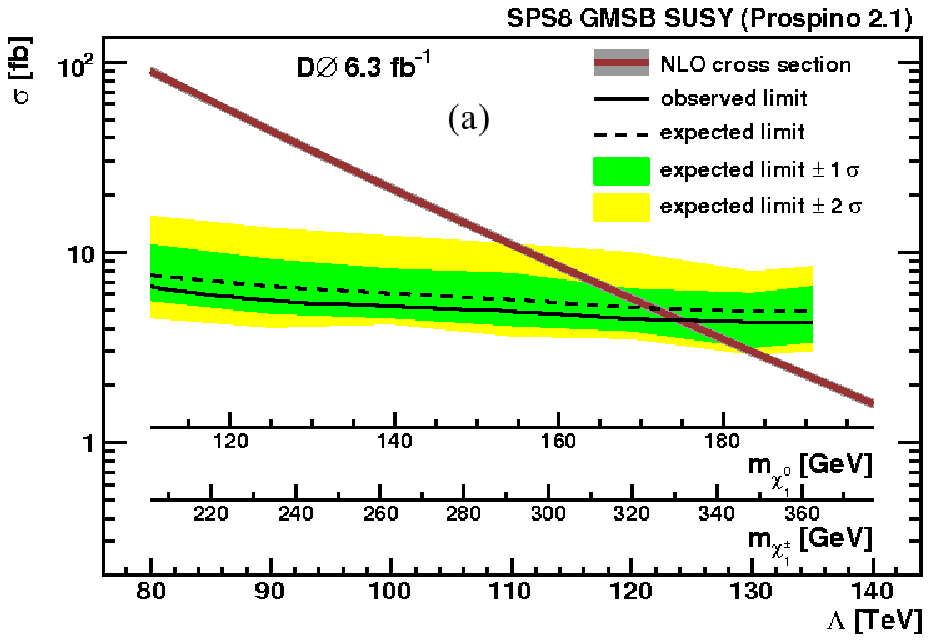}\hfill
\includegraphics[height=4.2cm]{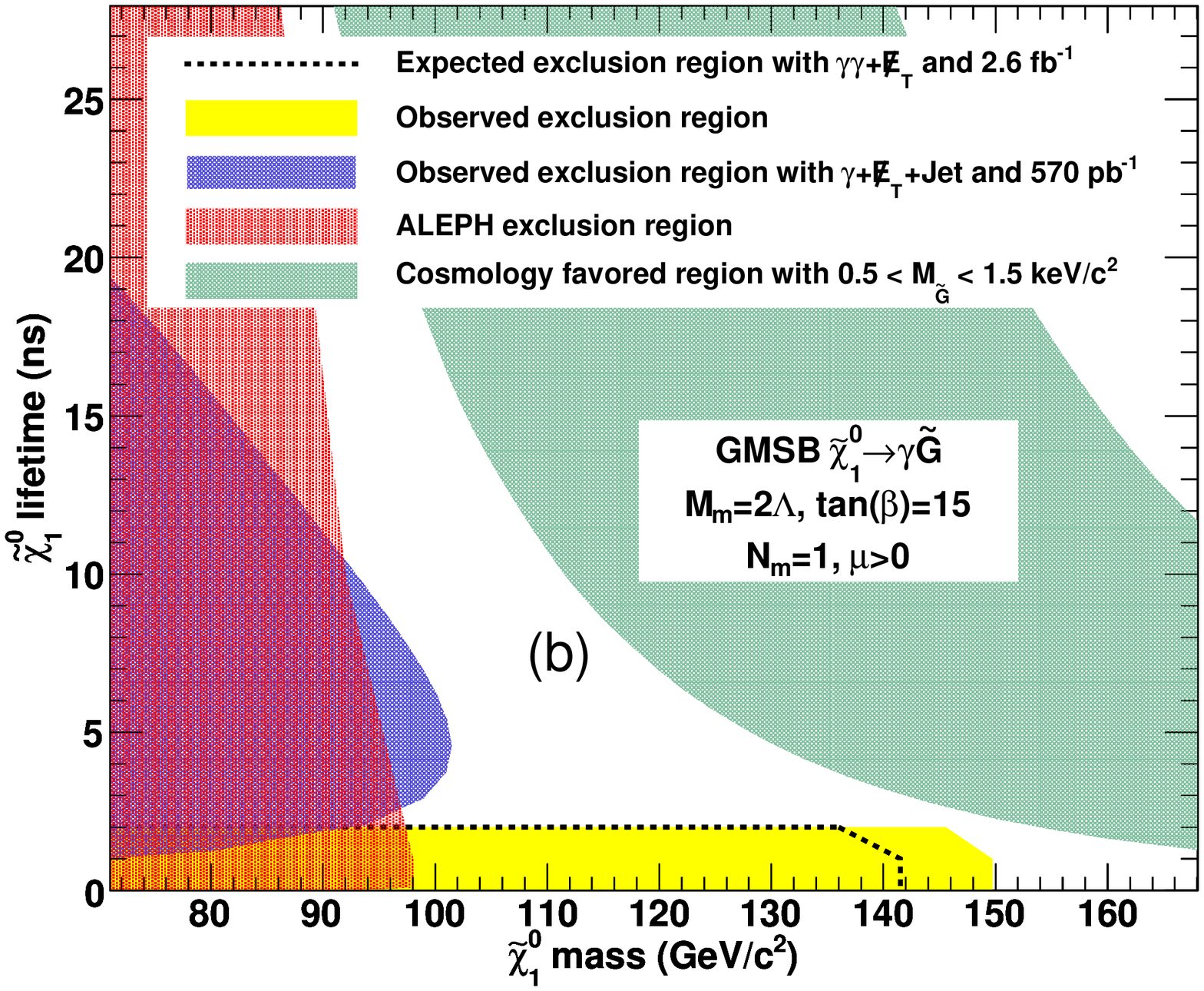}
\includegraphics[height=4.5cm]{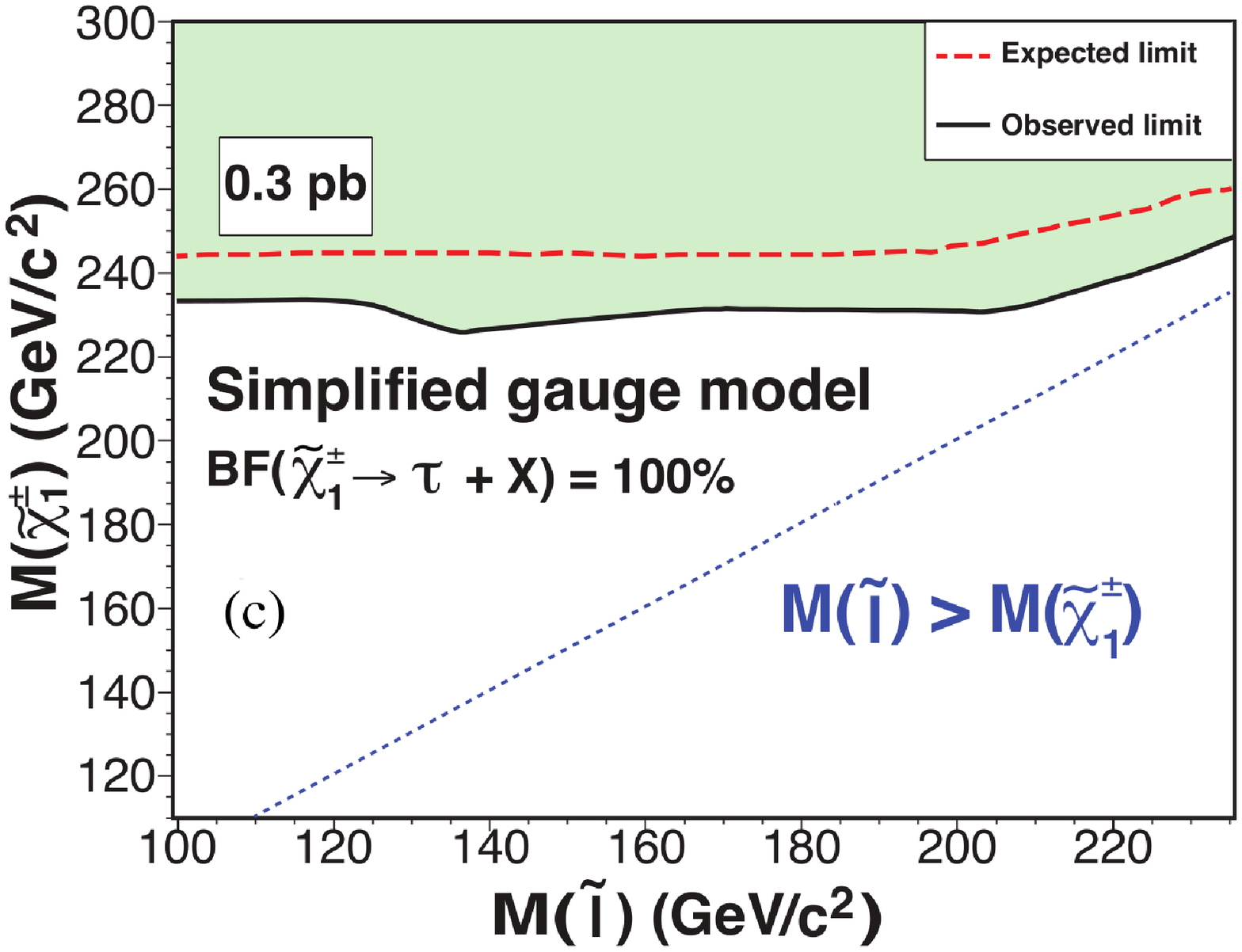}\hfill
\includegraphics[height=4.5cm]{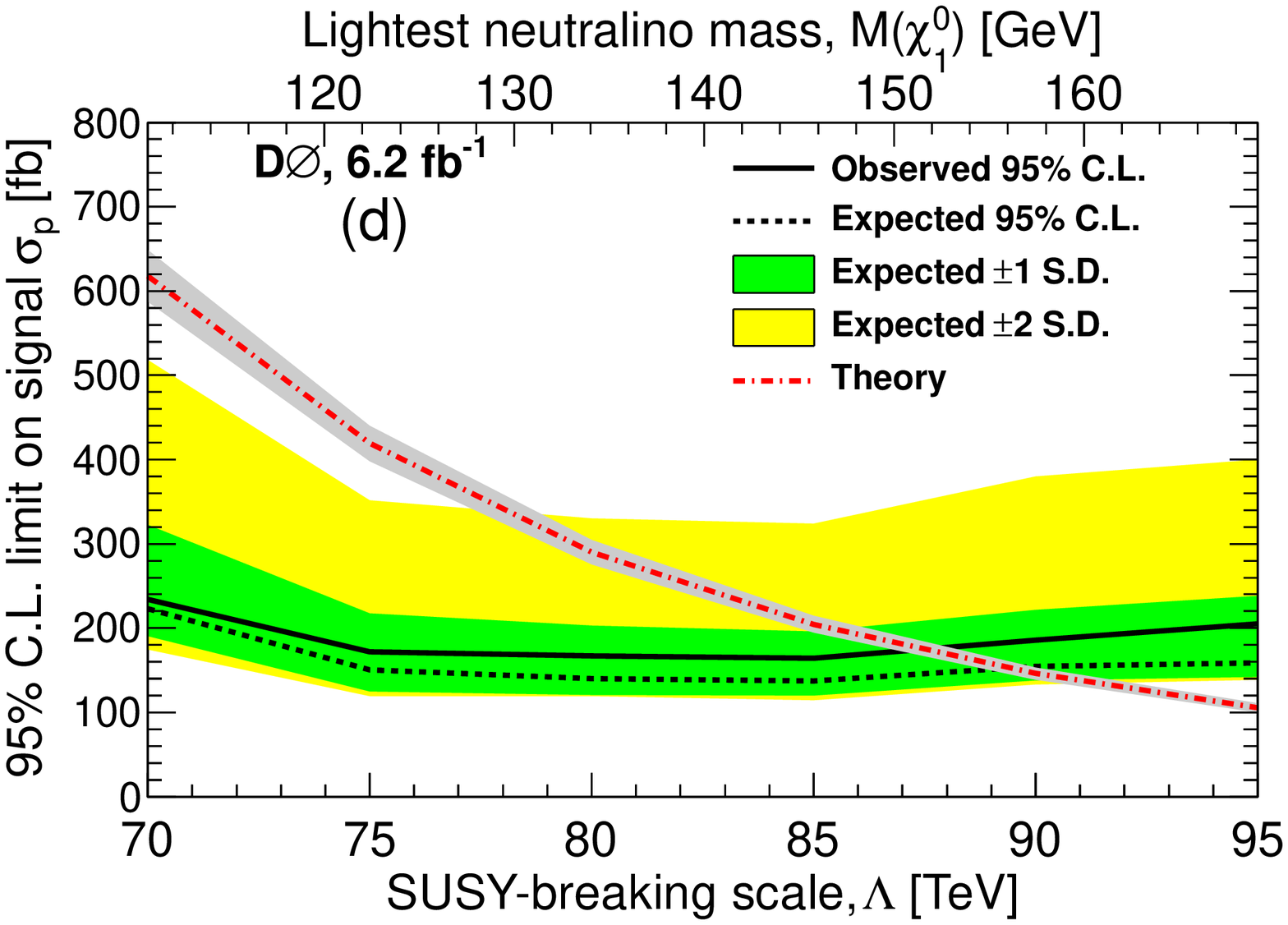}
\end{center}
\caption{(color online) Limits on GMSB scenarios. (a) The 95\% C.L.\ cross section upper limits from the 
 $\gamma\gamma+\met$ final state 
as a function of scale $\Lambda$, $m_{\chioneO}$ and 
$m_{\tilde{\chi}^{\pm}_1}$ from the D0 experiment. 
(b)
The  excluded regions in the $\tau_{\chioneO}$ vs. $m_{\chioneO}$ plane from the CDF experiment from the 
 $\gamma\gamma+\met$ and $\gamma_{\rm{delayed}}+\met+\rm{jet}$  searches.
 (c) The 
 excluded region in the $m_{\tilde{\chi}^{\pm}_1}$ vs. $m_{\tilde{l}}$ 
 plane from the CDF experiment in the search with $\tau$--leptons +\met,
and (d) the 95\% C.L.\ cross section  upper from 
 $Z\gamma+\met$ production as a function of 
$\Lambda$ and $m_{\chioneO}$ from the D0 experiment. 
\label{f5}}
\end{figure}

\subsection{\textbf{\textit{$R$--parity violation}}}\label{sec_susy3}

While one of the 
most attractive features of SUSY is its potential to
solve the dark matter problem, 
there is no inherent requirement for $R$--parity to be  conserved. 
If the restriction that $R$--parity conservation is released, then there is a 
great deal of variety in the new final states allowed from sparticle production.
For pragmatic reasons, efforts focused on
two different modes. The first was on 
pair production and decay of stops via $\sctop\to b\tau$~\cite{PhysRevLett.101.071802}
by the CDF experiment with limits
 shown in Fig.~\ref{fig_rpv}(a). However, with the restriction of $R$--parity removed, 
sparticles are no longer required to be produced in pairs. 
An example signal
of this type is single sneutrino production 
which decays via the
$\tilde{\nu}\to e\mu,e\tau$ or $\mu\tau$ final states.
A variety of searches were done at the CDF\cite{PhysRevLett.96.211802,PhysRevLett.105.191801}
and the D0\cite{PhysRevLett.100.241803,PhysRevLett.105.191802}  experiments.
No new physics was observed and limits were set with the results shown in Fig.~\ref{fig_rpv}(b,c). 
Other searches for RPV\cite{Abazov2006441,PhysRevLett.97.111801} also did not show any excess.
We also note that these same results can be interpreted in terms of other models, for example lepton flavor 
violating $Z'$ boson production and 
decay, and are described in section~\ref{sec_reson1}.

\begin{figure}[htb]
\begin{center}
\includegraphics[height=4.5cm]{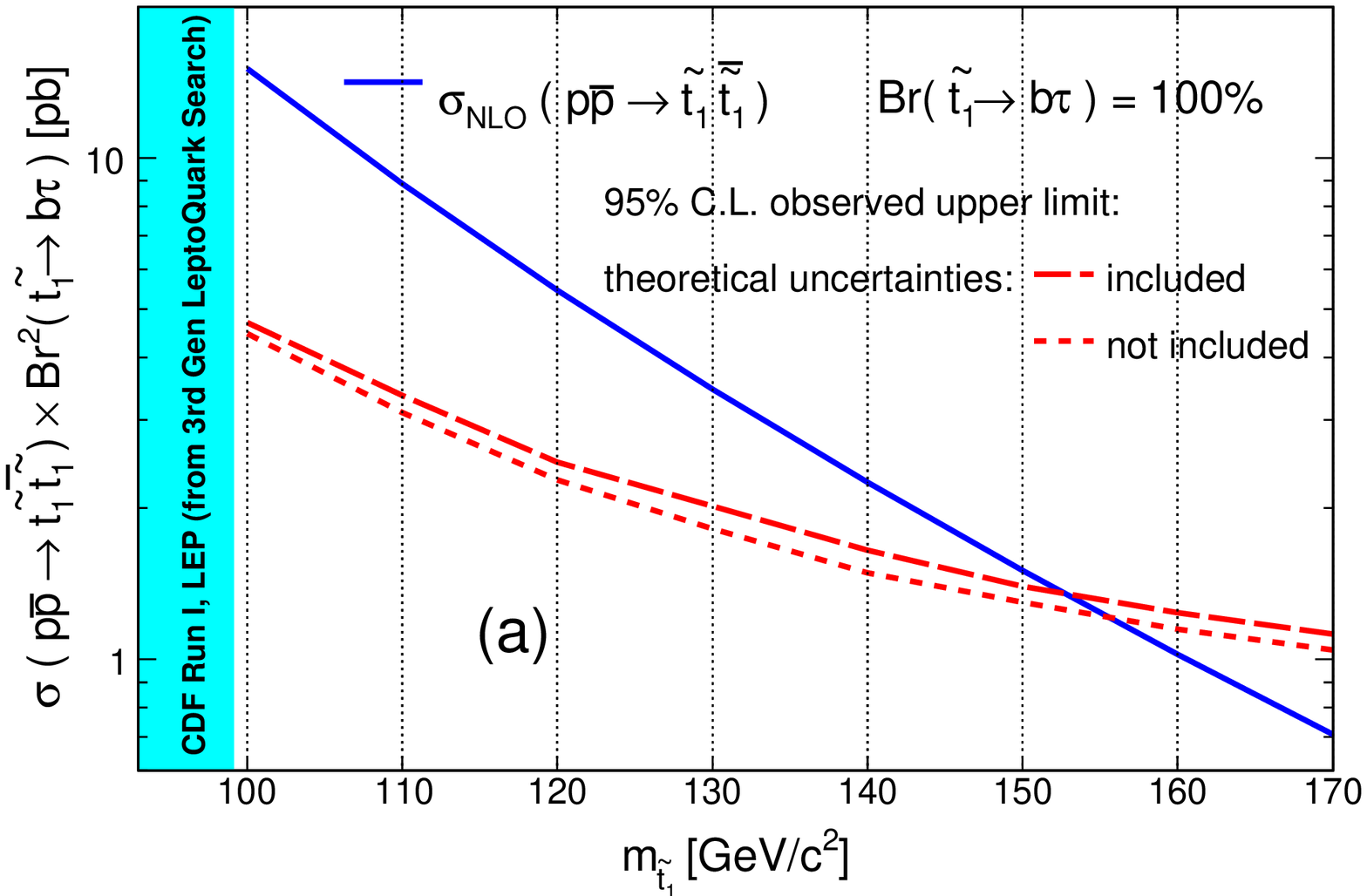}
\includegraphics[width=6cm]{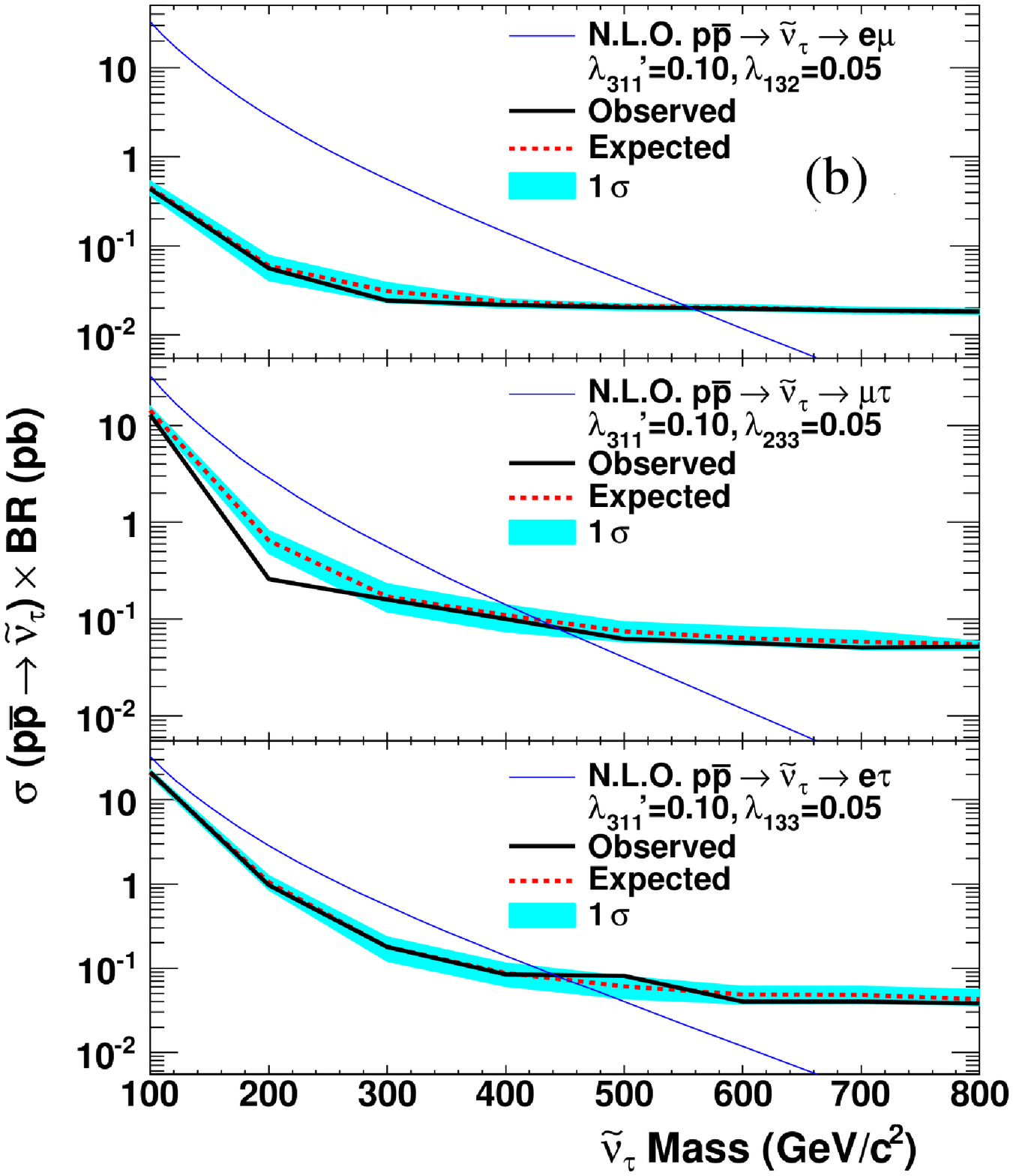}\vspace{0.5cm}
\includegraphics[height=4.5cm]{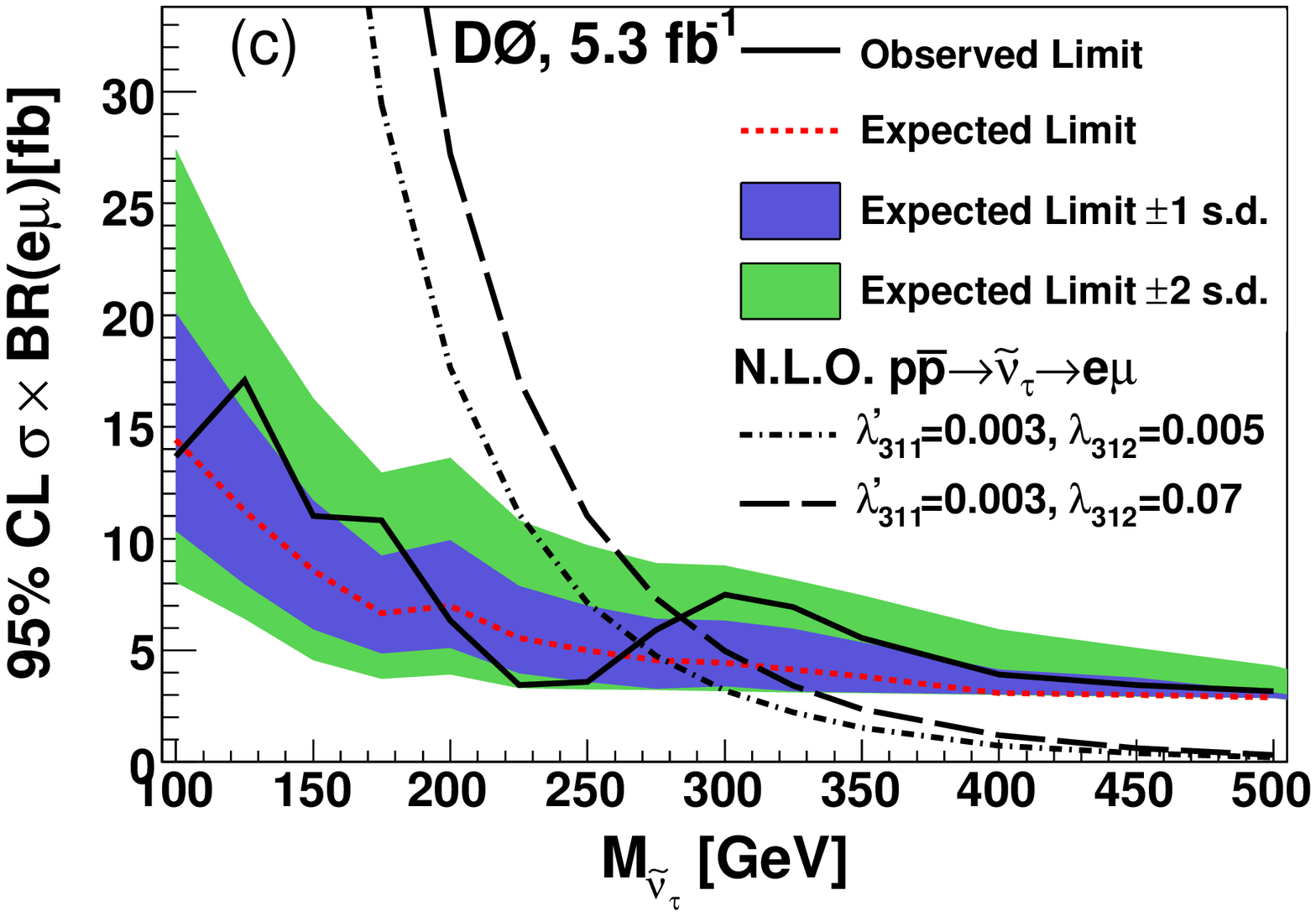}
\end{center}
\caption{(color online) Results from $R$--parity violation SUSY searches. (a) The
95\% C.L. cross section upper limit on pair--production and decay of 
 $\sctop\to b\tau$
from the CDF experiment. 
 Cross section limit results on (b)
$\tilde{\nu}$ in the 
$\tilde{\nu}\to e\mu,e\tau,\mu\tau$ final states 
 from the CDF experiment and (c) 
$\tilde{\nu}\to e\mu$ from the D0 experiment.
\label{fig_rpv}}
\end{figure}

\section{\textbf{Other BSM Searches}}\label{sec_nonsusy}

We next present results on 
both the classic Tevatron searches as well as a number of new types of 
searches that originated after the beginning of Run~II. These include 
resonances, hidden--valley model particles, long--lived particles, extra dimensions, dark matter, 
 as well as 
signature--based and model--independent searches.
Many of these models, like those containing an extended Higgs sector,
 are discussed in more detail 
in the Higgs boson chapter of this review\cite{HiggsReview}, 
 although some are referenced here for completeness.

\subsection{\textbf{\textit{Resonances}}}\label{sec_reson}

One of the primary analysis techniques to search for new particles,
which was developed long before the advent of colliders, 
 is to look for resonances 
in the invariant  mass distribution of two final state particles. This method can be used for a large number of different 
final states
and a signature of this type can arise from
new fermions and gauge bosons, excited fermions, leptoquarks, technicolor particles, and other models.
These results are presented next.

\subsubsection{\textit{New fermions and bosons}}\label{sec_reson1}

While there are many different models that predict  new fermions
 from extending the number of generations in the SM
 the experiments focused on searches for 
different types of heavy quarks that
 decay  to a vector boson, $V=W,~Z$, and 
a SM quark.
The CDF experiment searched for pair production and decay of
 fourth generation $b'$ quarks 
that decay exclusively via $b'\to bZ$\cite{PhysRevD.76.072006}.  
 The analysis was done in the $\ell\ell+3~\rm{jets}$
  final state. No 
significant excess is observed 
and 
$b'$ quarks are excluded with $m_{b'}<268$~GeV at 95\% C.L.\  (see Fig.~\ref{fig-vq}(a)). 
Another analysis by the D0 experiment searches for  
vector--like quarks, $Q$,
in single quark electroweak production in association with SM quarks\cite{PhysRevLett.106.081801}. 
At hadron colliders, electroweak production of
vector--like quarks can be significant, but depends on $m_Q$  and the coupling strength between 
the $Q$ and  
SM quarks, $\tilde{\kappa}_{qQ}$.
Single production and decay of $p\bar{p}\to qQ\to q(Vq)$ can produce an excess of events in the $V+2$~jet final state. 
Limits are set 
as a function of the various model parameters;
for $\tilde{\kappa}_{qQ}=1$ the process $Qq\to Wqq$ is excluded for a mass 
$m_Q<693$~GeV (see Fig.~\ref{fig-vq}(b)), and the process $Qq\to Zqq$
is excluded for a mass $m_Q<449$~GeV at 95\% C.L.
Other searches for fourth generation 
quark pair production 
 include decays to top quarks~\cite{PhysRevLett.104.091801,PhysRevLett.106.141803,PhysRevLett.107.082001,
PhysRevLett.108.211805,PhysRevLett.107.261801,PhysRevLett.106.191801,PhysRevLett.100.161803}. These encompass 
$b'\to tW$, $t'\to Wb$ and $Wq$. Similar searches for a
new heavy particle $T\to t+X$ where $X$ is an invisible particle
found no evidence of new physics (see Fig.~\ref{fig-vq}(c,d)). 

\begin{figure}[htb]
\begin{center}
\includegraphics[height=5.2cm]{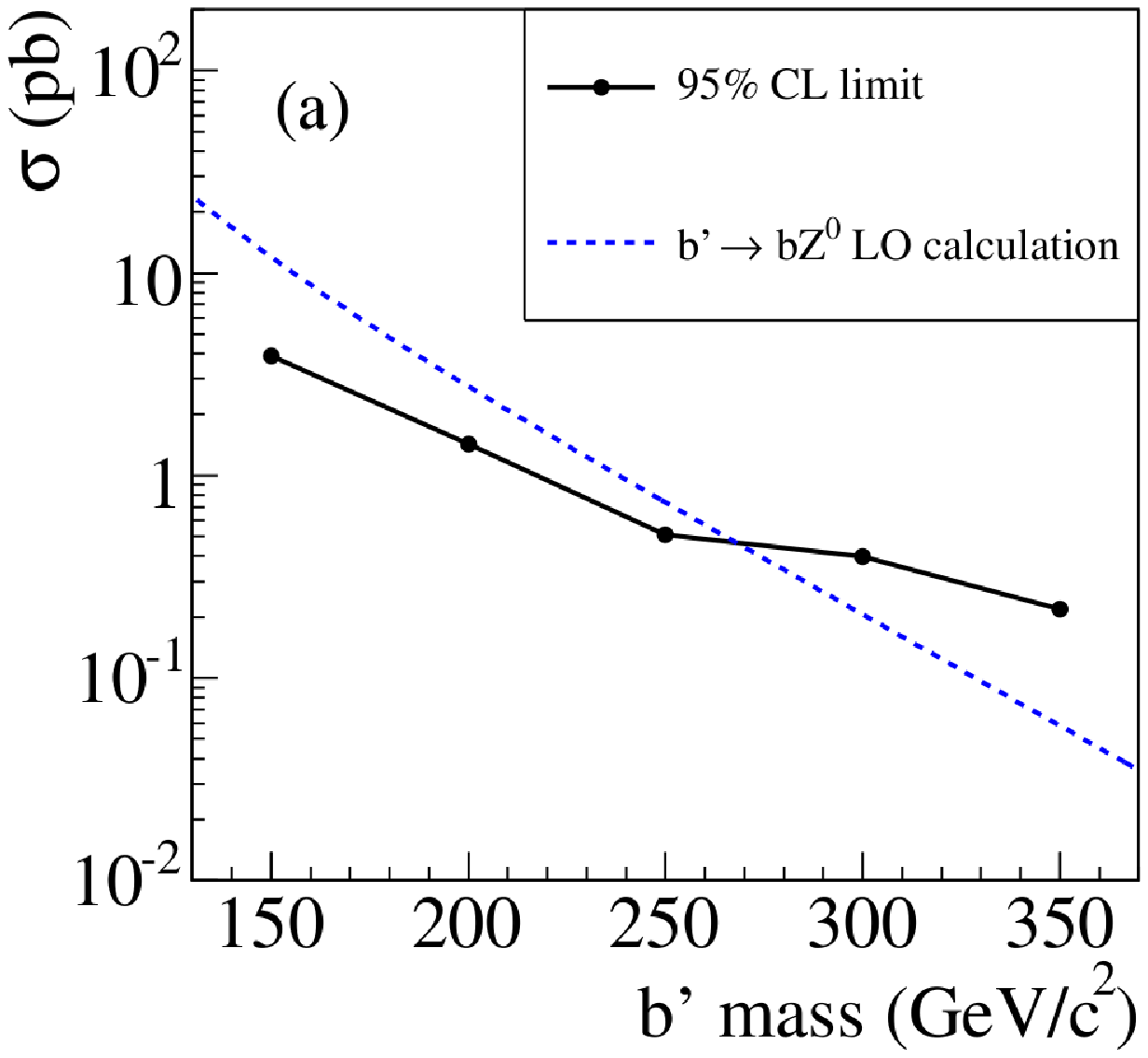}
\includegraphics[height=4.4cm]{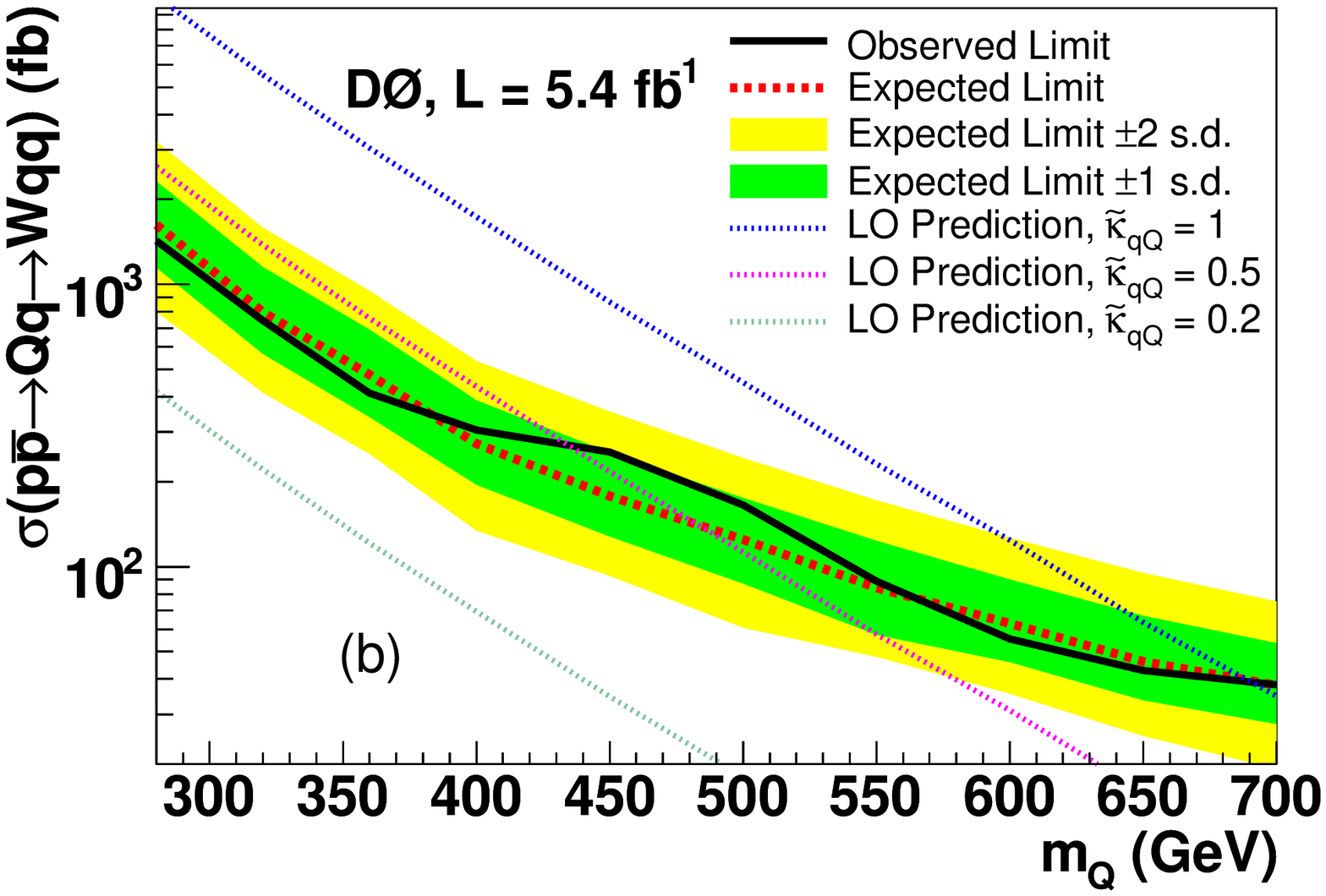}
\includegraphics[height=4.4cm]{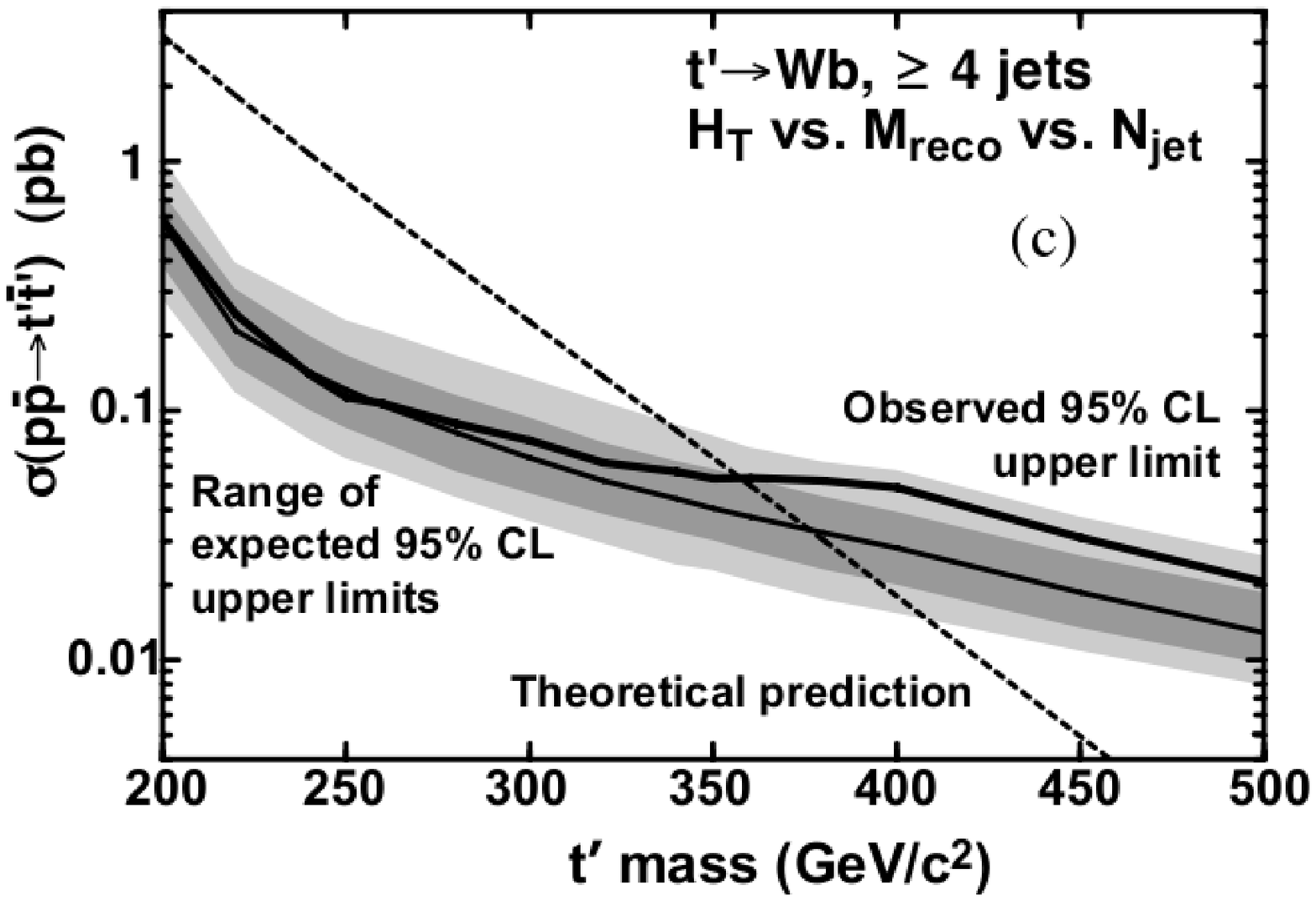}
\includegraphics[height=4.4cm]{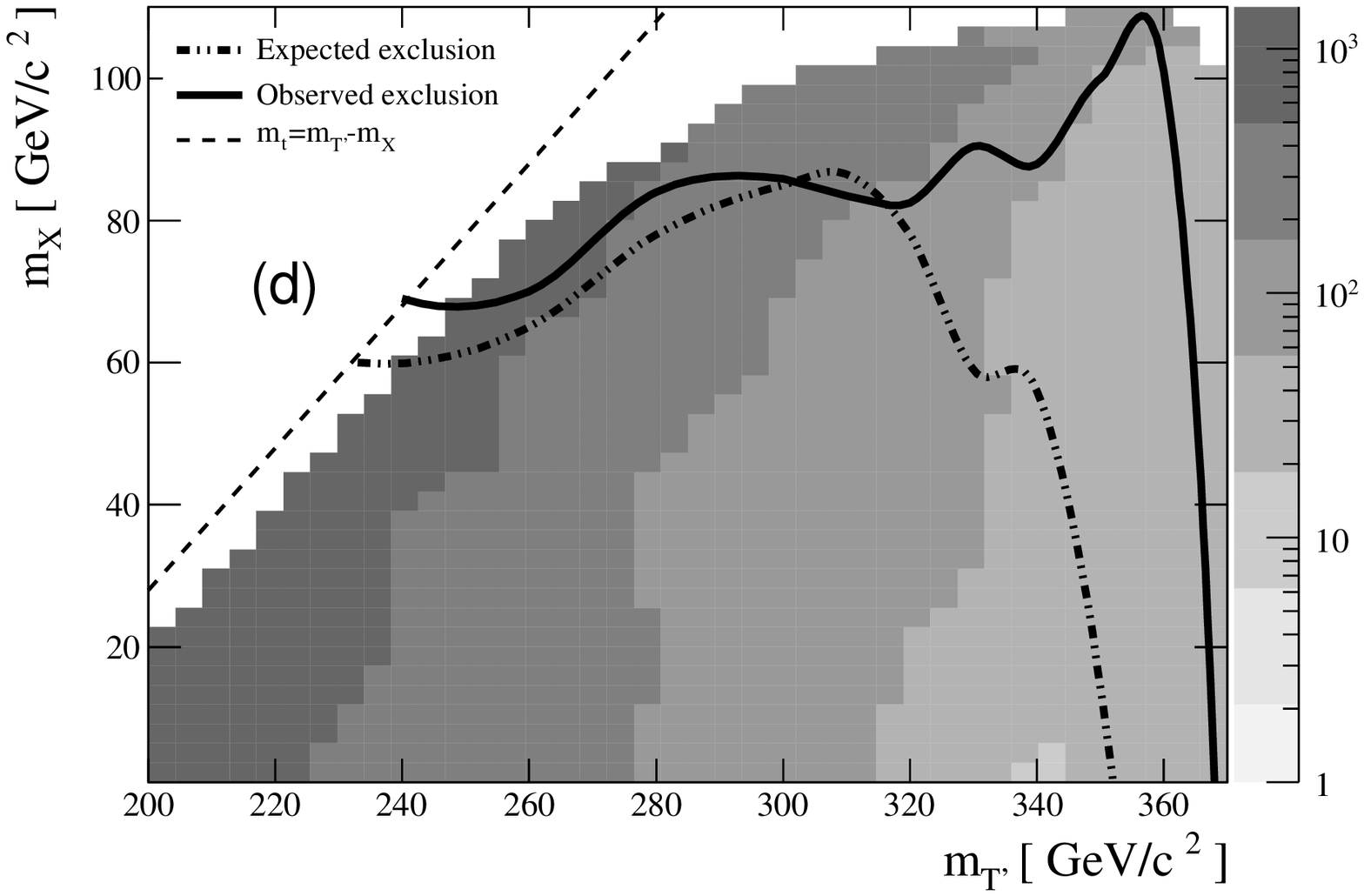}

\end{center}
\caption{(color online) (a) The 95\% C.L. 
cross section upper limits on pair production and decay of the $b'\to Zb$ as a function of $m_{b'}$ from the CDF experiment,
(b) the limits on a vector--like quark, $Q\to W+\rm{jet}$ as a function of 
$m_Q$ and for different coupling strengths with SM quarks, $\tilde{\kappa}_{qQ}$, from the D0 experiment,
(c) the limits on the $t'\to Wb$ as a function of $m_{t'}$, and (d) the limits in $m_T$ vs. $m_X$ in the search for a new heavy 
particle $T$ from the CDF experiment.
\label{fig-vq}}
\end{figure}

The new  gauge bosons predicted in 
 left--right symmetric models 
($SU(2)_L\times SU(2)_R$), grand unified theories (e.g. $E_6$),
or by the introduction of gauge groups beyond the SM
 are typically referred to as the $W'$ or $Z'$ bosons. 
Both the CDF and the D0 experiments 
searched for $W'$ bosons
in many different final states including 
$W'\to \ell\nu, tb$ and $WZ$.
The most common searches are the $W'\to e\nu$\cite{PhysRevD.75.091101,PhysRevLett.100.031804} 
and $W'\to\mu\nu$\cite{PhysRevD.83.031102} channels 
and no excess of events is observed. 
With the assumption that the \wpwz\  mode is
suppressed and that any additional generation of fermions can be ignored, the $W'$ boson is excluded 
for a mass $m_{W'}<1.12$~TeV; the results are
shown in Fig.~\ref{fig_wpenu}(a,b).
 Additional searches for $W'\to tb$\cite{PhysRevLett.103.041801,Abazov:2006aj,PhysRevLett.100.211803,Abazov:2011xs} 
 show no hints of new physics (see Fig.~\ref{fig_wpenu}(c,d)).
 Searches in the diboson final state are described with other diboson results below. 

\begin{figure}[htb]
\begin{center}
\includegraphics[height=5cm]{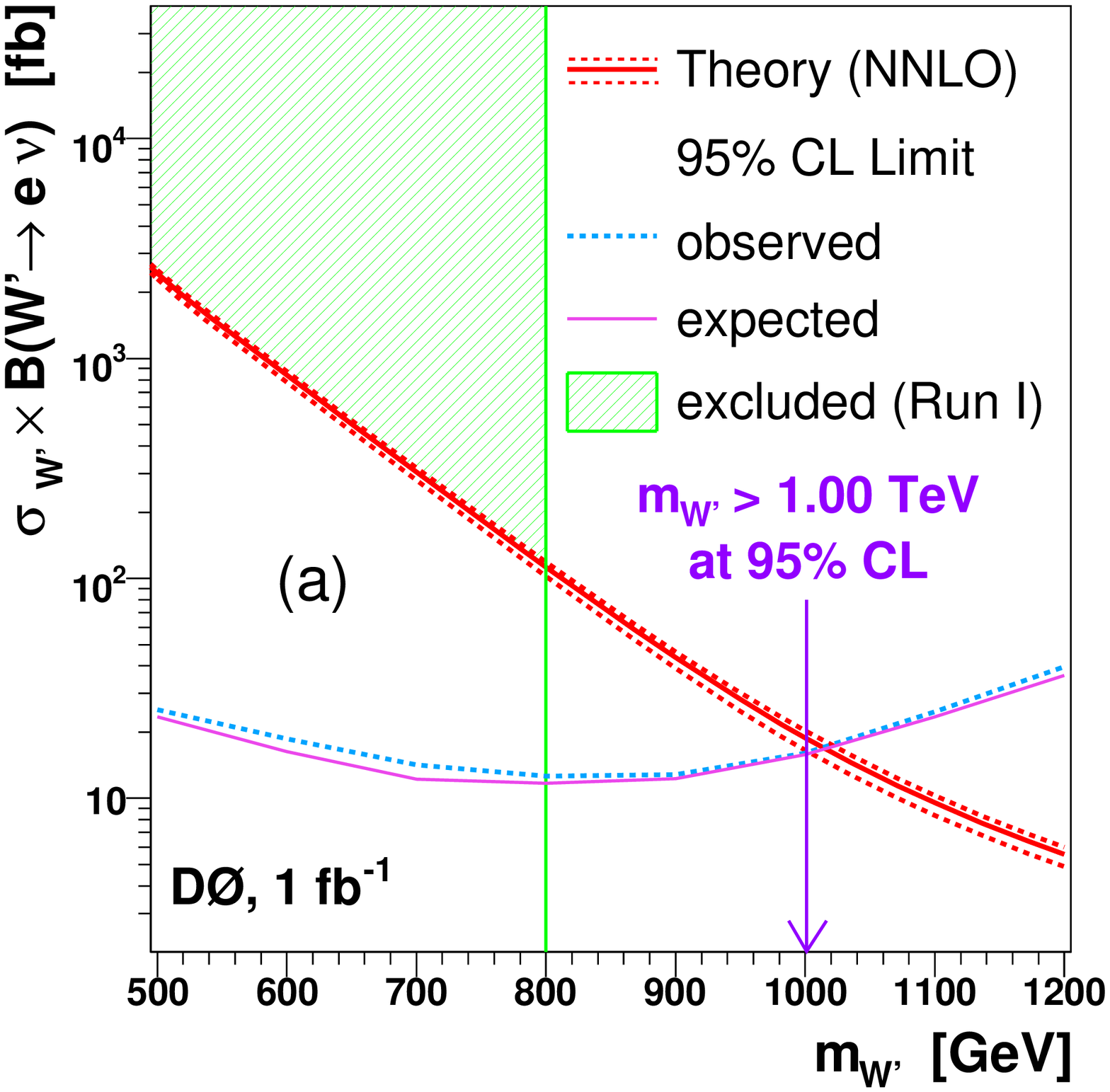}\hspace{1cm}
\includegraphics[height=5.3cm]{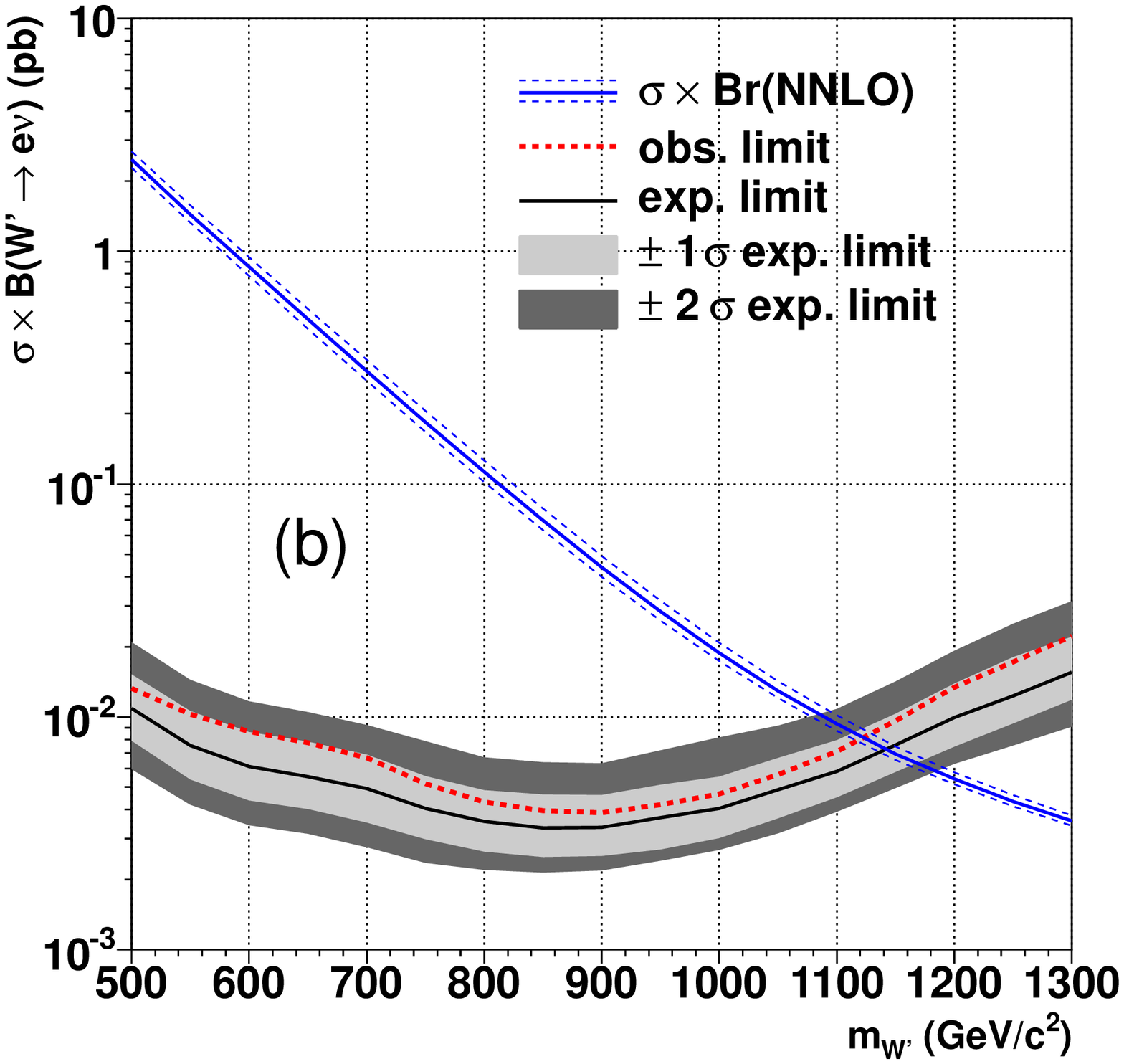}
\includegraphics[height=5cm]{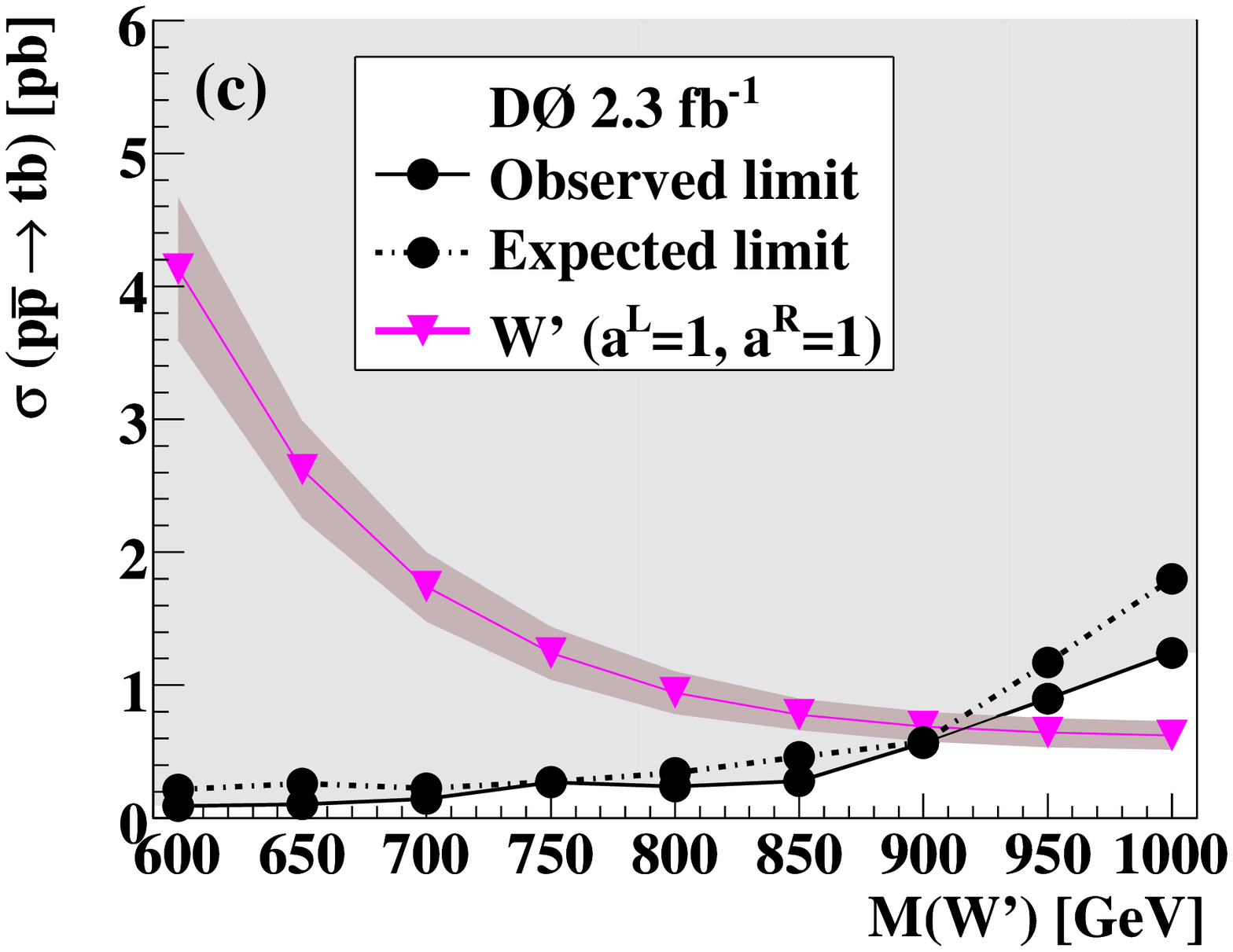}
\includegraphics[height=5cm]{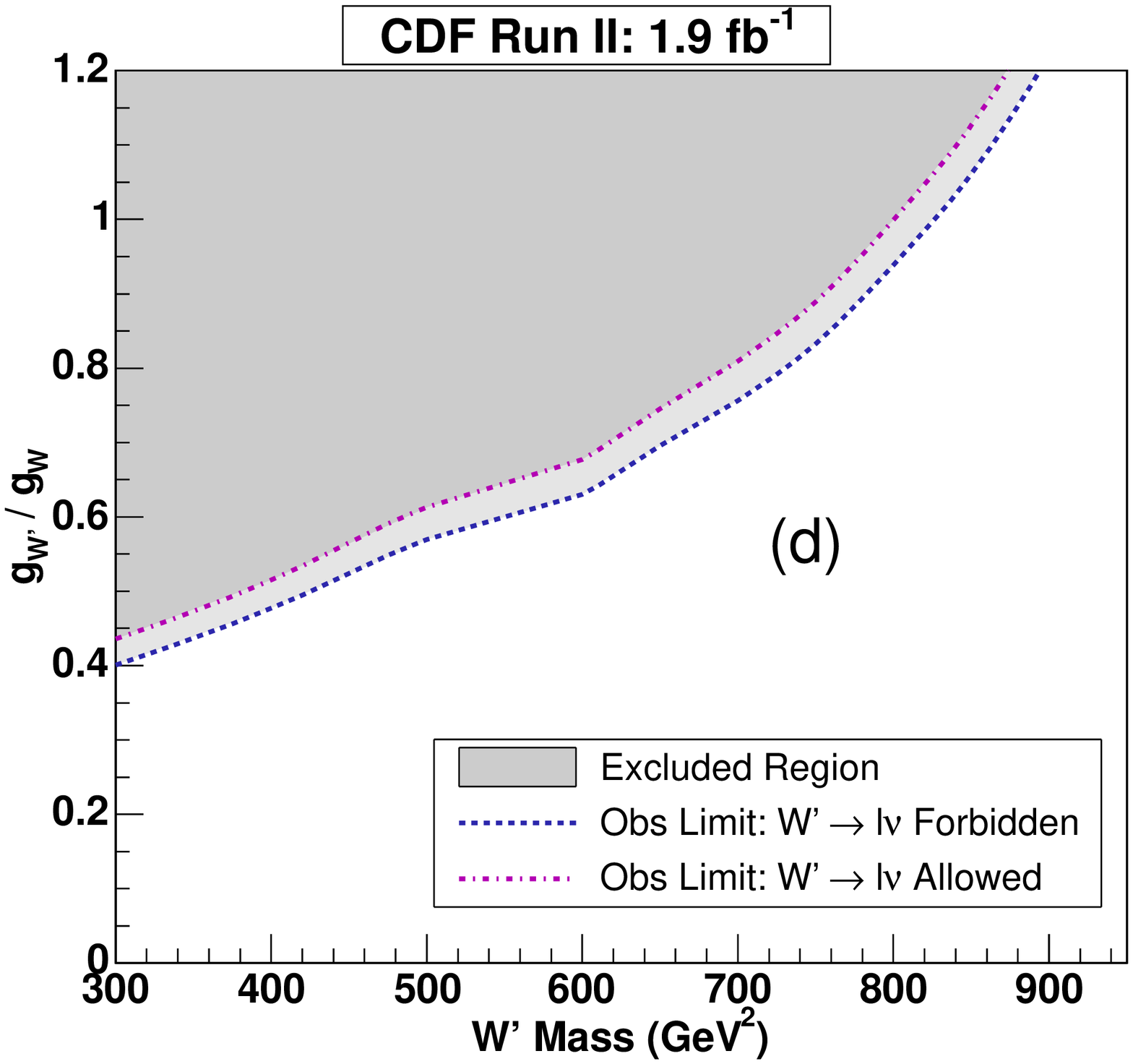}

\end{center}
\caption{(color online) The 95\% C.L.\ cross section upper limit on $W'\to e\nu$ process
as a function of the $M_{W'}$ from (a) the D0 and (b) the CDF experiments.
 The 95\% C.L.\ cross section upper limit on $W'\to tb$ process (c)
as a function of the $M_{W'}$ from the D0 experiment, and (d) in $g_{W'}/g_W$ vs. $M_{W'}$
from the CDF experiment.
\label{fig_wpenu}}
\end{figure}

A new $Z'$ boson will occur in  
theories where BSM gauge groups have an
 additional $U(1)$ gauge group.
The most common analysis is to search for a narrow resonance in the mass distribution for the 
$Z'\to\ell\ell, jj, t\bar{t}$ or $WW$. Both 
the D0\cite{Abazov:2010ti}
and the CDF\cite{PhysRevLett.106.121801, 
PhysRevLett.102.091805, PhysRevLett.102.031801} 
 experiment
looked for these signatures in dilepton final states.
Fig.~\ref{fig_zp}(a) shows the $M_{ee}$ distribution from the CDF experiment, 
exhibiting  a modest 
excess of events in data around $M_{Z'}\sim 240$~GeV; if
only SM physics is assumed in the search region, this excess has a significance of 2.5 s.d. 
The D0 experiment did not observe any significant excess as shown in 
Fig.~\ref{fig_zp}(b), and 95\% C.L.
upper limits on $\sigma\times BR(p\bar{p}\to Z'\to ee)$ for various models are set, varying between  
$M_{Z'}<772$~GeV and $M_{Z'}<1023$~GeV
as shown in Fig.~\ref{fig_zp}(c).
In both $Z'\to \mu\mu$ searches, no significant excess was observed,  with limits on 
production
of the $Z'$ boson assuming various models between $M_{Z'}<817$~GeV and $M_{Z'}<1071$~GeV,
as shown in Fig.~\ref{fig_zp}(d). 
Other searches in $Z'\to jj$ and $t\bar{t}$
found no excesses~\cite{PhysRevD.79.112002,PhysRevLett.110.121802,Abazov:2008ny,PhysRevD.85.051101} 
(see Fig.~\ref{fig_zp}(e,f)).
The decay $Z'\to WW$ is described with the other diboson searches below. 
Lepton flavor violating searches, for example $Z'\to e\mu, e\tau, \mu\tau$
are typically done 
in the context of $R$--parity violating SUSY, but have $Z'$ interpretations\cite{PhysRevLett.96.211802}. 

\begin{figure}[htb]
\begin{center}
\includegraphics[height=5cm]{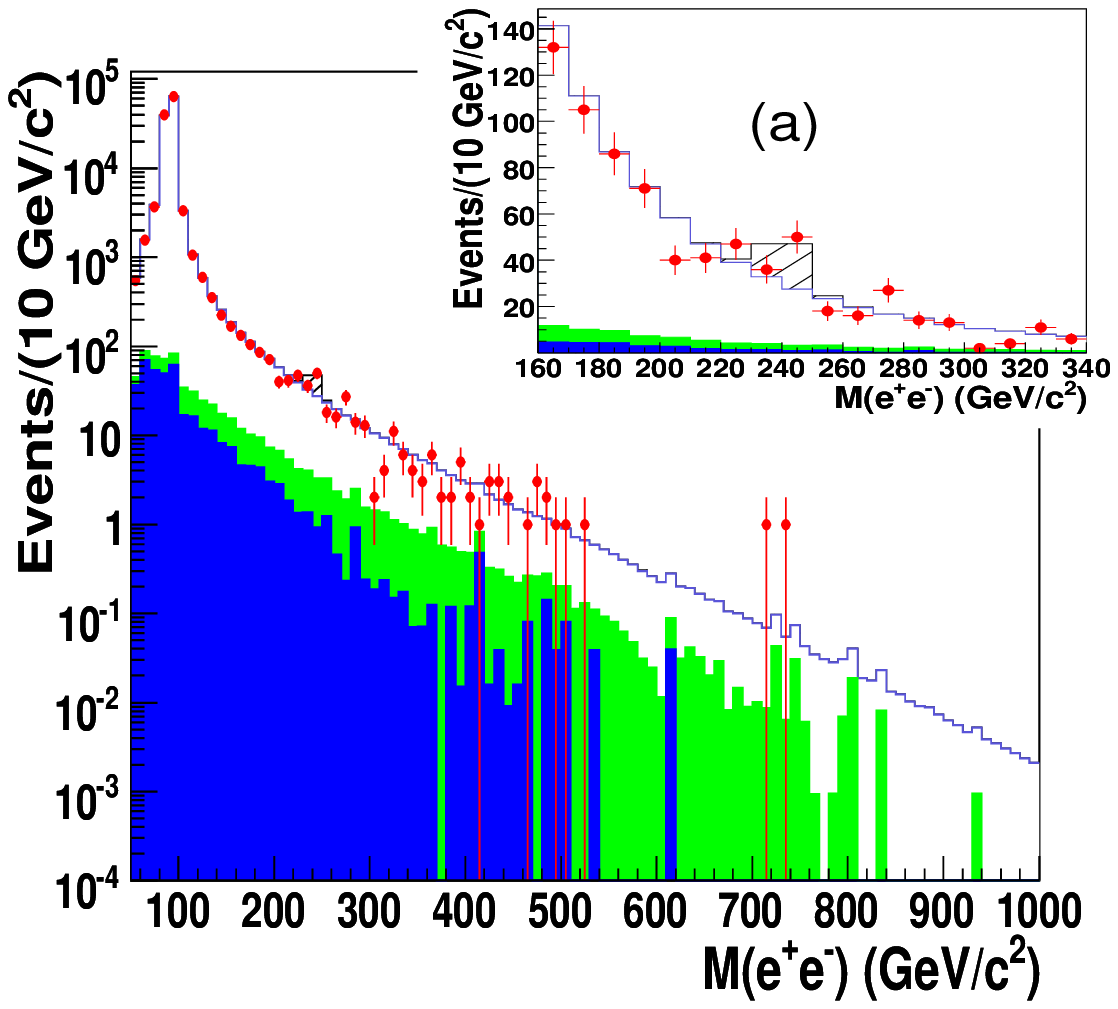}\hspace{1cm}
\includegraphics[height=5.2cm]{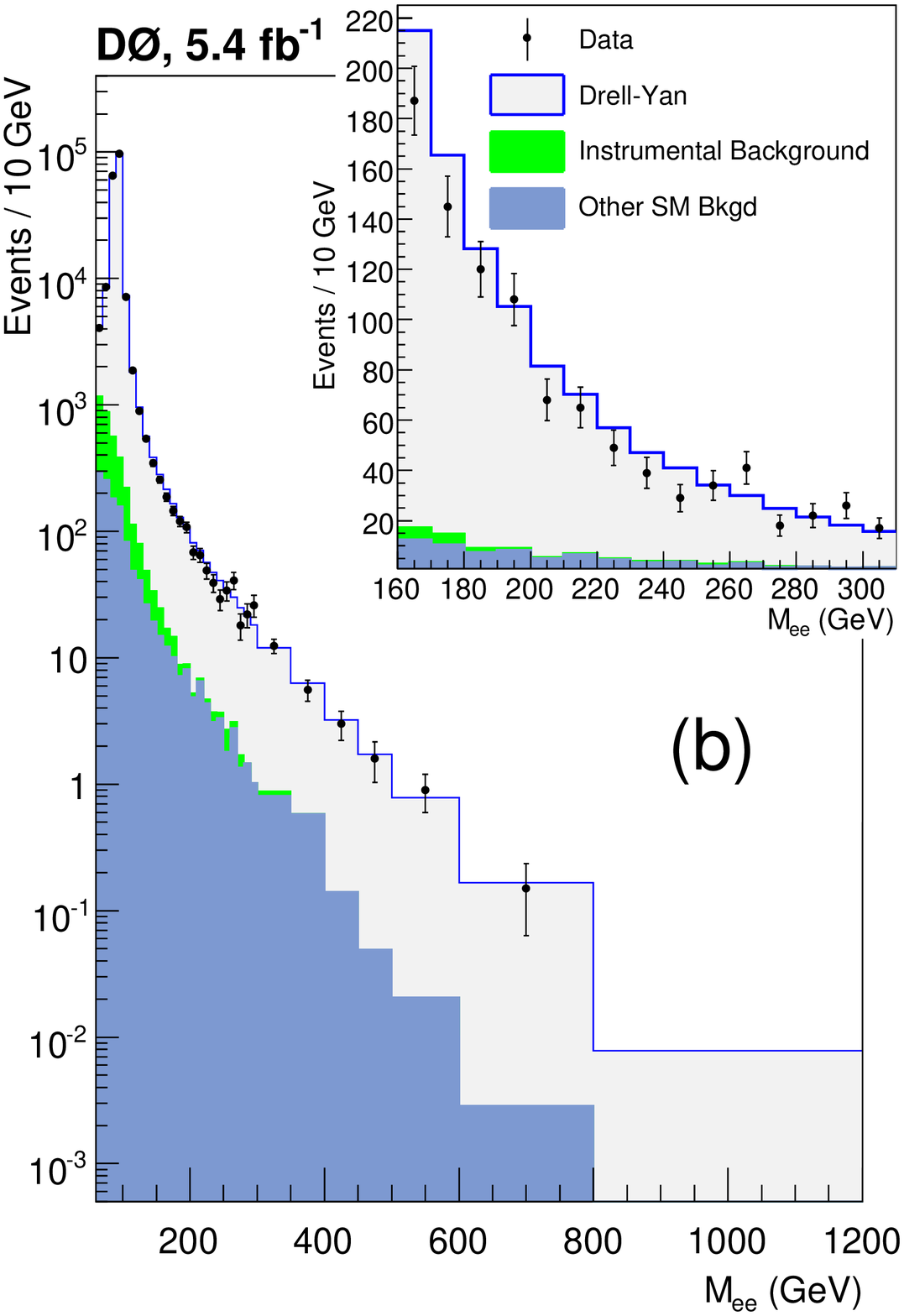}
\includegraphics[height=5cm]{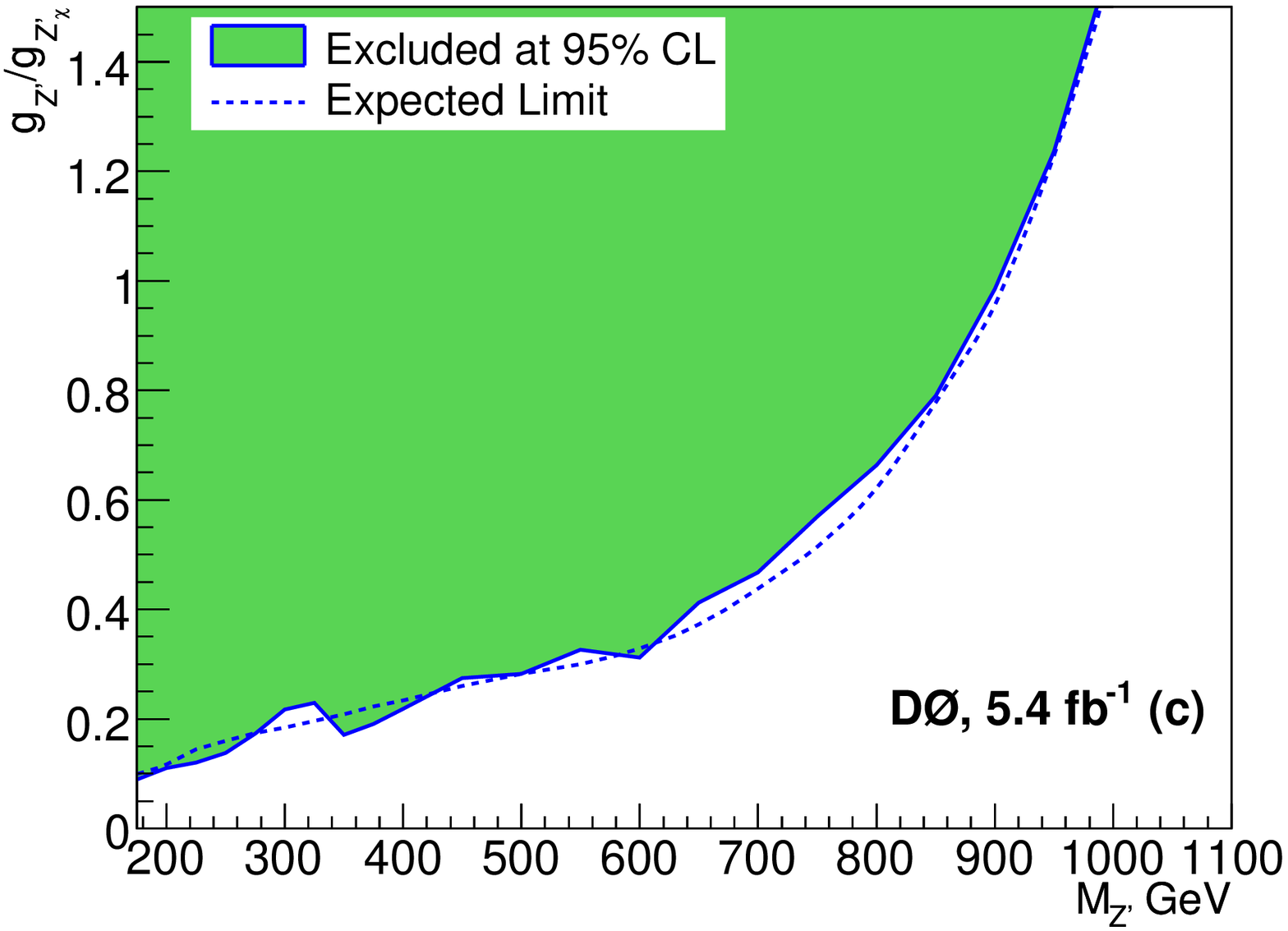}\hfill
\includegraphics[height=5.2cm]{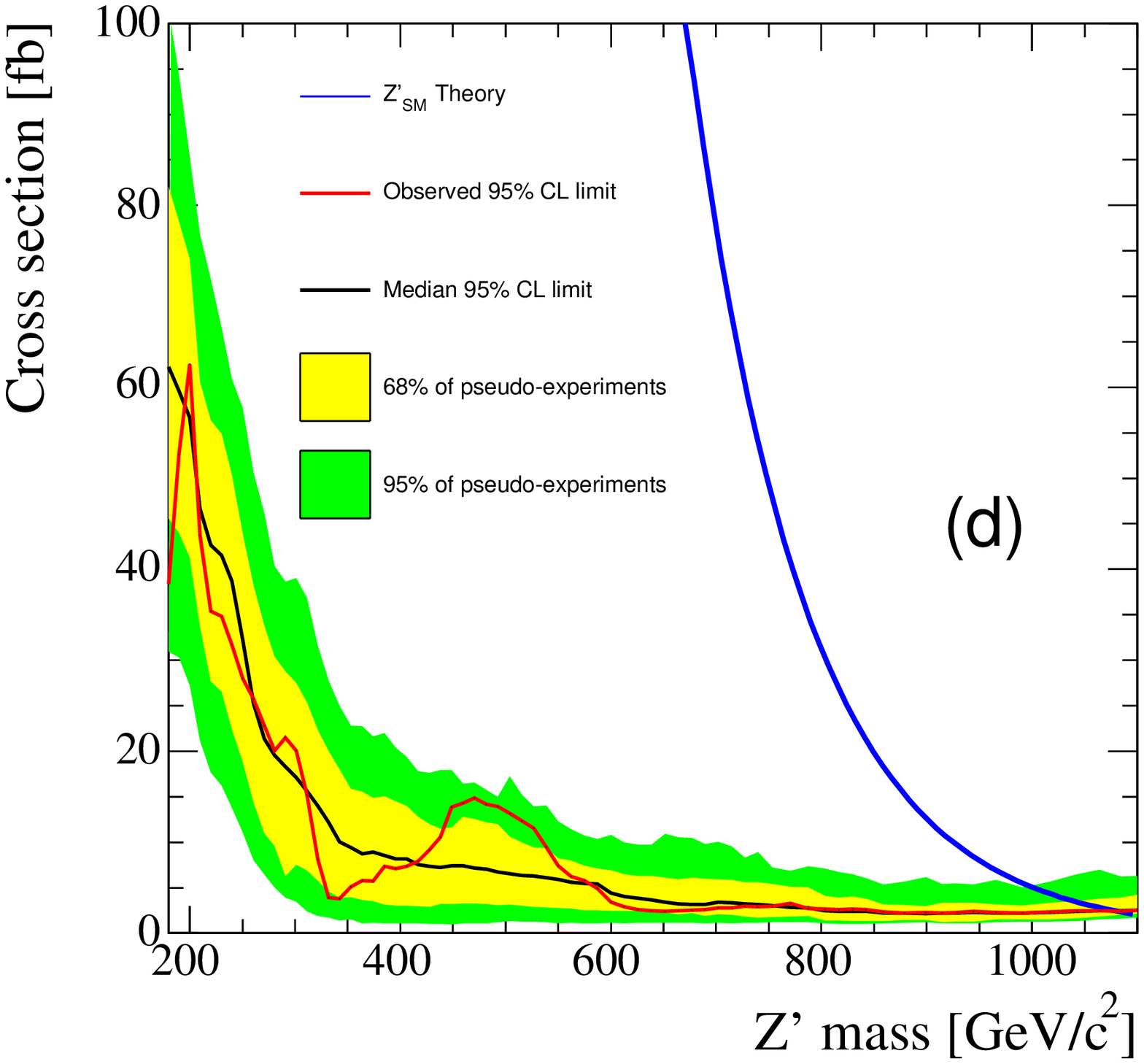}
\includegraphics[height=5cm]{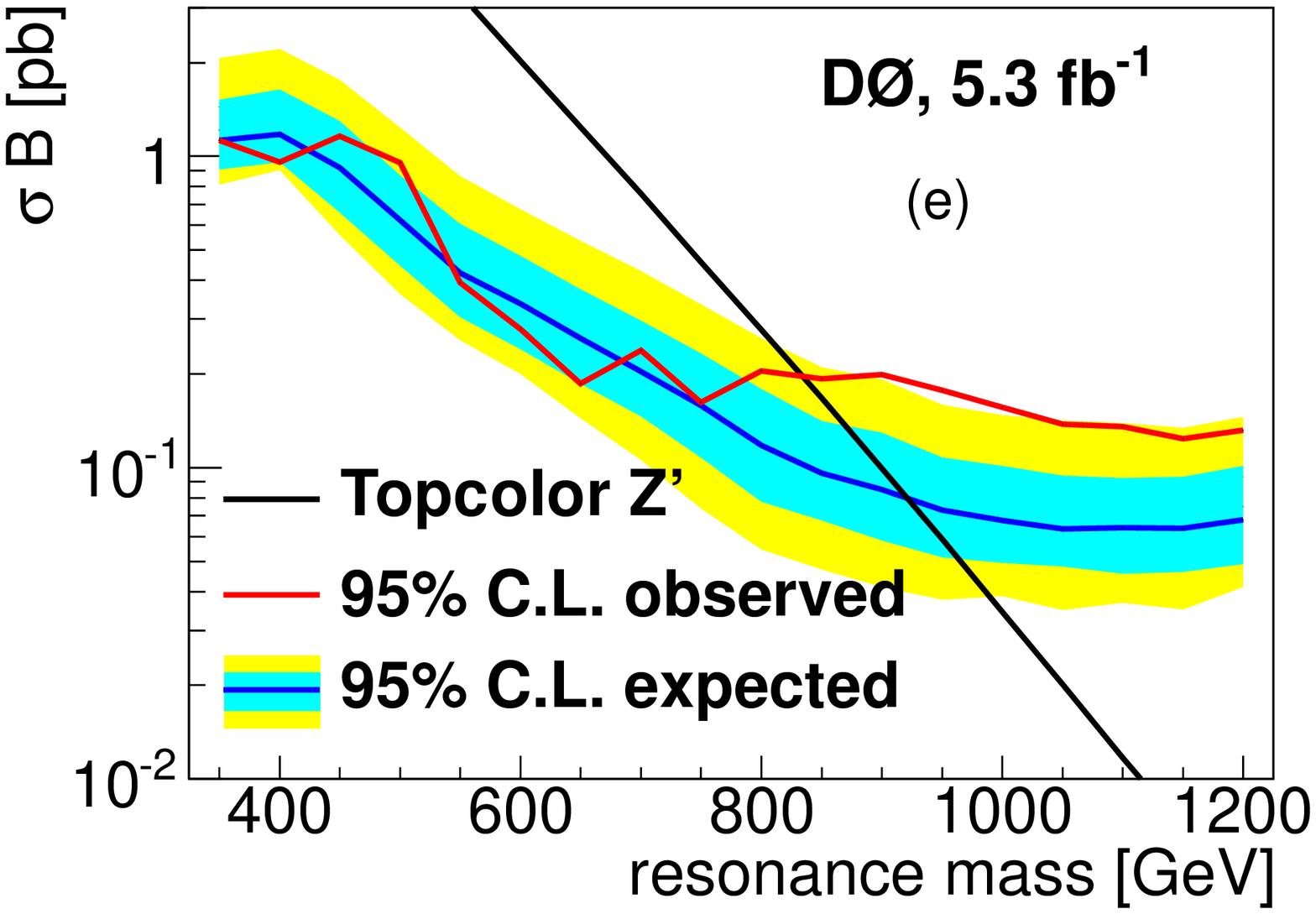}\hfill
\includegraphics[height=5.2cm]{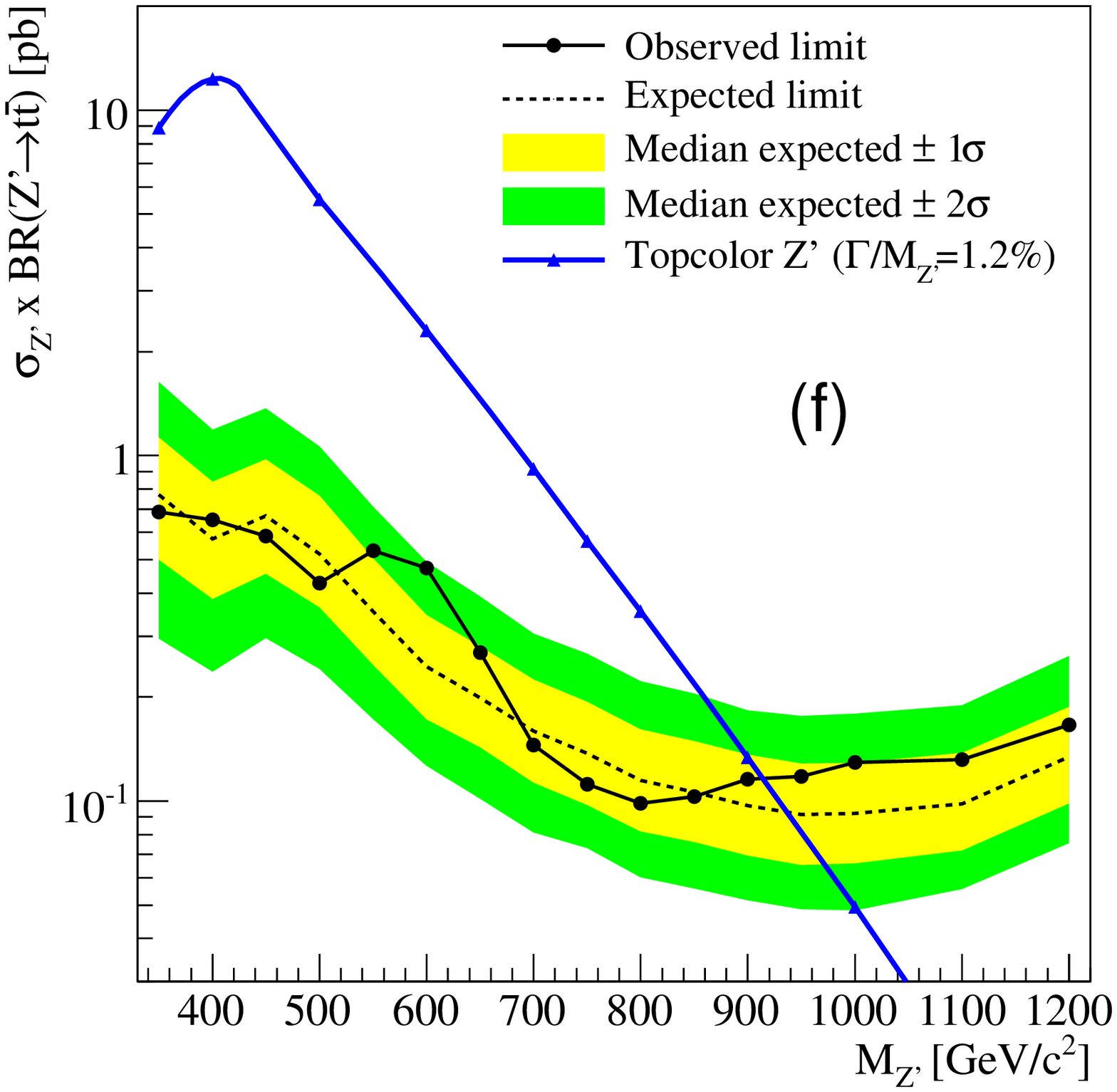}
\end{center}
\caption{(color online)  The dielectron invariant mass in the search for $Z'\to ee$ from
(a) the CDF experiment and
(b)  the D0 experiments. 
(c) The 95\% C.L. upper limits on the Z' couplings ratio ($g_{Z'}/g_{Z'_{\chi}}$) 
 as a function of $M_{Z'}$
from the D0 experiment. 
(d) The 95\% C.L. upper limits on the $\sigma\times{\cal BR}(Z'\to\mu\mu)$
as a function of the $M_{Z'}$ from the CDF experiment. 
 The 95\% C.L. upper limits on the $\sigma\times{\cal BR}(Z'\to tt)$
as a function of the $M_{Z'}$ from 
(e) the D0 experiment and 
(f) the CDF experiments.
\label{fig_zp}}
\end{figure}

The CDF and the D0 experiments also searched for 
resonances in the  
 $WW$, $WZ$ and $ZZ$ decay modes.  
 While these searches
can be analyzed as $Z'\to WW$, $Z'\to ZZ$  and $W'\to WZ$, other interpretations are possible.
 These are done at 
 the CDF experiment in $V'\to VV\to\ell+\met+\rm{jets}$\cite{PhysRevLett.104.241801}, and $X\to ZZ$\cite{PhysRevD.85.012008}
  with various 
$Z$ boson decays. The $X\to ZZ$ search showed a small excess in data in the low--sensitivity four--lepton channel,
which was not observed in the other two more sensitive 
$\ell\ell jj$ and $\ell\ell+\met$ final states\cite{PhysRevD.85.012008}; this result will be interpreted in the section~\ref{sec-EDres}.
The D0 experiment\cite{PhysRevLett.104.061801,PhysRevLett.107.011801}
 searched in a combined way for $W'$ decay to one, two or three leptons (assuming  
$W'\to WZ\to \ell\ell\ell\met, \ell\ell jj, \ell\ell\met$+jets).
In addition to the standard methods, a novel method is used (now adopted by the LHC) to investigate the 
possibility that the 
heavy $W'$ or $Z'$ boson is so massive that the decay bosons 
are highly boosted. 
For hadronically decaying 
vector bosons the two light quarks could get merged and produce a single 
broad jet with an invariant mass close to the $W$ or $Z$ boson mass. 
No significant excess over data was found in either experiment in any of these modes.
Limits on the $W'$ boson are between 180~GeV and 690~GeV, and on the 
$Z'$ boson between 242~GeV and 544~GeV as
  shown in Fig.~\ref{f9}.

\begin{figure}[htb]
\begin{center}
\includegraphics[height=4.2cm]{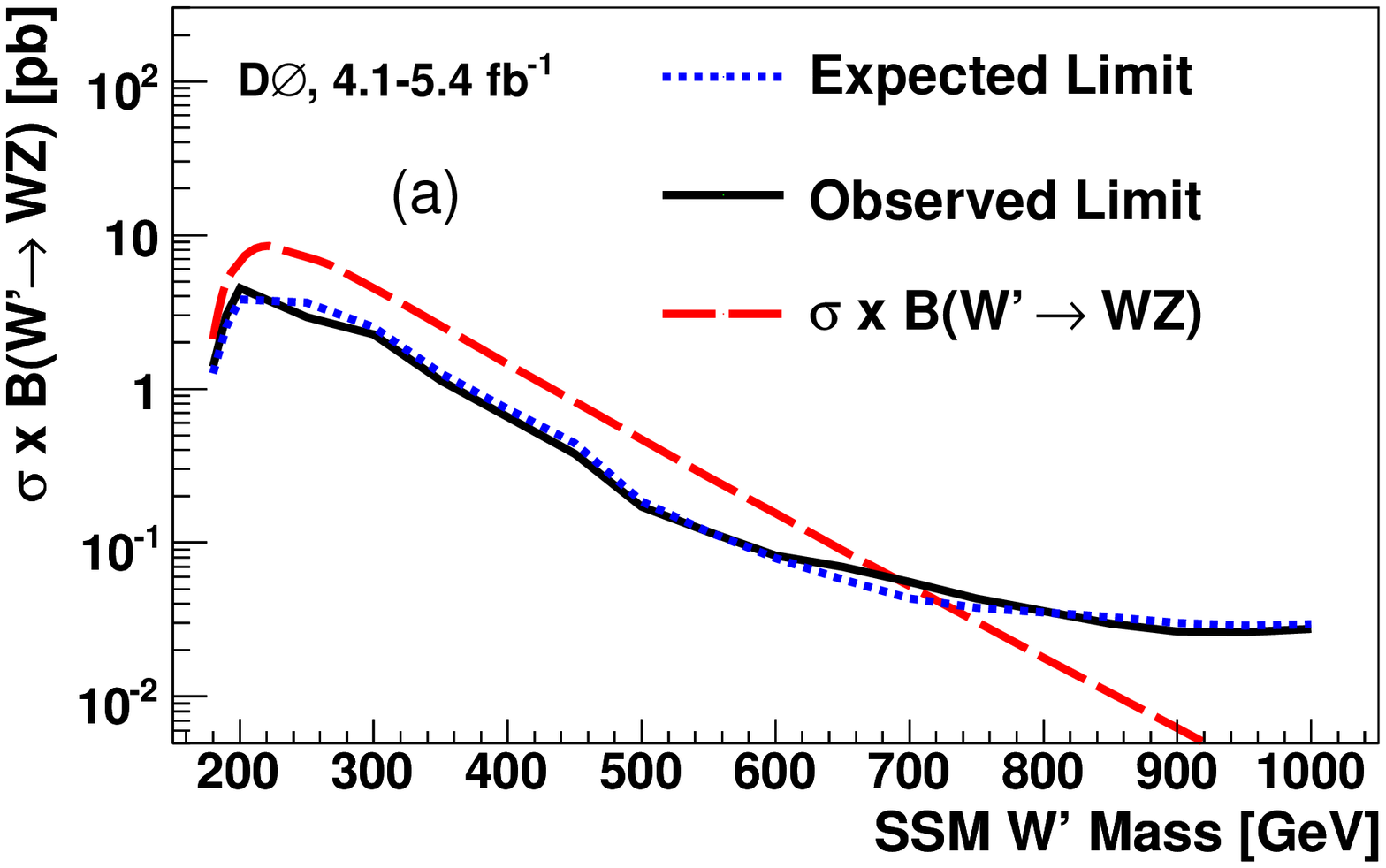}\hfill
\includegraphics[height=4.cm]{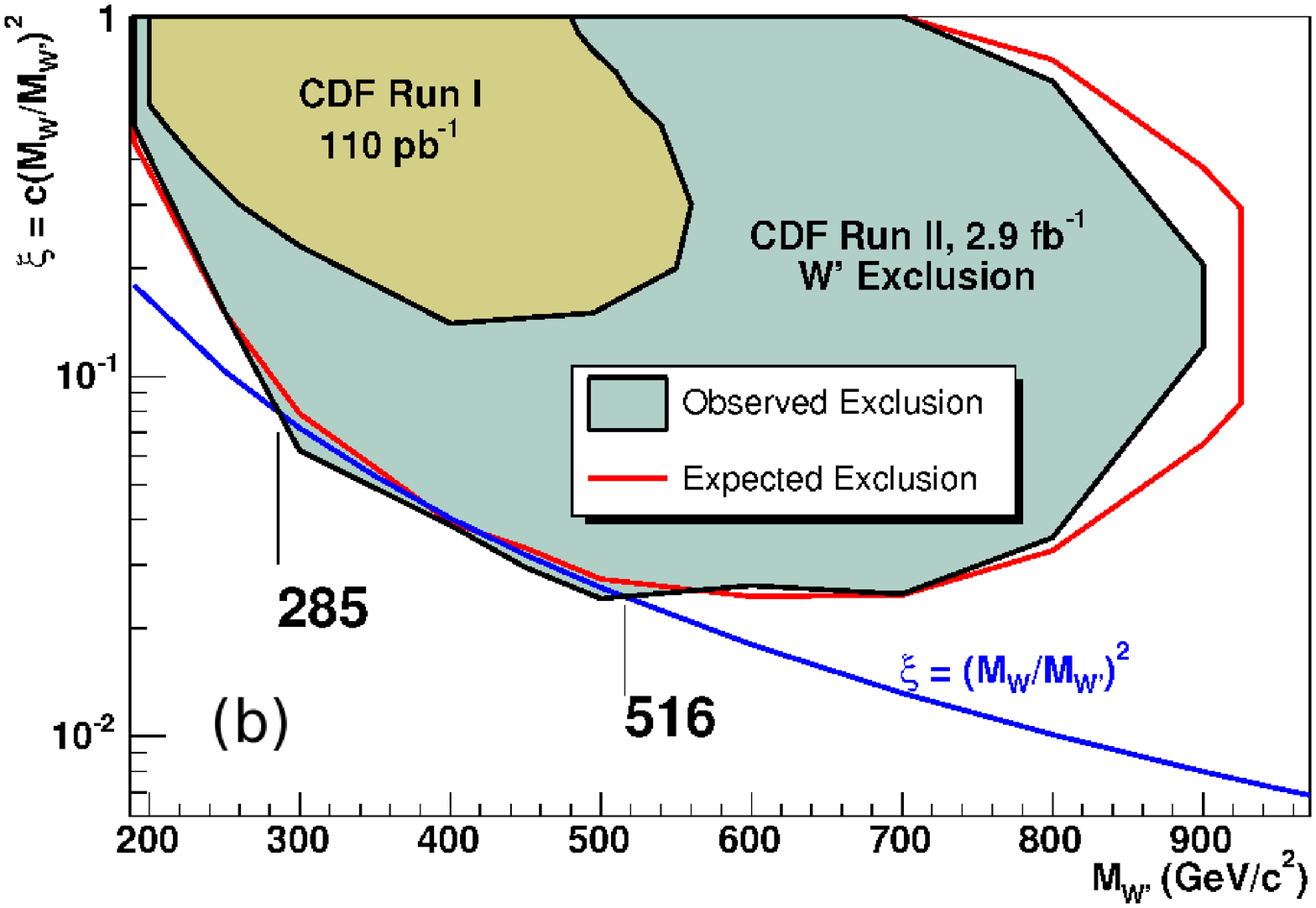}
\end{center}
\caption{(color online) (a) The 95\% C.L.\ upper limit on  
$\sigma\times{\cal BR}(W'\to WZ)$ from the D0 experiment and (b)
the exclusion region in the $\xi$ vs. $M_{W'}$ plane in the \wpwz\ search from the CDF experiment,
 where the parameter $\xi$ sets the coupling strength between the SM $W$ boson and any new $W'$ boson. \label{f9}}
\end{figure}

Finally, the 
 CDF experiment\cite{PhysRevLett.111.031802} 
 searched for both resonant and non--resonant production of pairs of strongly interacting particles,
each of which decays to a pair of jets, $p\bar{p}\to X\to (YY) \to (jj)(jj)$ and $p\bar{p}\to YY \to (jj)(jj)$. This search is particularly sensitive
at lower masses, where the LHC experiments expect high background rates.
No evidence of new particles is observed and 
results are interpreted as an exclusion of the $Y$ particle in both production scenarios 
with the results shown in Fig.~\ref{fig-4jRes}.
These results are directly applicable in axigluon models. 

\begin{figure}[htb]
\begin{center}
\includegraphics[height=4.5cm]{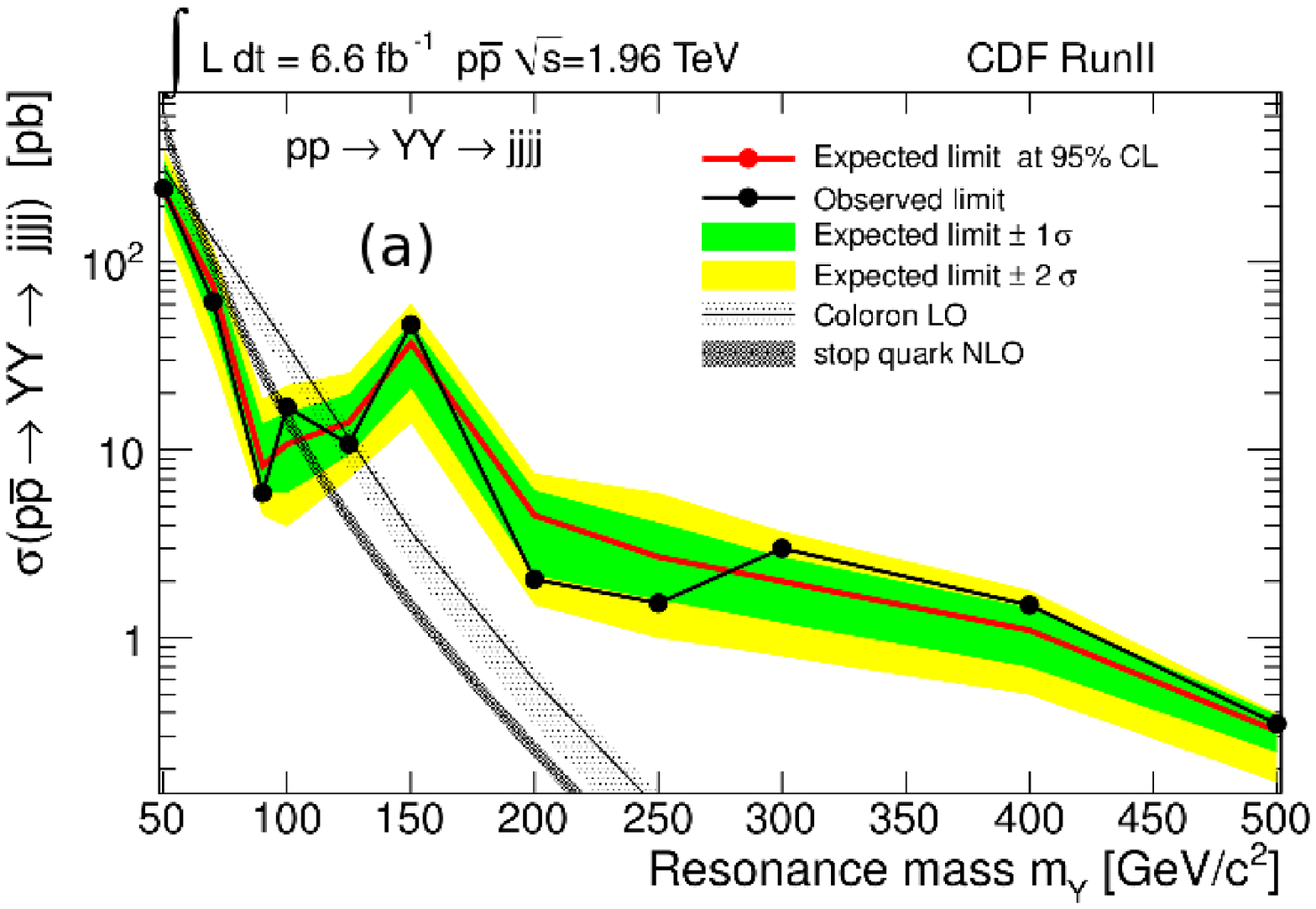}
\includegraphics[height=4.5cm]{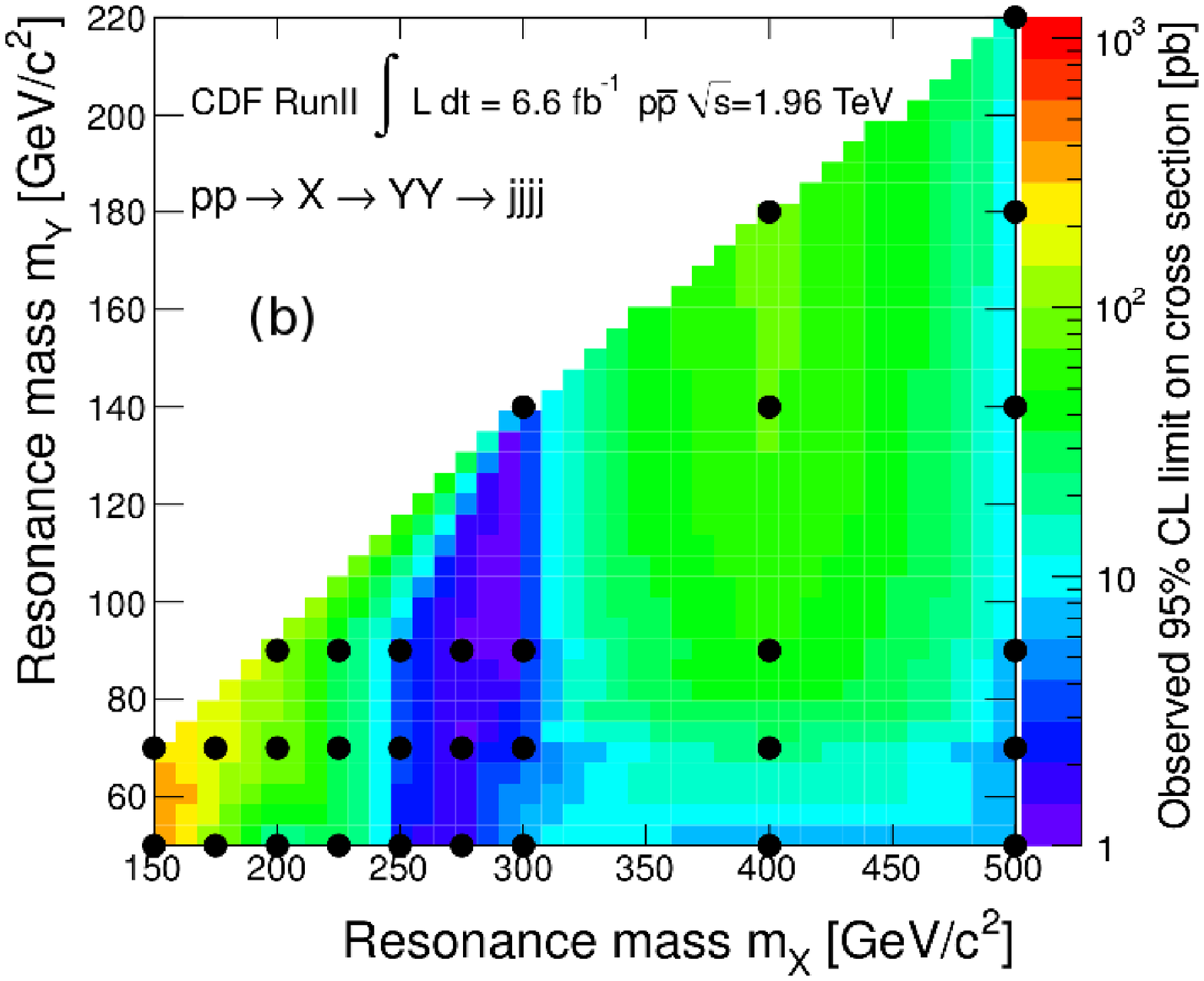}
\end{center}
\caption{(color online) The 95\% C.L.\ upper limits on (a) $\sigma(p\bar{p}\to YY\to jjjj)$ as a function of the $M_Y$ and (b)
$\sigma(p\bar{p}\to X\to YY\to jjjj)$ in the $M_Y$ vs. $M_X$ plane from the CDF experiment.
\label{fig-4jRes}}
\end{figure}

 \subsubsection{\textit{Excited fermions}}\label{sec_reson2}

The search for compositeness focuses on excited states of the SM fermions. 
Both excited electrons and excited muons, as well as excited quarks are searched for in the 
$e^*\to e\gamma$\cite{PhysRevD.77.091102,PhysRevLett.94.101802}
 and $\mu^*\to\mu\gamma$\cite{PhysRevLett.97.191802,PhysRevD.73.111102},
and $q^*\to q\gamma$~or~$qW$~\cite{PhysRevLett.72.3004},
$q^*\to qZ$\cite{PhysRevD.74.011104}, and $q^*\to qg$ modes\cite{PhysRevD.79.112002}. Results are interpreted as exclusion limits in 
contact interaction model
for a mass $m<876(853)$~GeV for $e^*(\mu^*)$, 
and in a gauge--mediated model
for a mass $m<430(410)$~GeV for $e^*(\mu^*)$.

\subsubsection{\textit{Leptoquarks}}\label{sec_reson3}

Leptoquarks can exist as either vectors or scalars and can be produced in either 
 pair production or single production modes.
In all cases, the $LQ$ can decay  
 to $\ell q$ or $\nu q'$ where $\ell=e,\mu,\tau$,  and the 
 parameter $\beta$ defines the branching fraction for $LQ\to\ell q$.
 Due to experimental constraints on flavour changing
neutral currents\cite{Beringer:1900zz},
it is assumed that $LQ$s only couple to fermions of the same generation.
Both the CDF\cite{Aaltonen:2007rb, PhysRevD.73.051102,PhysRevD.72.051107,PhysRevD.71.112001}
 and the D0\cite{PhysRevD.71.071104, Abazov:2009ab, PhysRevD.84.071104,Abazov:2006vc, Abazov:2006ej, Abazov:2008np, 
PhysRevLett.99.061801, PhysRevLett.101.241802, Abazov:2010wq, Abazov:2006wp, Abazov:2008at}
 experiments focused on pair production of leptoquarks of all generations
in the   
$\ell q\ell q$, $\ell q\nu q'$ and $\nu q\nu q$
 final states. In all cases, no excesses were observed. Since vector and scalar 
 production are similar, limits can be set on both using the same results, but we begin by reporting the 
 results in the scalar $LQ$ case. 
For $\beta=0$, when both $LQ$ decay to $\nu q$, it is not possible to 
distinguish between the search for the first and second generation,
and a common limit for a mass $m_{LQ}< 214$~GeV is set; 
$b$--tagging allows for a specific 
search for the third generation
in $bb+\met$, and the limit is $m_{LQ}<247$~GeV.
Searches with charged leptons
are done in the $\ell+\rm{jets}+\met$ and 
  $\ell\ell+\rm{jets}$
 final states; the third generation search often
requires 
the hadronic decays of  $\tau$--lepton accompanied with $b$--jets.
For $\beta=0.5$ the first generation $LQ$ is excluded for a mass $m_{LQ}< 326$~GeV, the second generation 
for a mass $m_{LQ}<270$~GeV, and the third generation
for a mass $m_{LQ}<207$~GeV; 
for $\beta=1$ the first generation $LQ$ is excluded for a mass $m_{LQ}<199$~GeV, 
the second generation for a mass $m_{LQ}< 316$~GeV, and the third generation
for a mass $m_{LQ}<210$~GeV.
Figure~\ref{lq}(a) shows the exclusion region in the 
$\beta$ vs. $M_{LQ}$ plane in the search for the first generation scalar $LQ$ pairs.
Reinterpreting the data in terms of first generation 
 vector $LQ$ model, 
Fig.~\ref{lq}(b) shows the limits for 
three different assumptions about the couplings. For the third generation, with $\beta=1$, vector
 $LQ$s are excluded for a mass $m_{LQ}<317$~GeV and $m_{LQ}<251$~GeV at 95\% C.L.\
with two different assumptions about couplings\cite{Aaltonen:2007rb}.

\begin{figure}[htb]
\begin{center}
\includegraphics[height=4.8cm]{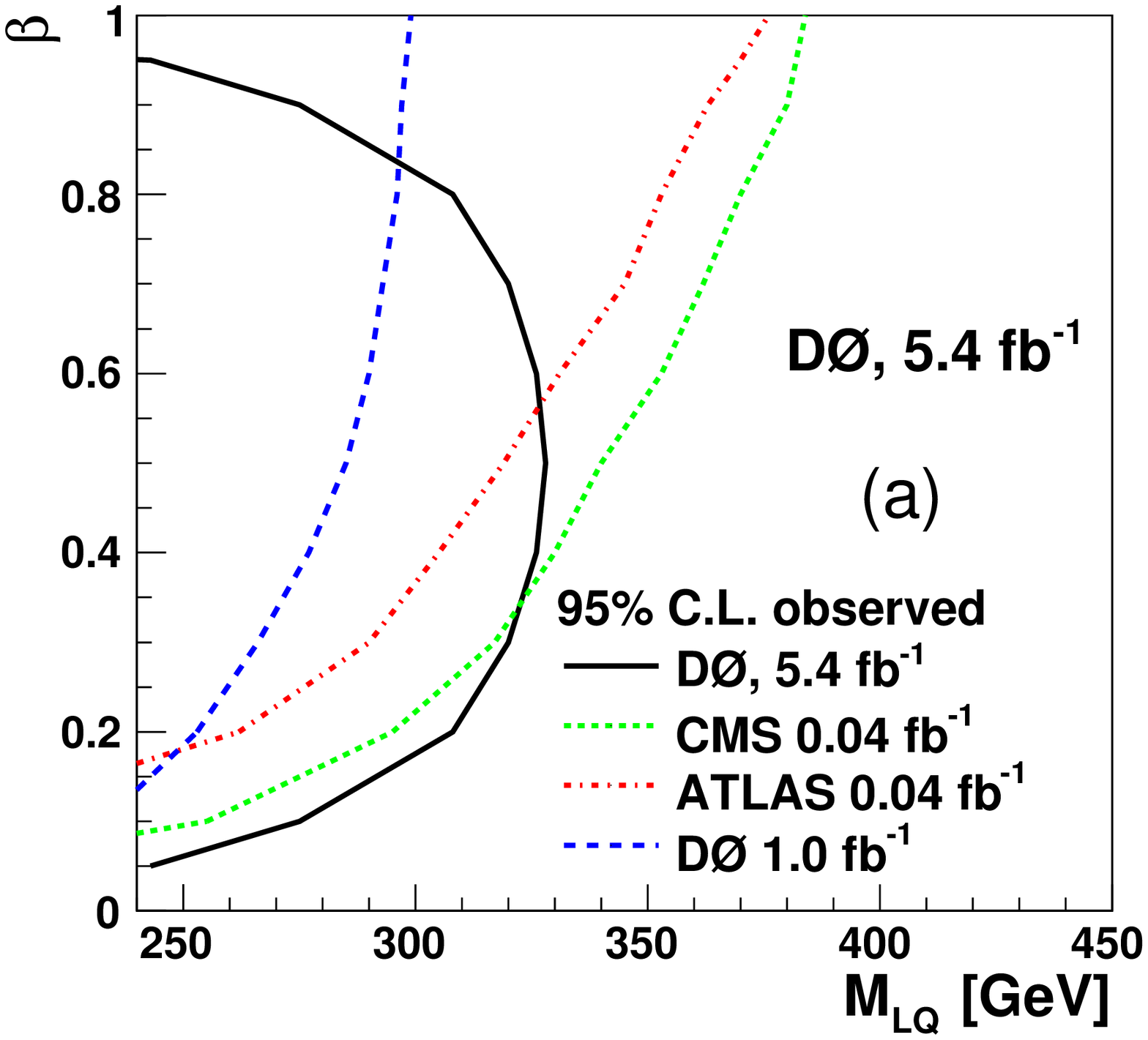}\hfill
\includegraphics[height=4.8cm]{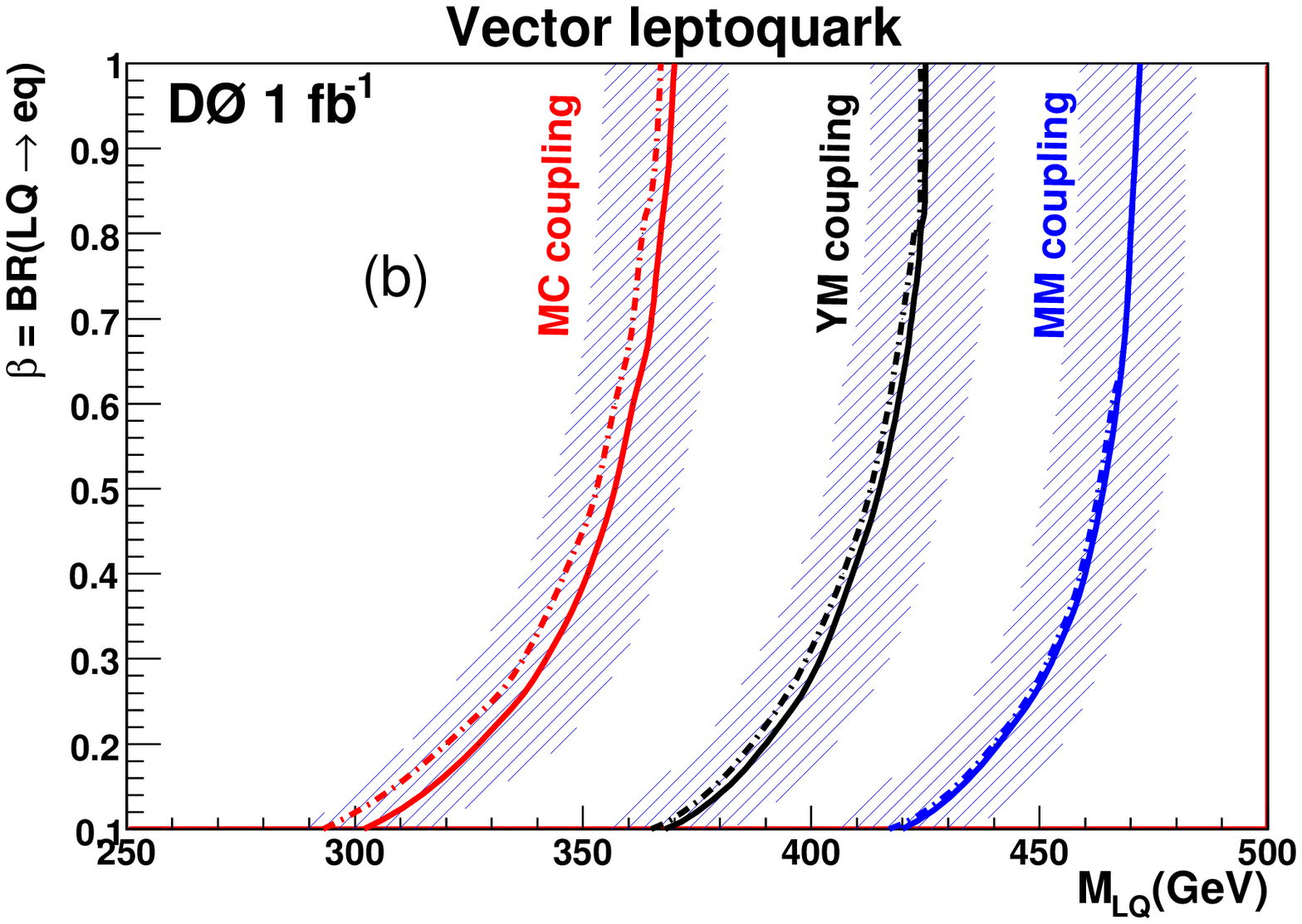}
\end{center}
\caption{(color online) (a) The exclusion region in the $\beta$ vs. $M_{LQ}$ plane from
 the search for first generation scalar $LQ$ pairs. 
(b) The exclusion regions in the same plane but for 
 first generation vector $LQ$ pairs. Both results are from the D0 experiment.\label{lq}}
\end{figure}

\subsubsection{\textit{Technicolor}}\label{sec_reson4}

Much of technicolor phenomenology is driven by the technicolor strawman model\cite{Lane:2002sm}.
In this model, the most promising signature
 is the production and decay of a 
  technicolor $\rho_T$, which  can decay via $\rho_T\to\pi_T+W\to bbW$
  (where $\pi_T$ is a technipion) or $\rho_T\to WZ$ depending on the masses of the particles involved. 
Other 
new particles such as a techniomega,  $\omega_T$, can be produced and decay via 
 $\omega_T\to\gamma+\pi_T\to\gamma bb$.
No evidence for new physics is observed\cite{PhysRevLett.98.221801,PhysRevLett.104.111802,PhysRevLett.104.061801}.
Some of the expected and observed 95\% C.L.\ excluded
regions are shown in Fig.~\ref{fig-TC}.

\begin{figure}[htb]
\begin{center}
\includegraphics[height=4.7cm]{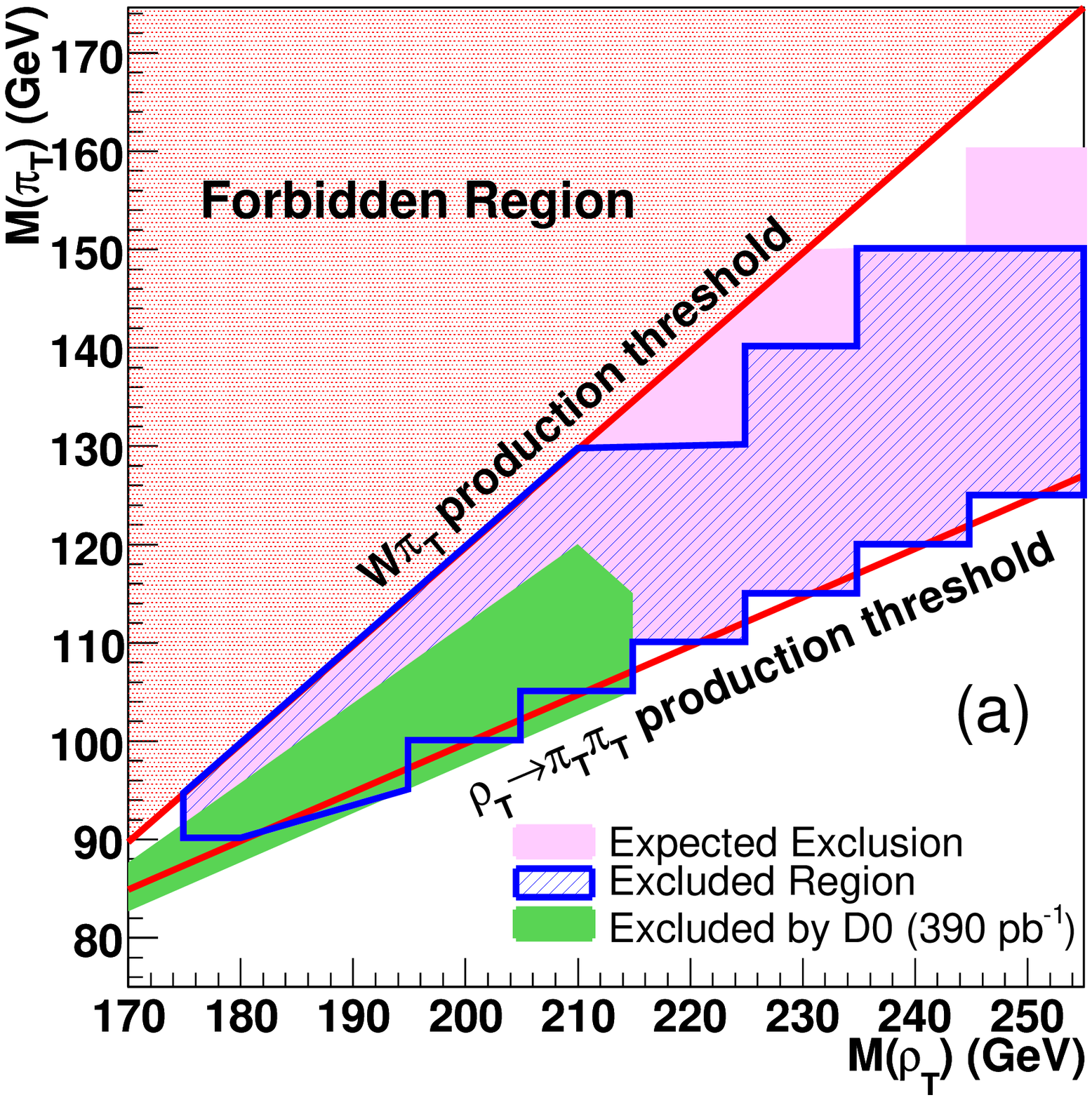}\hfill
\includegraphics[height=4.8cm]{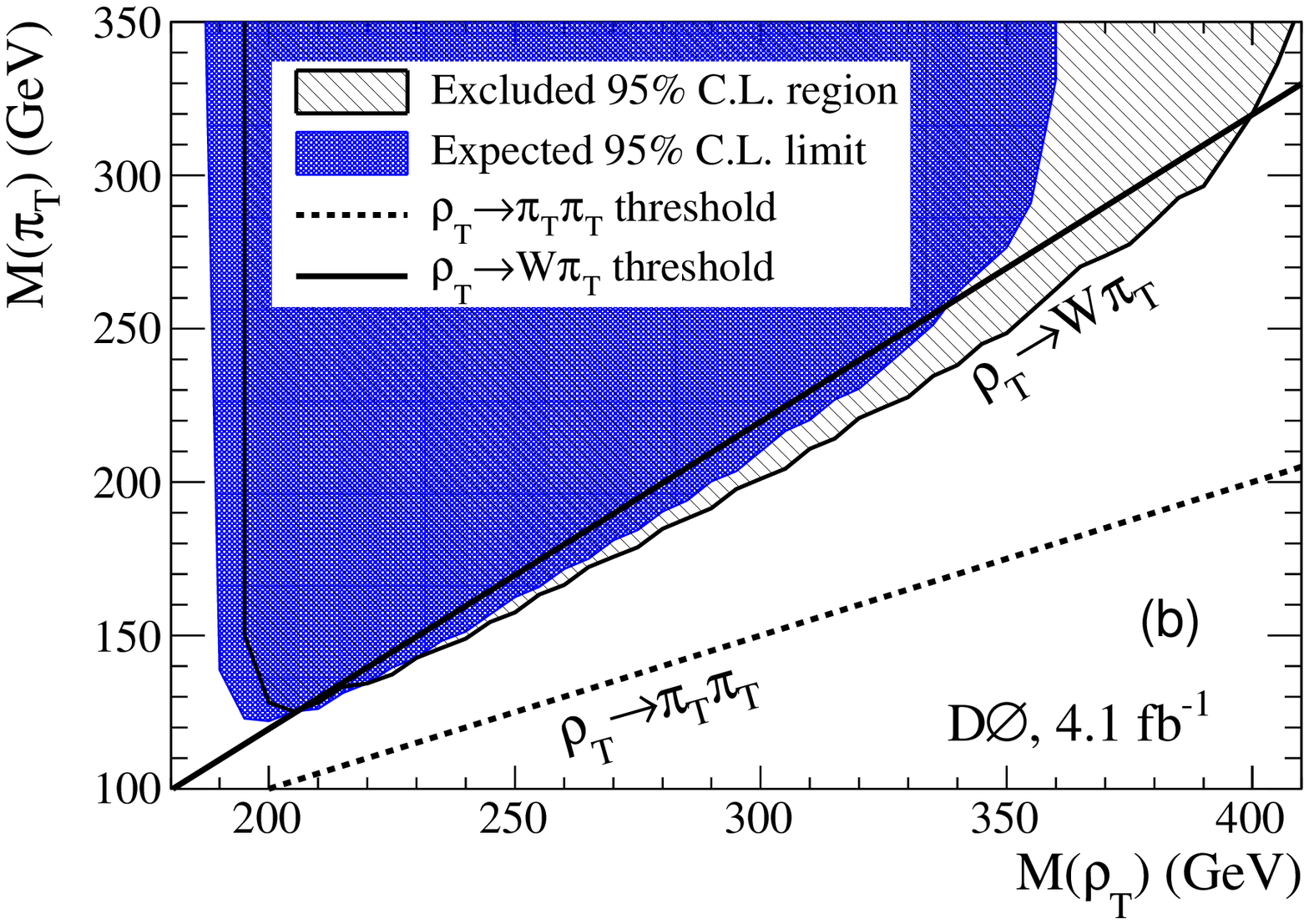}
\end{center}
\caption{(color online) Results for technicolor models. (a) The exclusion region 
in the $M_{\pi_T}$ vs. $M_{\rho_T}$ plane from  the $bb+W$ final state from the CDF experiment and (b)
 the results for
  the $\rho_T\to WZ$ final state from the D0 experiment.
\label{fig-TC}}
\end{figure}

\subsubsection{\textit{Other resonance searches}}\label{sec_reson5}

We next mention a few resonance searches
that do not fall
in to any of the above categories. 
The SM predicts the branching ratio $Z\to\pi^0\gamma$ to be between $10^{-12}$ and $10^{-9}$.
The CDF experiment\cite{PhysRevLett.112.111803} searched for this rare process  by looking for a narrow resonance
with $m\sim 90$~GeV in the $\gamma\gamma$ invariant mass spectrum; This search is extended to include the
quantum mechanically forbidden processes $Z\to\gamma\gamma$ and $Z\to\pi^0\pi^0$.
No significant excess in the data was found, and 95\% C.L.\  upper bounds on the branching ratios are determined:
${\cal BR}(Z\to\pi^0\gamma)<2.01\times10^{-5}$,
${\cal BR}(Z\to\gamma\gamma)<1.46\times10^{-5}$, and
${\cal BR}(Z\to\pi^0\pi^0)<1.52\times10^{-5}$,
which remain
most stringent in the world.

In 2011 the CDF experiment~\cite{PhysRevLett.106.171801} created a world--wide stir when it 
reported an excess of events in the invariant mass
distribution of jet pairs produced in association with a $W$ boson  in the leptonic final state. 
 The observed excess, with a dijet invariant mass between 120 and 160 GeV, was
at the 3.2 s.d.\ level, appeared to have a Gaussian shape (as expected from a new particle due 
to mass resolution effects) and gave a production cross section at the 4~pb level
 as shown in Fig.~\ref{fig-dijetbump}(a,b).
 Following this lead, the D0 experiment~\cite{PhysRevLett.107.011804} investigated 
 this final state, but 
no excess was found as shown in Fig.~\ref{fig-dijetbump}(c);
  limits were set that excluded new particle production above 1.9~pb. 
The final resolution of this potential excess came when 
 the CDF experiment published the result 
in the jets$+\met$ final state\cite{PhysRevD.88.092004} and an updated version~\cite{PhysRevD.89.092001} of the 
leptonic analysis, where a number of  
systematic effects were investigated
and taken into account including improved
understanding of the detector response to quarks and gluons separately, 
 and modeling of instrumental backgrounds.
In these searches 
there is no indication of an excess
 and the final results are shown in 
Fig.~\ref{fig-dijetbump}(d). The  95\% upper limit was set
on the production cross section of the new particle at 0.9~pb.
This story underscores the need for two experiments.

\begin{figure}[htb]
\begin{center}
\vspace{-0.97cm}
\includegraphics[height=5.2cm]{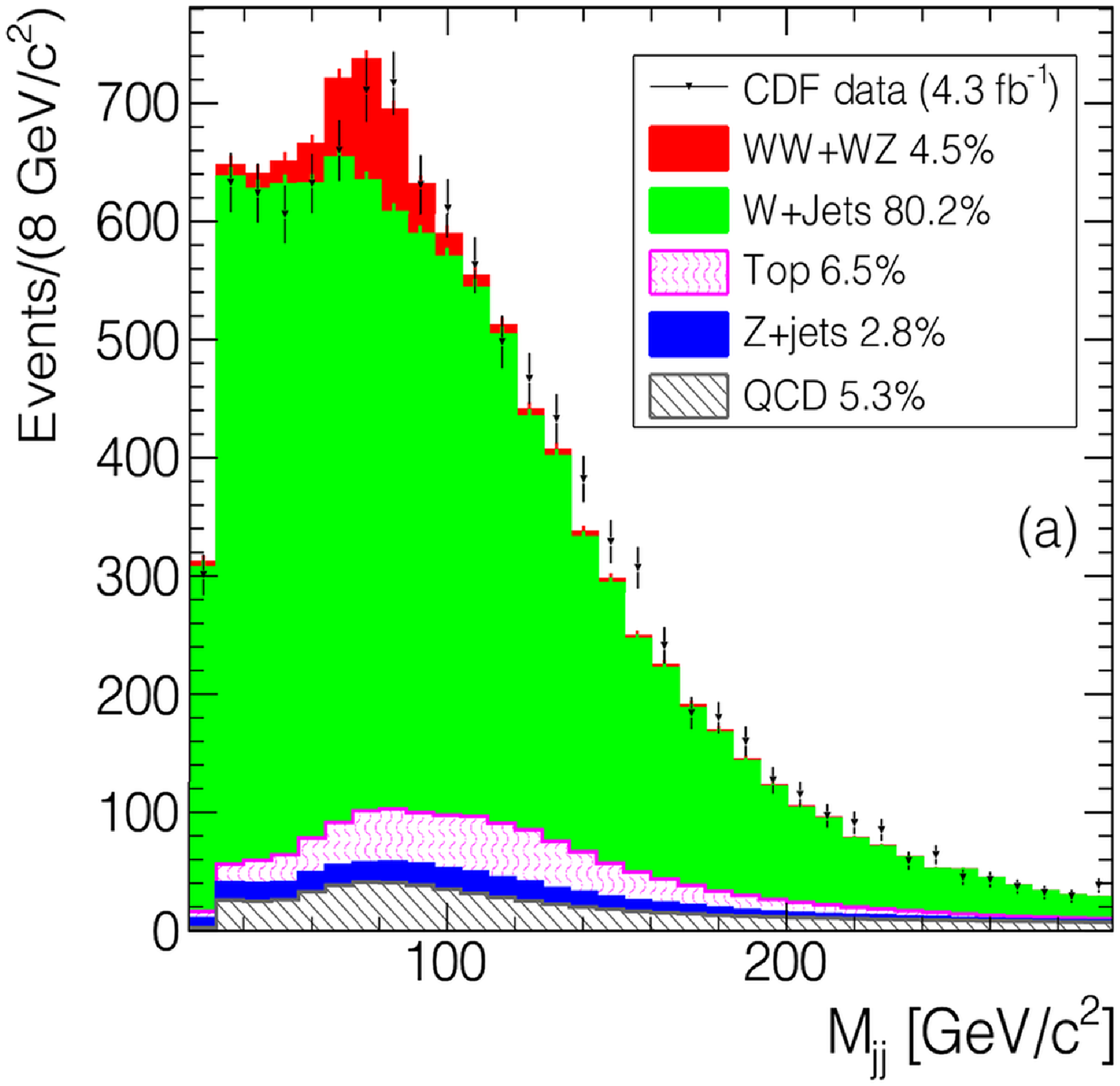}\hspace{1cm}
\includegraphics[height=5.2cm]{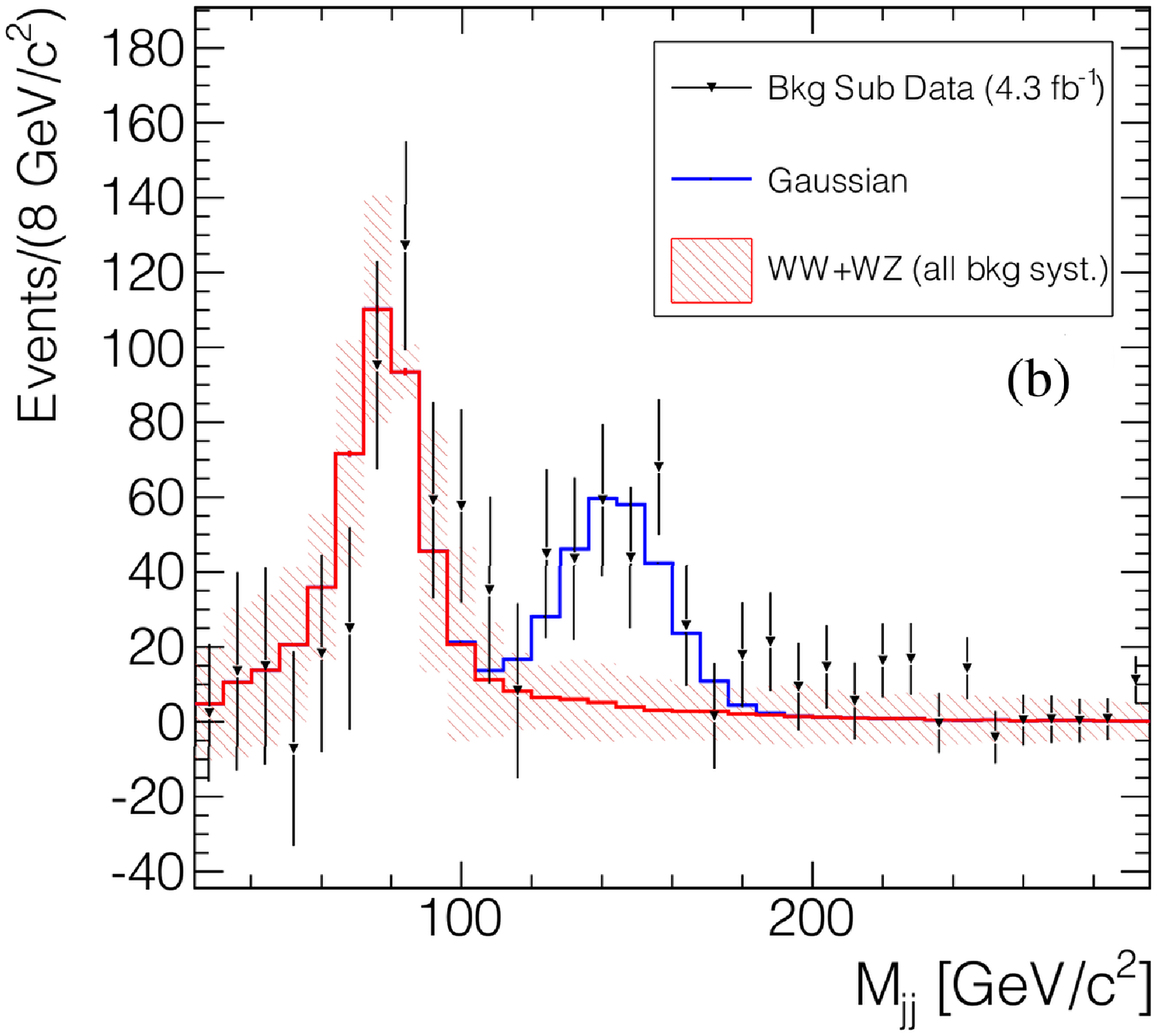}
\includegraphics[height=4.2cm]{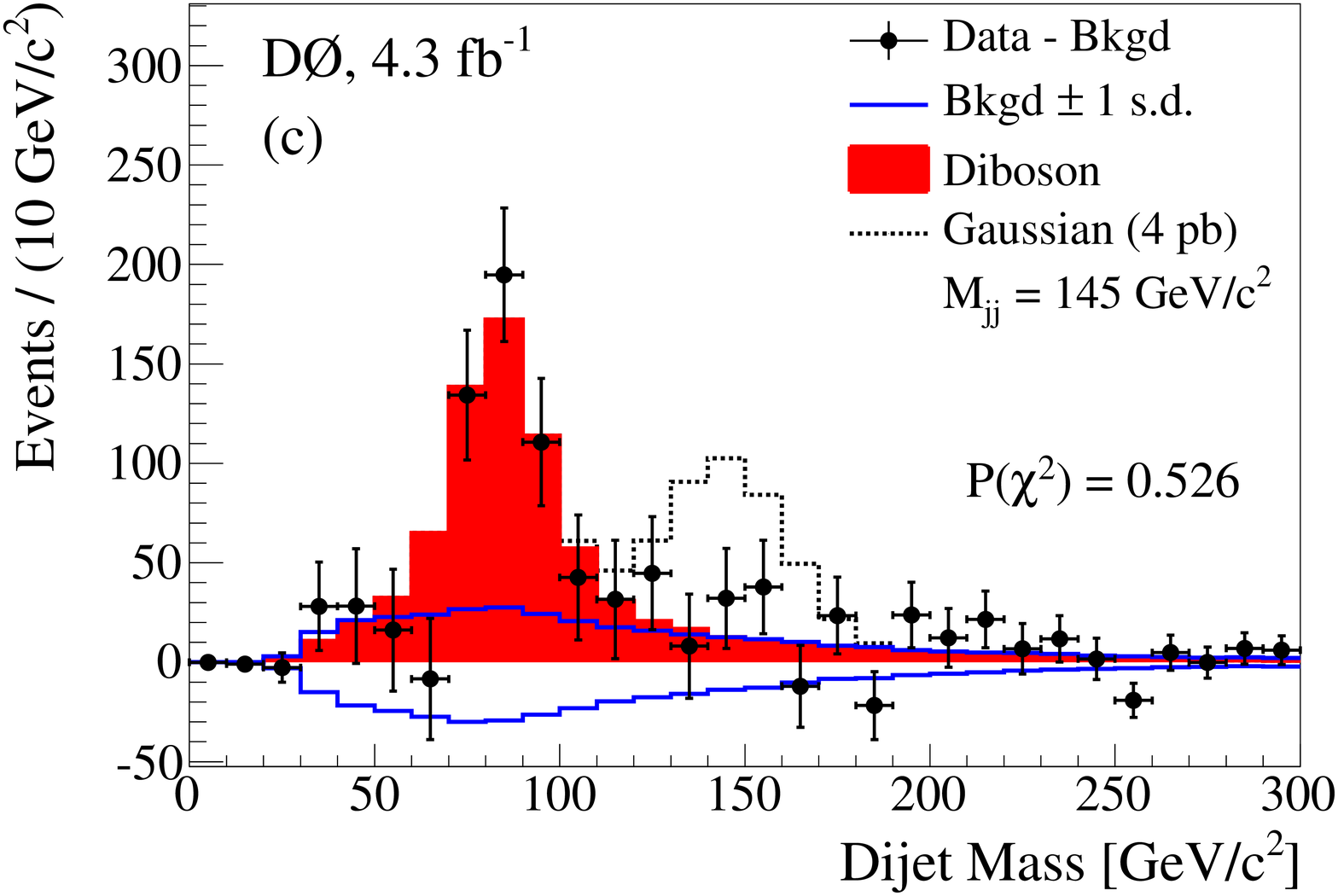}
\includegraphics[height=4.3cm]{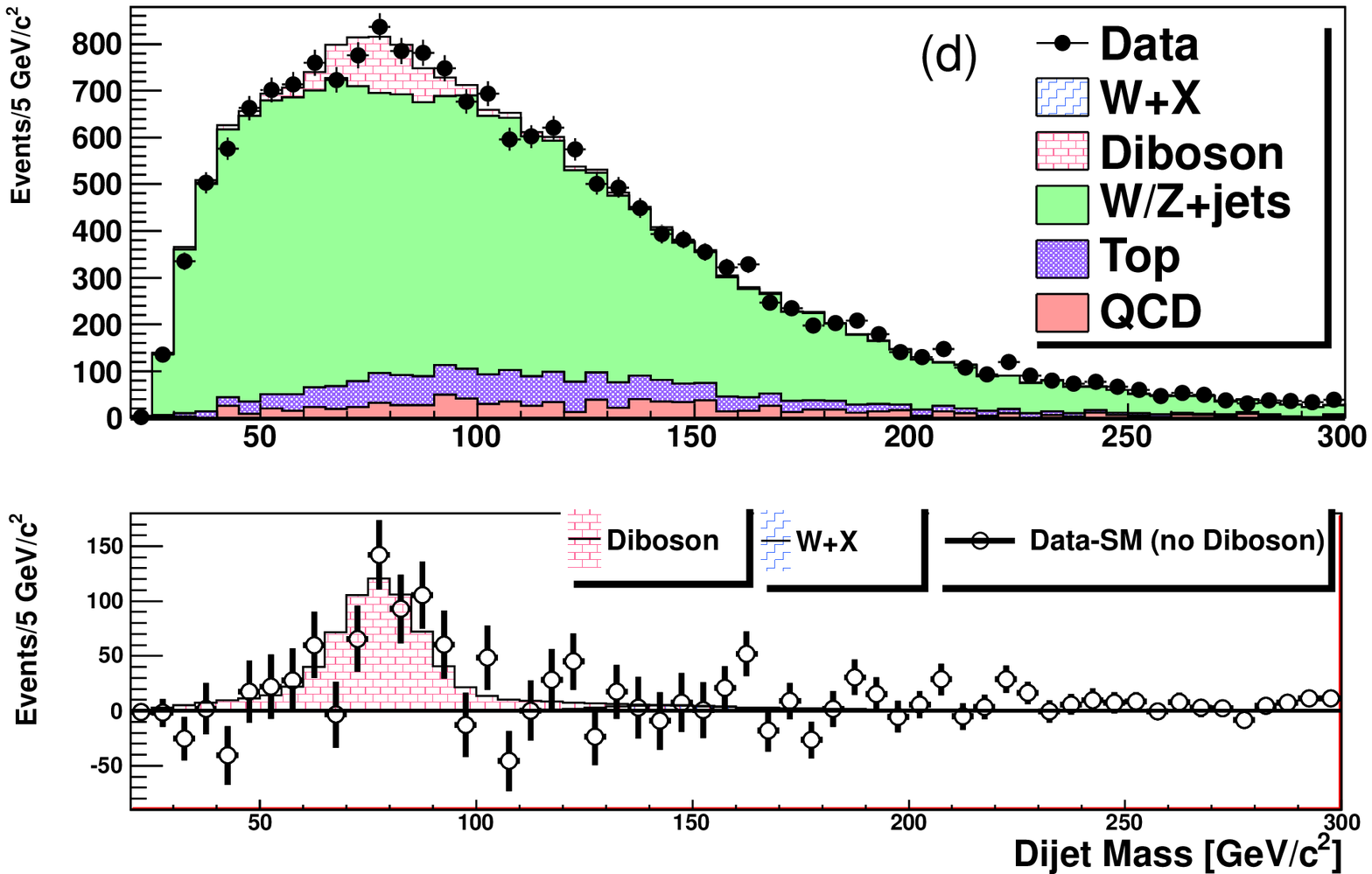}
\end{center}
\caption{(color online) The dijet invariant mass
 in $W$+dijet events  in the $\ell+\met$+2jet final state: (a) the data plotted on top of the known SM processes showing an excess
 around 140 GeV, and  (b) the same data with 
 the SM backgrounds (except $WW$ and $WZ$) subtracted, and using a  
 Gaussian component fit for the excess region
 from the CDF experiment. 
 (c)
The results in the same final state from the D0 experiment. 
(d) The updated
CDF analysis showing  that the original excess  was due to detector and analysis effects.
\label{fig-dijetbump}}
\end{figure}

\subsection[\textbf{\textit{Hidden--valley models, CHAMPS and long--lived particles}}]
{\textbf{\textit{Hidden--valley models, CHAMPS and other long--lived particles}}}\label{sec_LL}

There are many different types of long--lived particles predicted in new models. A
few have already been described in  the GMSB section, but there
 are others
such as hidden--valley model particles, CHAMPS (typically in SUSY models), monopoles, stopped gluinos and quirks which are described next.

\subsubsection{\textit{Hidden--valley/dark photons}}\label{sec_LL1}
 
 Hidden valley (HV) models provide a framework for studying the phenomenology 
 of secluded sectors, but make no specific predictions.
The D0 experiment performed variety of different searches. The first
 analysis\cite{PhysRevLett.103.071801} searched for 
Higgs boson production and decay 
into a pair of neutral long--lived HV particles that each decay to a $b\bar{b}$ pair.
The search is for pairs of 
 very--displaced vertices in the tracking detector, with radii in the range between 
1.6–-20~cm from the beam axis. No excess is 
found and limits are set as shown in Fig~\ref{fig_hv}(a).

HV models can also include SUSY.
 In one search 
 gaugino pairs can be produced and 
 decay into HV particles, 
 in particular a new light gauge boson (known as a dark photon) which in turn decays via fermion pairs, 
 and HV (or dark) neutralinos which escape the detector and produce large \met. This final state
 includes a photon, two spatially close leptons and large \met. 
 Since there is no evidence of dark photons, limits are set~\cite{PhysRevLett.103.081802} and the results 
 are shown in Fig.~\ref{fig_hv}(b).
A complementary search for 
 gaugino pair production is done by searching for a pair of dark photons, a pair of dark neutralinos and 
 other SM particles in the final state. These events have the unique final state of a 
 pair of isolated ``jets'' of charged leptons, so--called leptonic jets, produced in association
 with a large amount of \met. Again, no 
 evidence was found~\cite{PhysRevLett.105.211802} and limits are shown in 
Fig.~\ref{fig_hv}(c). 
 Finally, searches can be done for long--lived particles in HV  with pairs of electrons or photons
 in the final state
with results from $b'\to Zq\to eeq$\cite{PhysRevLett.101.111802}
shown in Fig.~\ref{fig_hv}(d).

\begin{figure}[htb]
\begin{center}
\includegraphics[height=4.3cm]{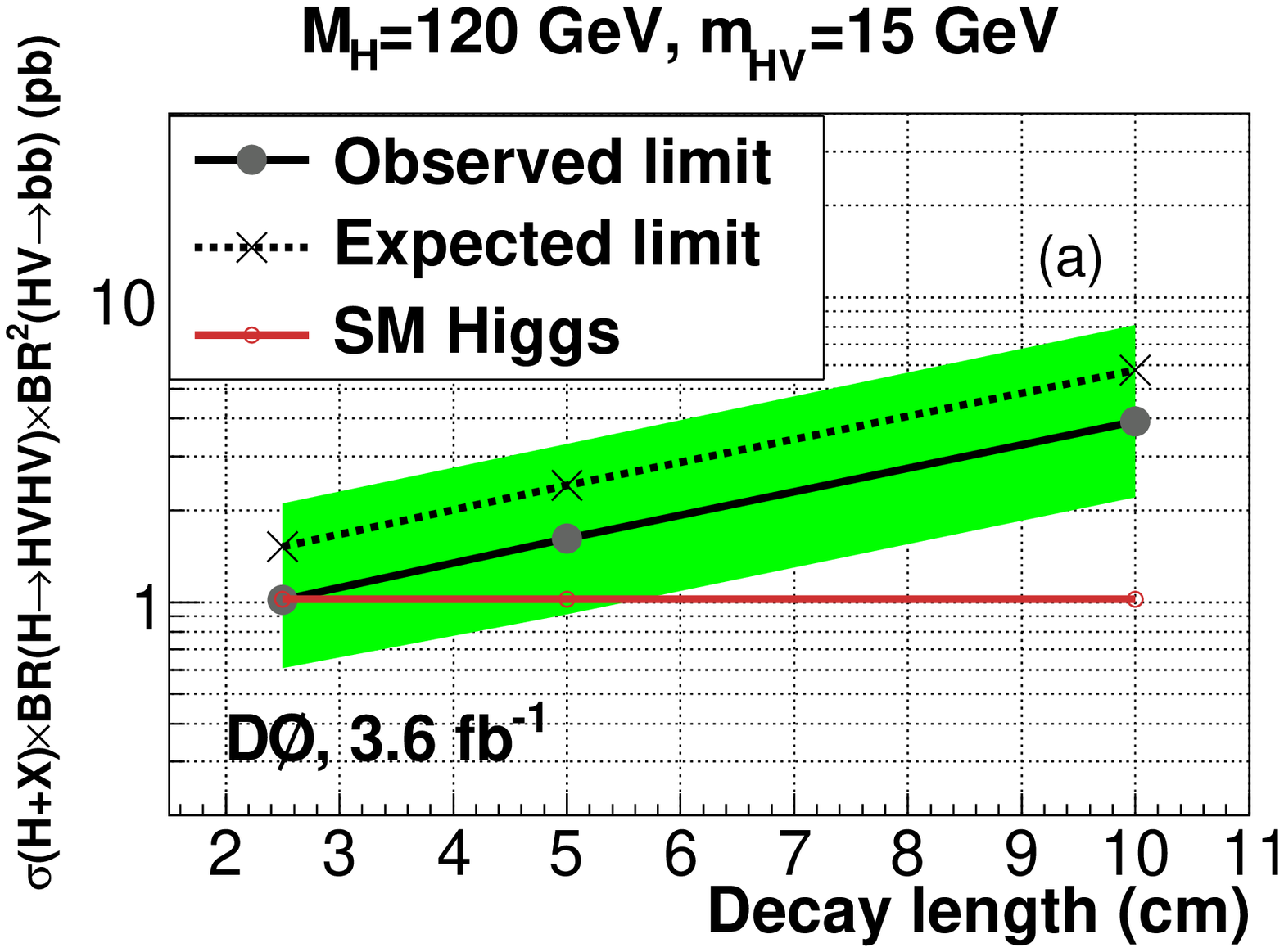}
\includegraphics[height=4.3cm]{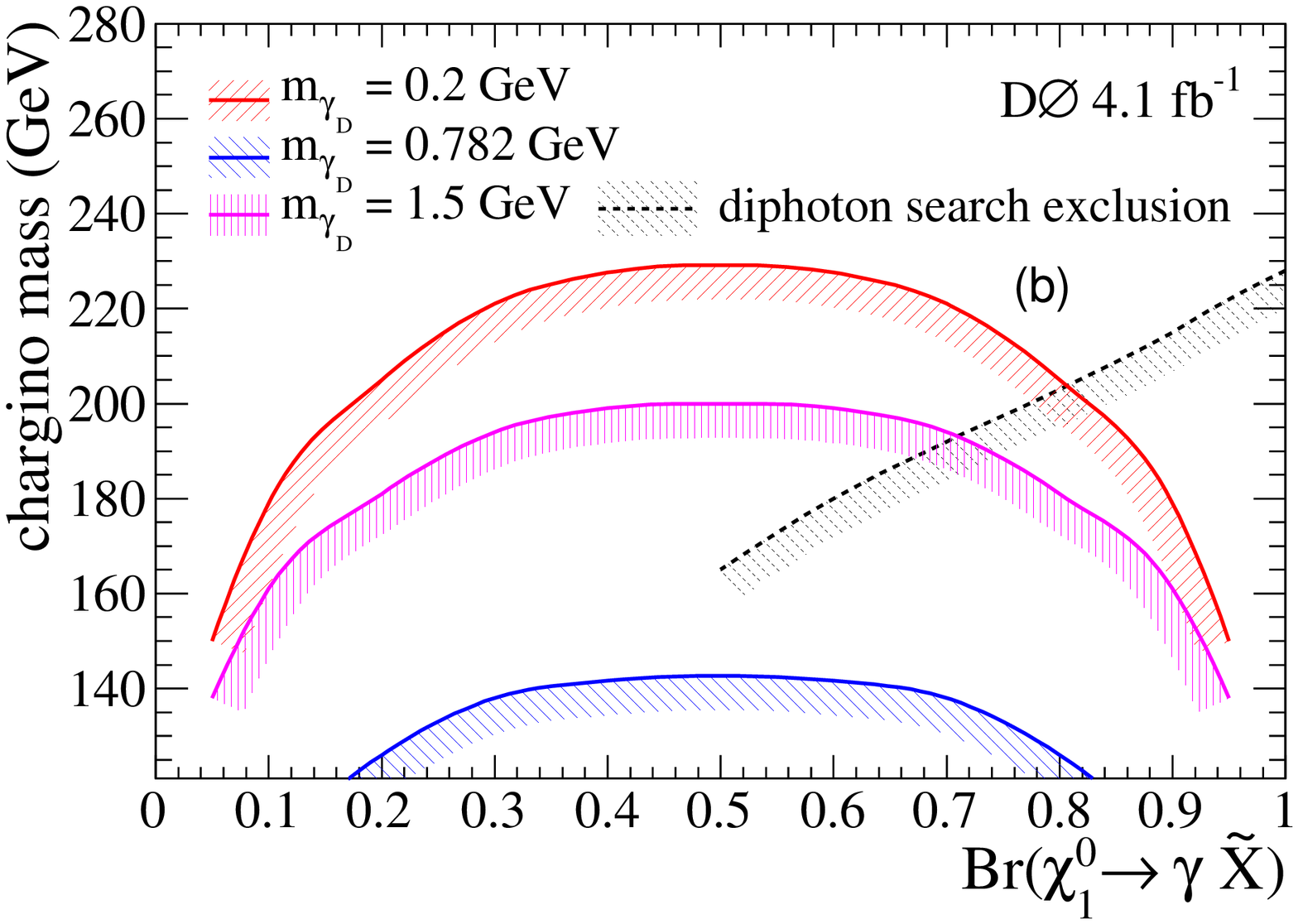}
\includegraphics[height=4.25cm]{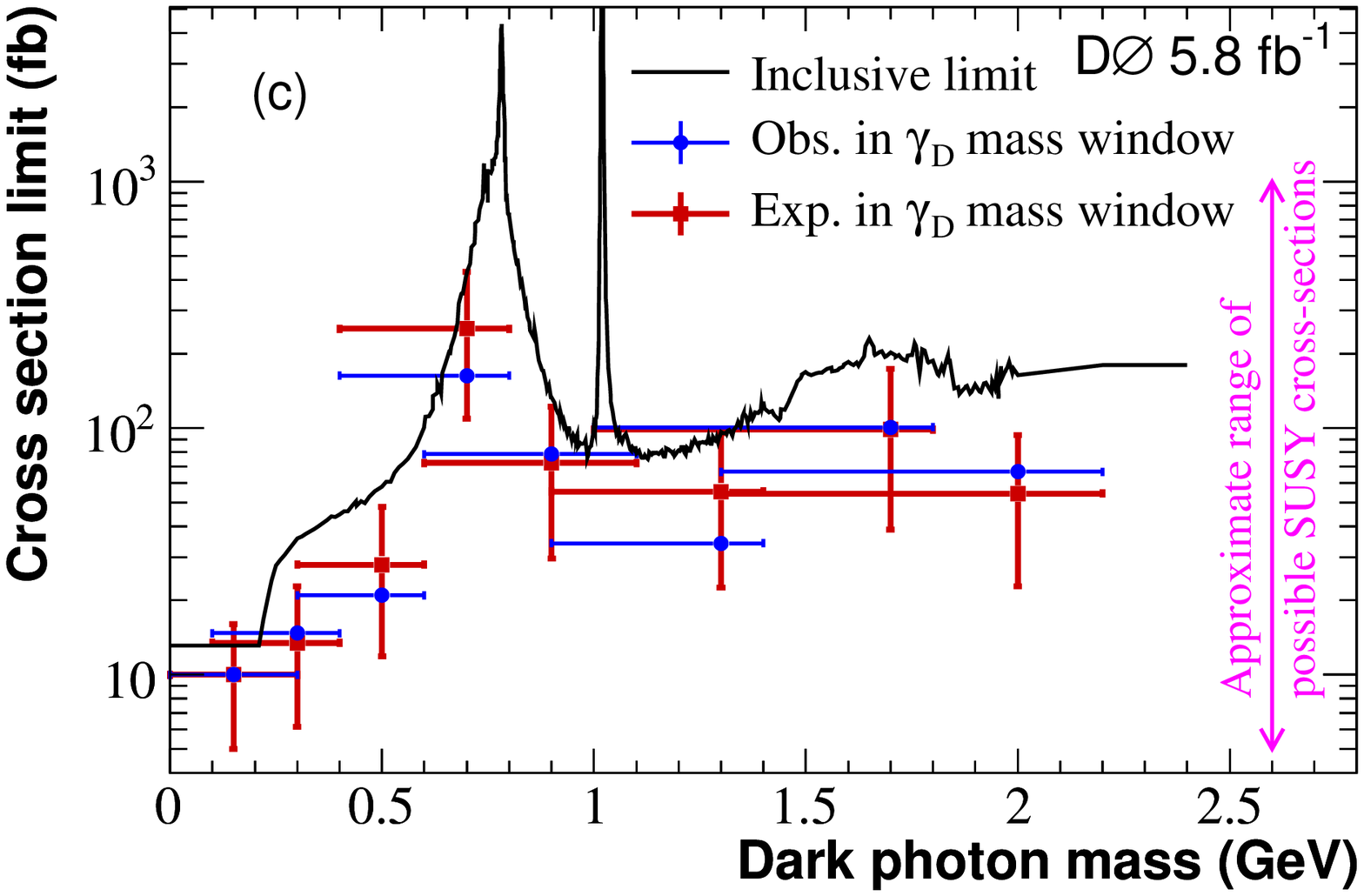}
\includegraphics[height=4.25cm]{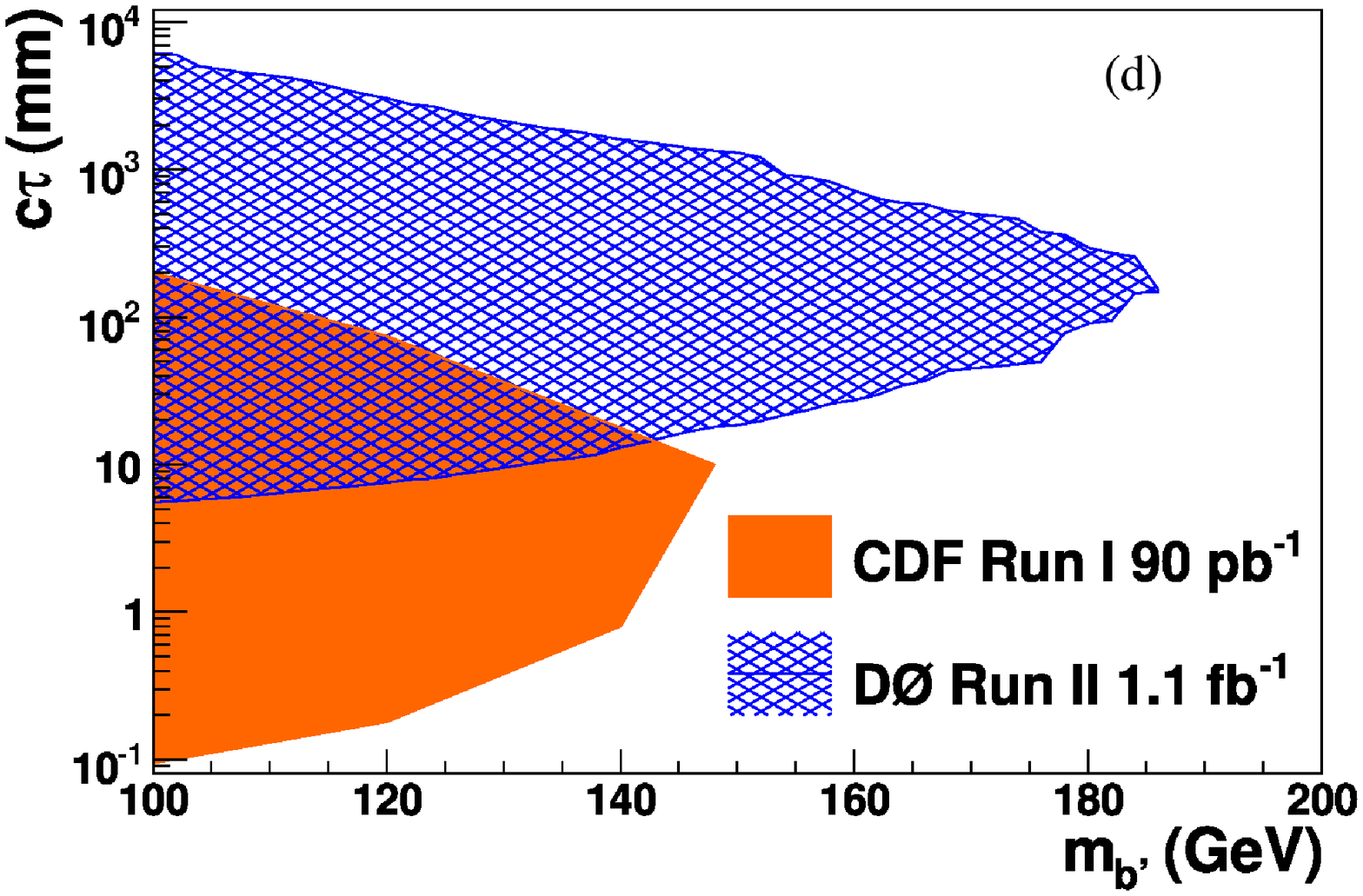}
\end{center}
\caption{(color online) Results from searches for the new particles from hidden--valley models
 from the D0 experiment. 
(a) The 
  95\% C.L. upper limits on the $\sigma\times{\cal BR}$ of the 
$H+X\to HV HV+X\to b\bar{b} b\bar{b}+X$ as a function of the decay length.
(b) The excluded region in the chargino mass vs. 
 the ${\cal BR}$ of the $\chioneO$ into a photon
for different dark photon masses from the search with a photon in final state. 
(c)
 The 95\% C.L. upper limit on  $\sigma$ as a function on the 
dark photon mass in the search with leptonic--jets in final state. (d) The 
excluded region in the $c\tau_{b'}$ vs. $m_{b'}$ plane in the search for long--lived
$b'\to Zq\to eeq$. \label{fig_hv}}
\end{figure}

\subsubsection{\textit{Charged massive stable particles}}\label{sec_LL2}

Searches for CHAMPS are typically done by examining events for the presence of
 a single charged particle 
that behaves like a
``heavy muon'' in that it only interacts as a minimum ionizing particle as it traverses the detector.
These particles can be produced directly (often as pairs) or as decay products of other particles. 
Both the CDF\cite{PhysRevLett.103.021802}
and the D0\cite{PhysRevLett.102.161802, 
 PhysRevLett.108.121802, PhysRevD.87.052011} experiments found no evidence for CHAMPS.
Results are 
typically interpreted in SUSY models with limits as  
 shown in Fig.~\ref{fig_ch1}(a,b) where the production mechanisms are  
 gaugino--like, stop--like , or $\tilde{\tau}$--like CHAMPS.

\begin{figure}[htb]
\begin{center}
\includegraphics[height=5.2cm]{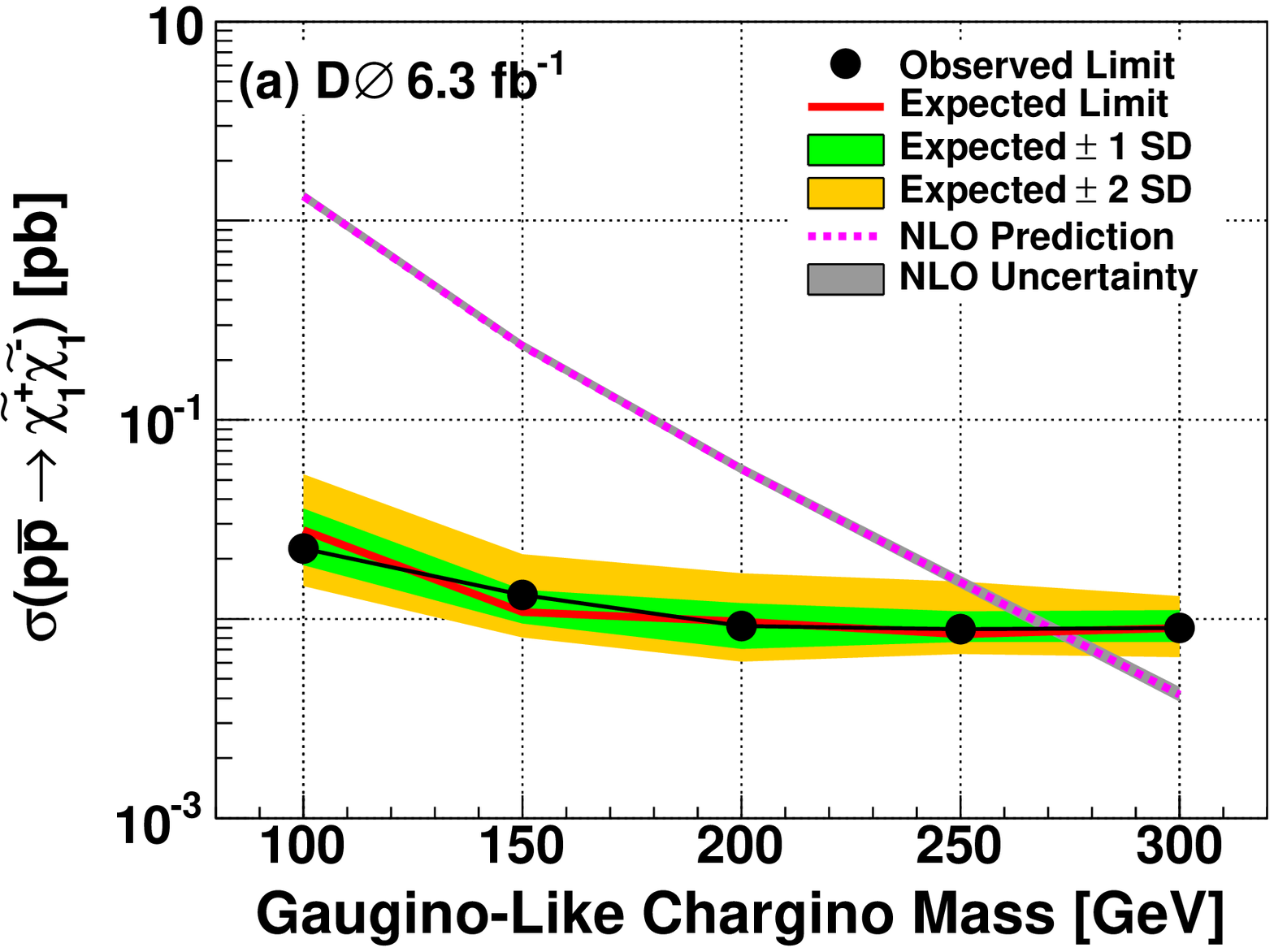}
\includegraphics[height=4.8cm]{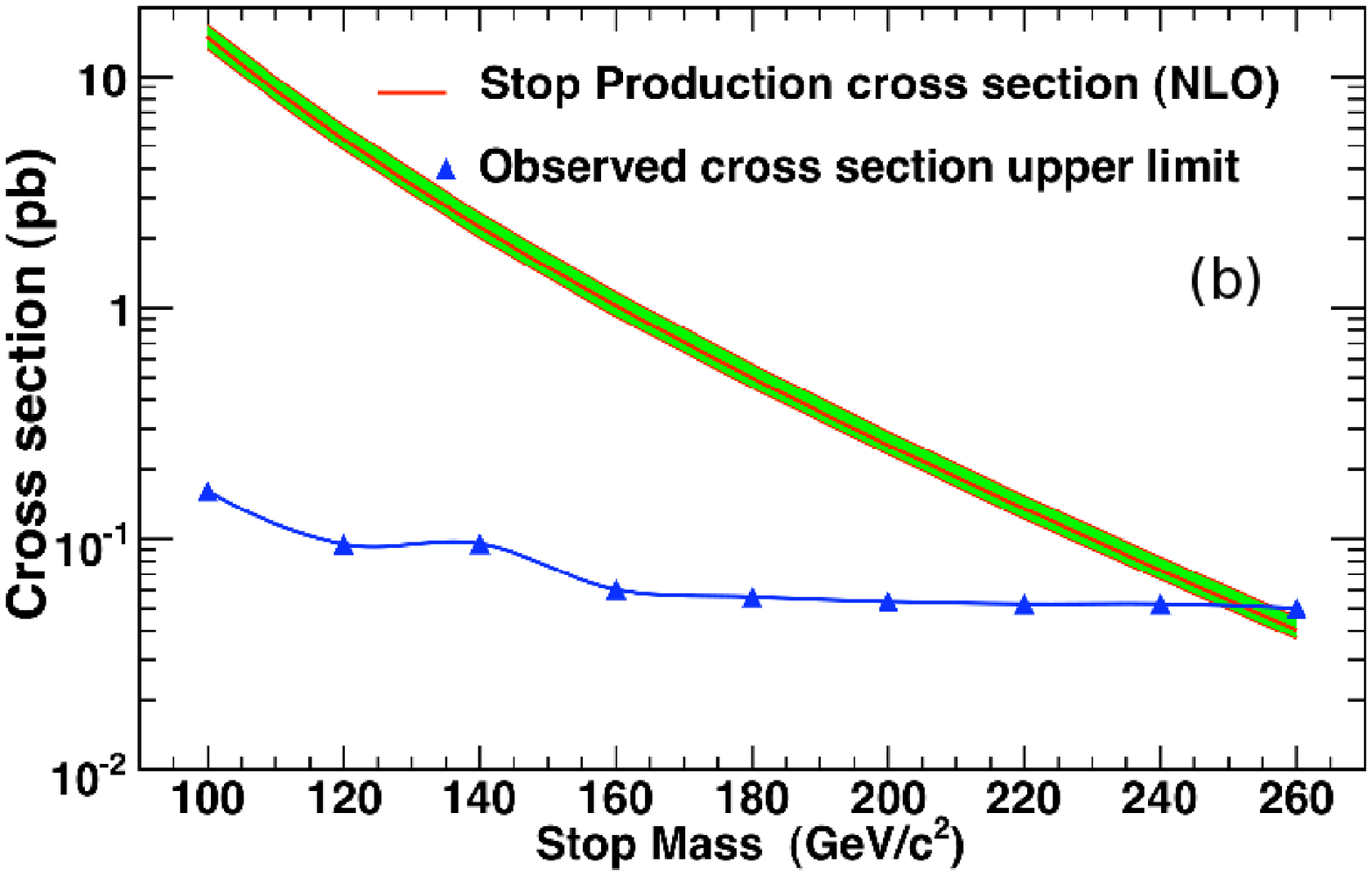}
\includegraphics[height=5.2cm]{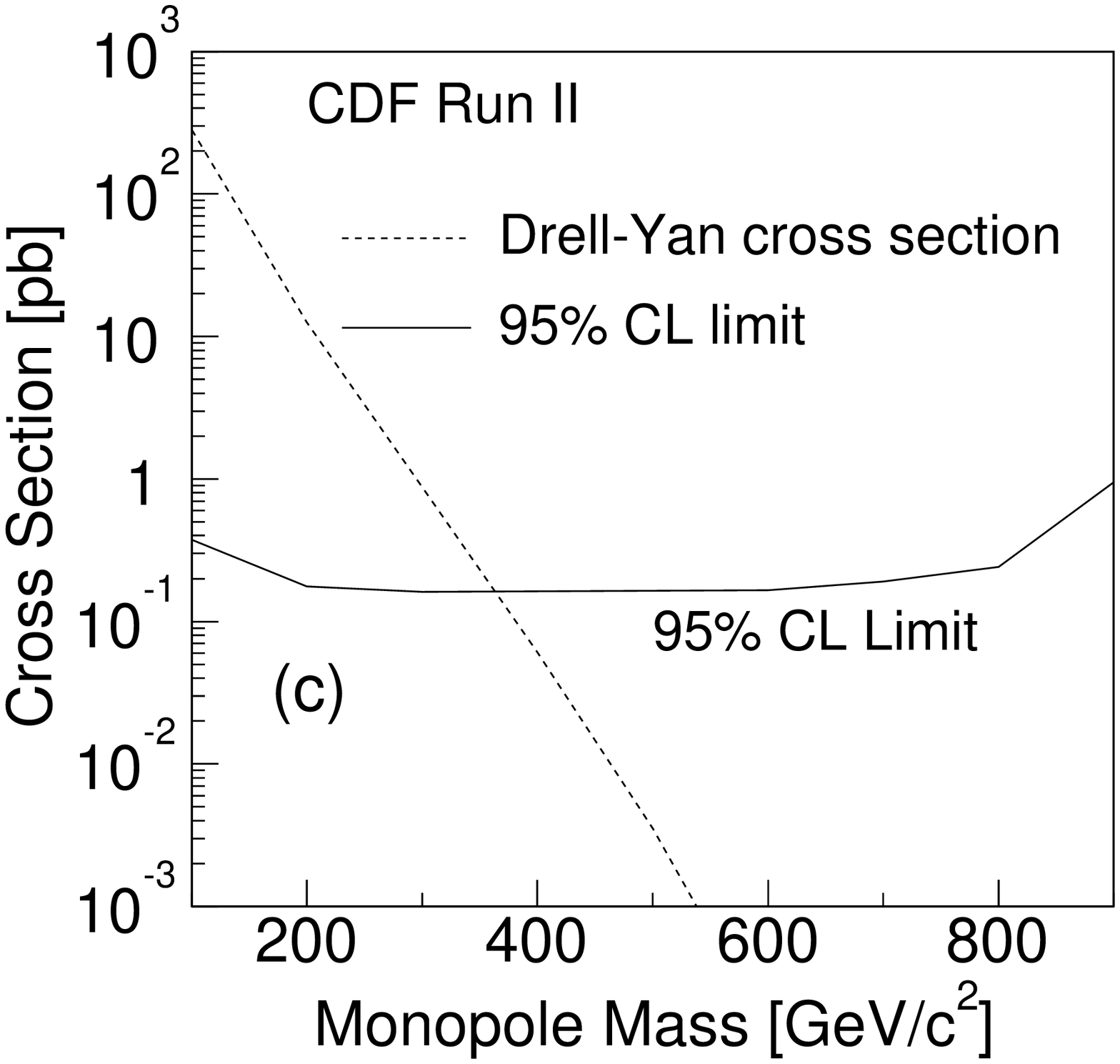}

\end{center}
\caption{(color online) Results for long--lived particle searches.
The 95\% C.L.\ cross section upper limits for  (a)   gaugino--like charginos
  CHAMPS from the D0 experiment and for (b)  top--squark CHAMPS 
 from the CDF experiment. (c) The 95\% C.L. cross section upper limit as a function of magnetic monopole mass
 from the CDF experiment.
\label{fig_ch1}}
\end{figure}

\subsubsection[\textit{Monopoles, stopped gluinos and quirks}]
{\textit{Other searches for long--lived  particles -- monopoles, stopped gluinos and quirks}}

Pairs of Dirac magnetic monopoles, if they exist in nature, are predicted to 
 be directly produced in collisions. 
Because of their large mass and magnetic charge they will move differently 
through the magnetic field and can be identified by their late time--of--arrival at the outer parts of the detector. 
Searches for monopoles in Run~II have been done at the 
CDF detector~\cite{PhysRevLett.96.201801} with no evidence of new production. 
Monopoles, assuming simple models of Drell--Yan style production,  are excluded at 95\% C.L. for masses smaller than 360~GeV
(see Fig~\ref{fig_ch1}(c)).

Searches for stopped gluinos are done by looking for 
$R$--hadrons that get trapped in the calorimeter. They can then 
 decay up to 
100 hours after their production. The search is done  at the D0 experiment by looking for deposits of energy in the calorimeter which are not 
 synchronized with an accelerator bunch crossing. 
Results\cite{PhysRevLett.99.131801} are shown in the Fig.~\ref{fig-Quirk}(a). 

The Fermilab neutrino experiment NuTeV observed an excess of 
dimuon events~\cite{Adams:2001ska} that could be interpreted as SUSY models with R--parity violation~\cite{Martin:1997ns} 
or HV models~\cite{Strassler:2006im}.
A follow--up analysis at the D0 experiment searched for pair-production of neutral particles each travelling 
 for at least 5~cm before decaying into a pair of muons\cite{PhysRevLett.97.161802}.
No evidence is found and limits are set with results shown in the Fig.~\ref{fig-Quirk}(b).

 New particles known as quirks, $Q$
which are strongly interacting under their own $SU(N)$ force,
 can be pair 
 produced at hadron colliders if they also carry SM charges. 
In addition to the quirk mass, 
the strength of the new $SU(N)$ gauge coupling, infracolor (which becomes 
strong at the scale $\Lambda$) is important phenomenologically. 
In a case when $\Lambda << m_Q\simeq 0.1 - 1$~TeV, breaking of the infracolor string is exponentially
suppressed due to the large value of the ratio $m_Q/\Lambda$, and the quirk--antiquirk pair stays connected by
the infracolor string like a rubber band that can stretch
to macroscopic length proportional to $m_Q/\Lambda^2$. 
 The D0 experiment~\cite{PhysRevLett.105.211803}
searched for cases where the extra gauge group 
 is $SU(2), SU(3)$ or $SU(5)$. 
In these scenarios, 
we have the unusual signature that the individual quirks ionize atoms in the tracking chamber, 
but the macroscopic distance between the quirk and anti-quark 
provide a neutral charged object that does not change direction as it traverses the detector. Thus, it can be 
reconstructed as a slow, highly ionizing, high $p_T$ track that decays after a few cm. 
Since there could be a high $p_T$ jet from initial state radiation, the signature will consist of 
this special type of track, one jet, and large \met\ aligned with the track. 
No evidence for quirks are found and
the results are shown in Fig.~\ref{fig-Quirk}(c).

\begin{figure}[htb]
\begin{center}
\includegraphics[height=4.3cm]{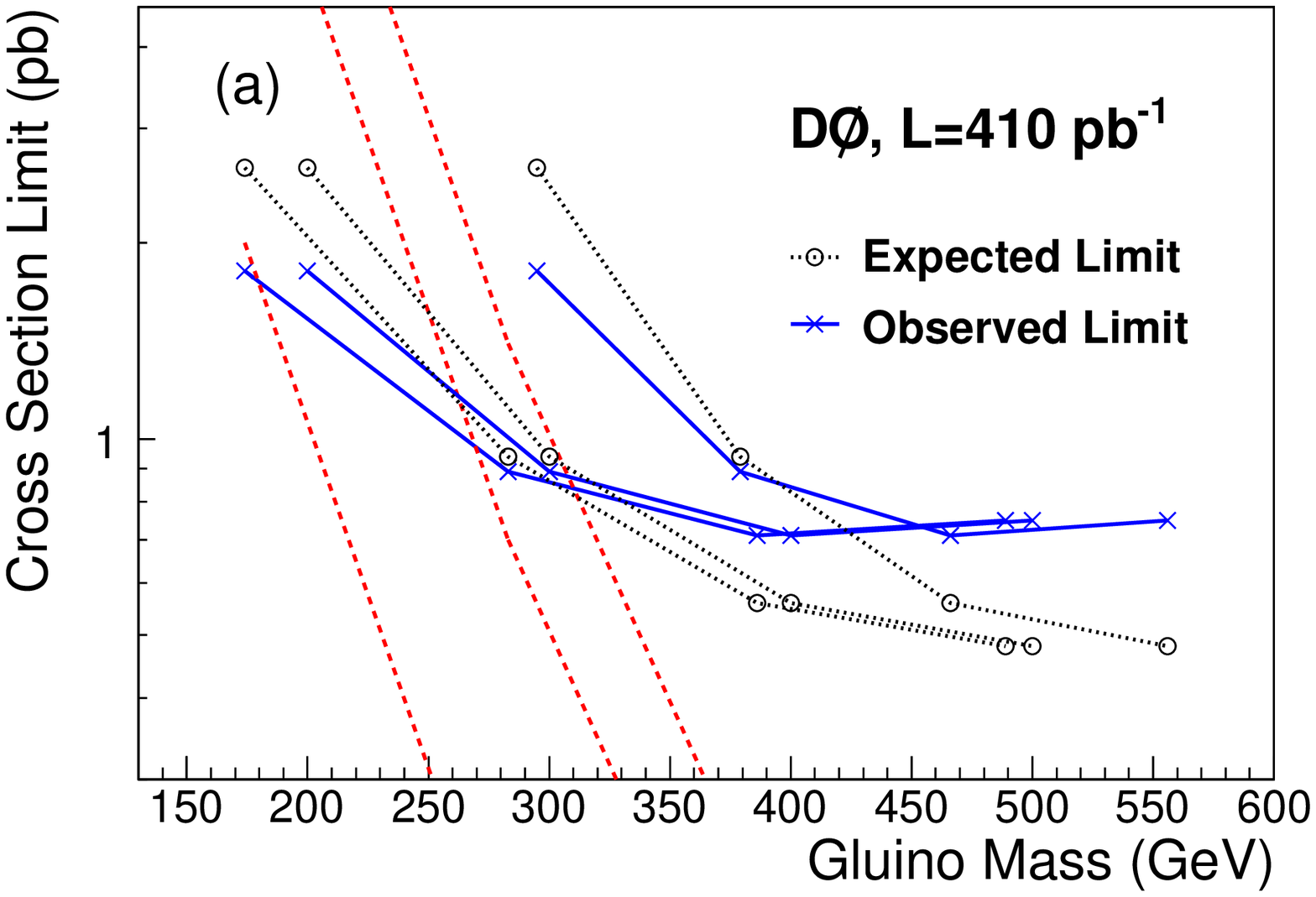}
\includegraphics[height=4.1cm]{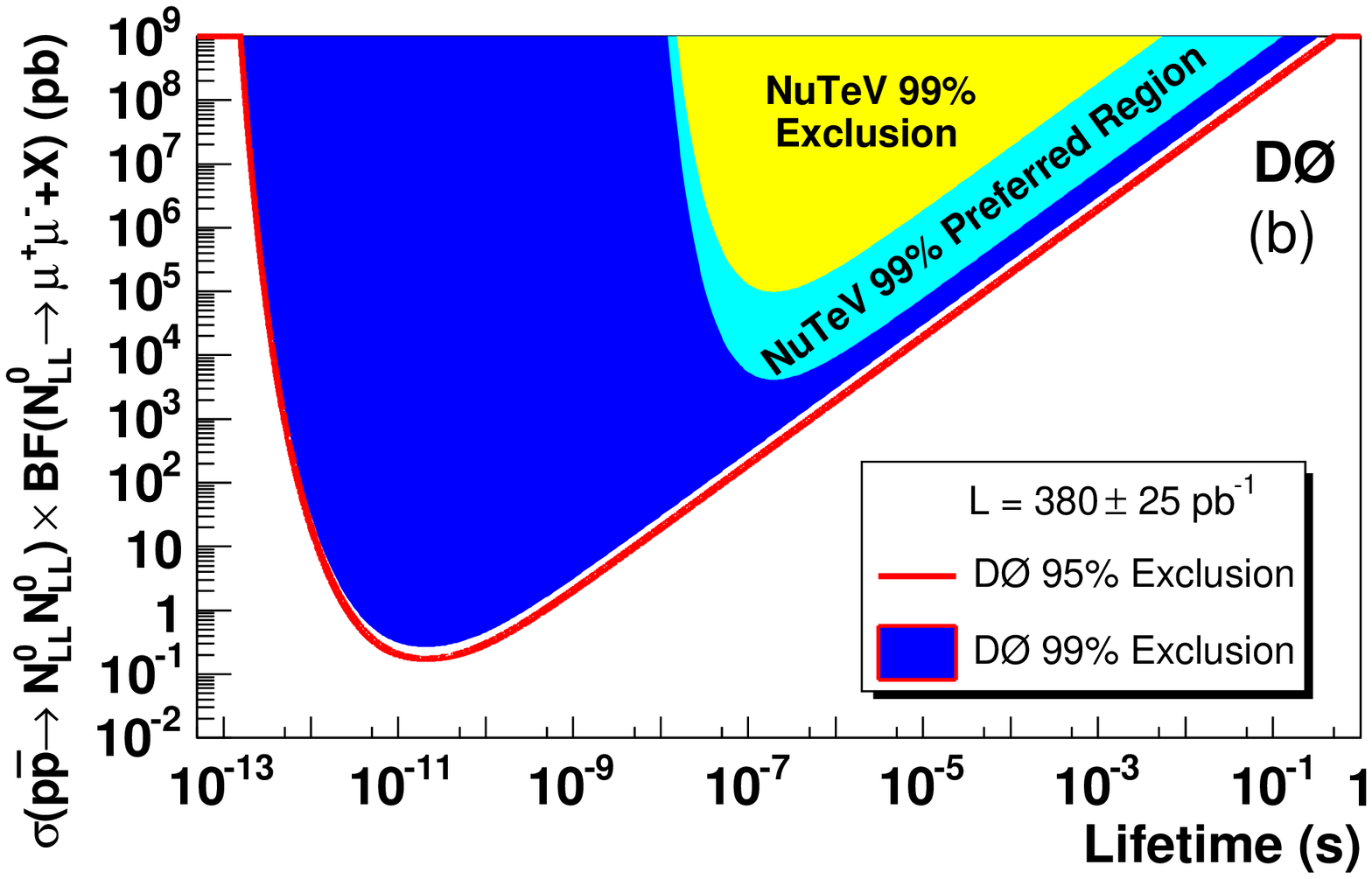}
\includegraphics[height=4.3cm]{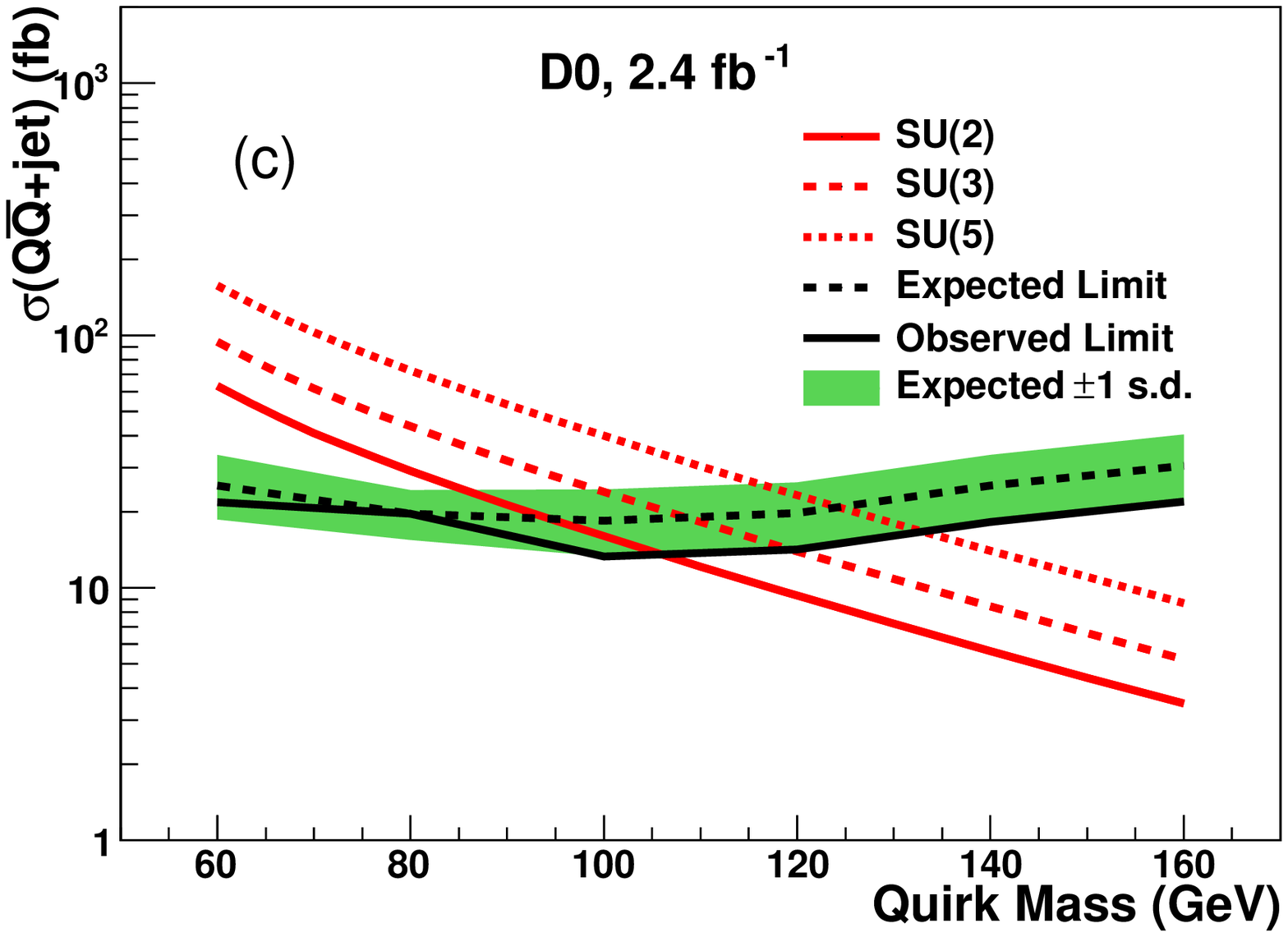}
\end{center}
\caption{(color online) (a) The 95\% C.L.\ cross section upper limits on  
  stopped gluinos
with the assumption that ${\cal BR}(\tilde{g}\to g\chioneO)=1$  from the D0 experiment. 
(b) The 95\% C.L. upper limits 
on the $\sigma\times{\cal BR}$ of 
 pair production of neutral long--lived particles decaying to
pairs of muons as a function of their lifetime.
(c) The 95\% C.L. cross section upper 
limits for pair of quirks and a jet (from initial state radiation)
as a function of the mass of the quirk. 
 \label{fig-Quirk}}
\end{figure}

\subsection{\textbf{\textit{Extra dimensions and dark matter}}}\label{sec-EDres}

While many models of extra dimensions are constrained by experiment\cite{Beringer:1900zz}, there is significant room to allow the 
possibility of new particles and interactions. On the one hand many LED and UED models predict 
new particles, 
which will
 ``leave'' the detector 
without interacting, or they will interfere with SM processes  in various final states;
evidence for the models in former case would show up in ways that are similar
to dark matter searches. On the other hand, excited KK modes of the 
graviton  which are localized on the SM brane,  spin--2 particles $G^*$, 
could produce resonances in $ee$, $\gamma\gamma$, $WW$ and $ZZ$ final states that are readily searched for. 

The D0 and the CDF experiments searched for extra dimensions in a number of ways. 
We begin with a description of the searches for 
LED models in which Kaluza--Klein (KK) gravitons are directly produced but 
immediately disappear.
In this case the gravitons are often produced with high transverse momentum and 
in association 
with a quark, a gluon, or a photon, giving rise to either monojet or monophoton final states
with a large \met\ due to the escaping graviton. 
No evidence of new physics is 
observed~\cite{PhysRevLett.101.011601,PhysRevLett.101.181602,PhysRevLett.97.171802} 
and results in the $M_D$ vs. $N_D$ plane, where $M_D$ 
the fundamental Planck scale in the $(4 + n)$--dimensional space--time
 and $N_D$
is number of extra dimensions,
 are shown in Fig.~\ref{fig-ED}(a).
Other models indicate that 
evidence can be inferred in fermion and/or boson final states
from the  interference between 
 KK gravitons and SM diagram terms.
The D0 experiment~\cite{PhysRevLett.102.051601,PhysRevLett.95.161602} investigated these signatures in the $ee$, 
$\gamma\gamma$ and $\mu\mu$ final states by searching for deviations in  the correlation between the invariant mass 
 and the angular distribution of the pairs from SM--only predictions. 
Results from $ee$ and $\gamma\gamma$ search are shown in Fig.~\ref{fig-ED}(b).

The searches for  UED processes typically focus on the 
 production and decay of KK particles, denoted here with a *. 
 Typically production begins with KK gluons ($g^*$) or quarks ($q^*$) and decay via 
$q^* \rightarrow qZ^* \rightarrow q (l l^*) \rightarrow ql (l \gamma^*)$. 
In the case with only one extra dimension, minimal UED (mUED), the $\gamma^*$ is stable 
and is a dark matter candidate.
This can result in a
final state which includes two leptons (same--sign or opposite sign) as well as a SM jet and \met.
The D0 experiment~\cite{PhysRevLett.108.131802} searched for mUED in the same--sign lepton
 final state 
and excluded $R_c^{-1}$ up to 260~GeV,
where $R_c$ is the radius of the compact dimension.
This limit corresponds to a mass of 317 GeV of the lightest KK quark.
When additional extra dimensions exist
the $\gamma^*$ decays via $\gamma^*\to\gamma+G$, where $G$ is graviton, yielding the $\gamma\gamma+\met$
 final state from pair production of $\gamma^*$. 
 For the D0 search for UED~\cite{PhysRevLett.105.221802}, a model with six extra dimensions, 
 a fundamental Planck scale of 5 TeV, and ${\cal BR}(\gamma^*\to\gamma G)\approx 1$
yielded  a limit of 
 $R_c^{-1}<477$~GeV along with the other results shown in Fig.~\ref{fig-ED}(c).

\begin{figure}[htb]
\begin{center}
\includegraphics[height=4.4cm]{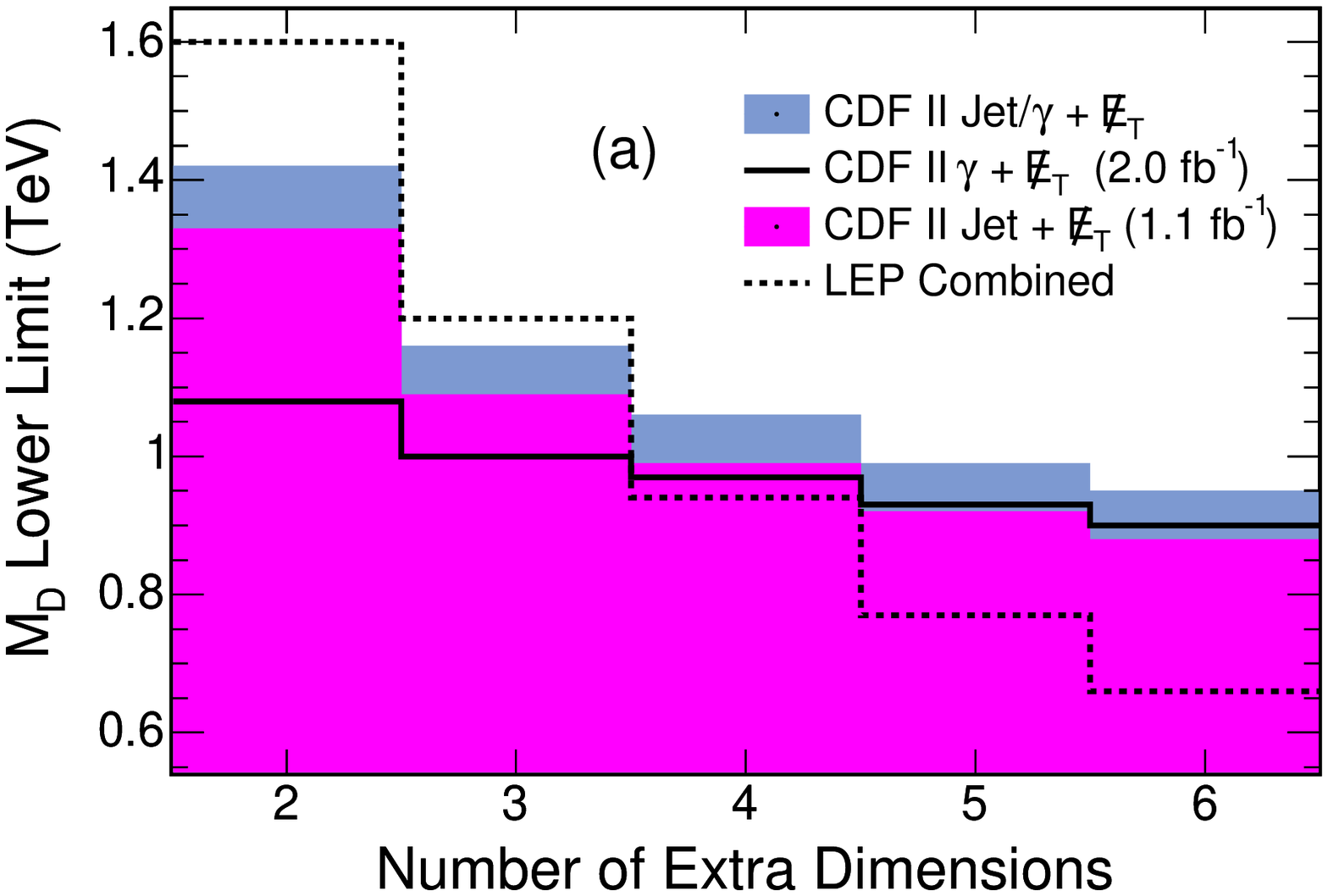}
\includegraphics[height=4.3cm]{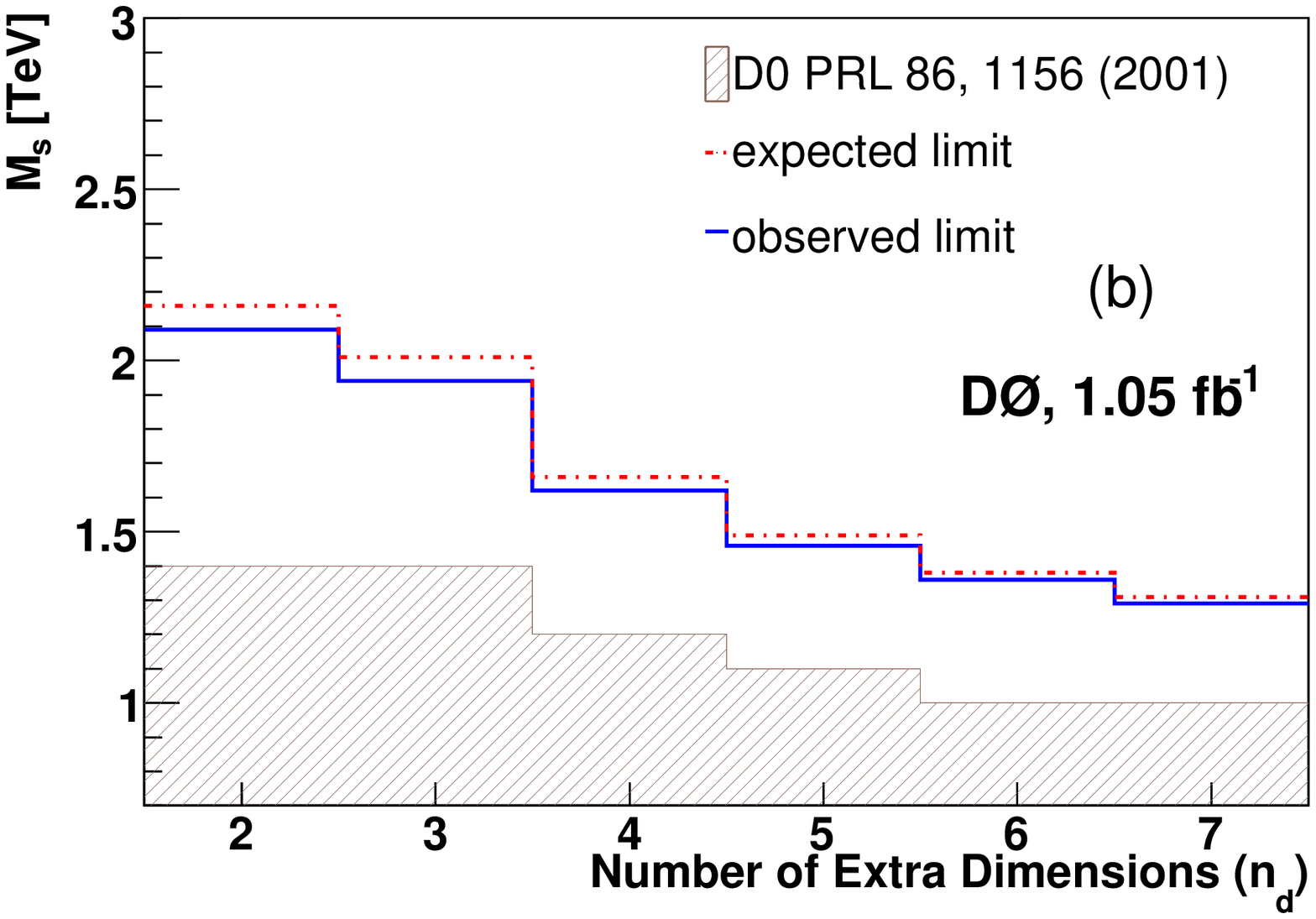}
\includegraphics[height=4.3cm]{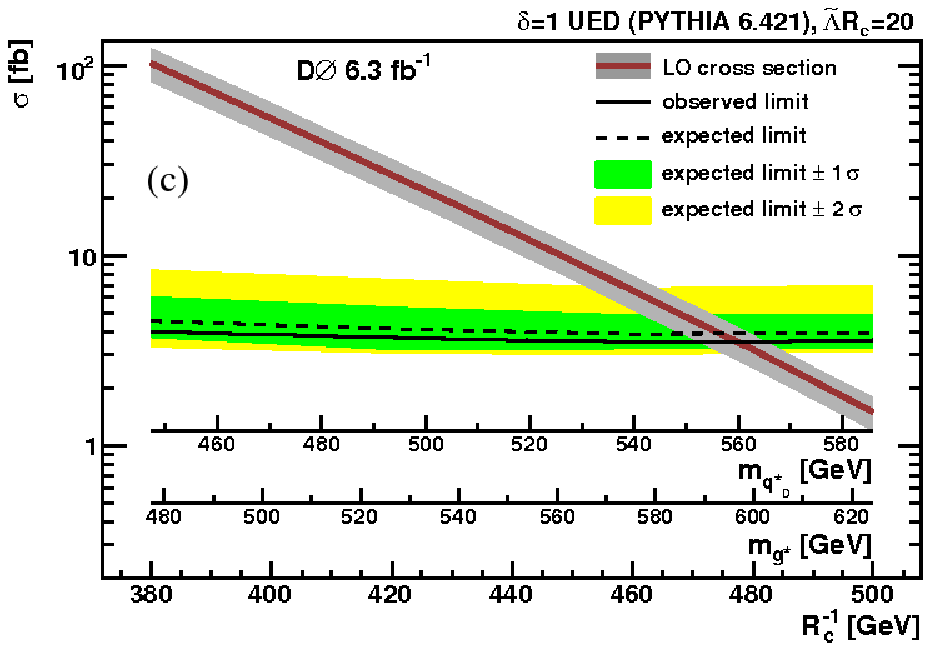}
\end{center}
\caption{(color online) Limits on extra dimensions from the CDF and D0 experiments. 
(a) The 
excluded region in the  
$M_D$ vs. 
$N_D$ plane in a search for LED from
 the combined monophoton and monojet final states
from the CDF experiment. 
(b)The limits from the 
  $ee$ and $\gamma\gamma$ final states from the D0 experiment.
 (c) The 95\% C.L.\ cross section upper limits for a UED model in the $\gamma\gamma+\met$ final state 
from the D0 experiment.
 \label{fig-ED}}
\end{figure}

Excited KK
modes of the graviton, $G^*$,  which are localized on the SM
brane, are predicted in 
 Randal--Sundrum (RS) models
 with a warped spacetime metric.
  Two parameters determine graviton
couplings and widths: the constant $k/\overline{M}_{PL}$, where $k$ is
the curvature scale of the extra dimension, and $\overline{M}_{PL}= M_{PL}/\sqrt{8}$ 
is the reduced Planck scale, and graviton excitation, $M_1$.
 The D0~\cite{PhysRevLett.95.091801,PhysRevLett.100.091802,PhysRevLett.104.241802} and the 
 CDF experiments~\cite{PhysRevLett.107.051801,PhysRevLett.102.031801,PhysRevD.83.011102}
searched for evidence of single  
 RS graviton
production and decay via $G^*\to \ell\ell$ or $VV$.
 No significant excess of events in the dilepton ($e$ or $\mu$) or $\gamma\gamma$ 
 was found and limits
are shown in Fig.~\ref{f8}(a,b).
The results from the searches for the other diboson resonances described in the previous section
 were also interpreted as limits on the RS graviton 
 production~\cite{PhysRevD.85.012008,PhysRevD.83.112008,PhysRevLett.104.241801,PhysRevLett.107.011801}.
Limits from the $G^*\to WW$ are set 
for a mass $m_{G^*}<754$~GeV when $k/\overline{M}_{PL}=0.1$.
 In $G^*\to ZZ$  
an excess of events is observed in low yield four--lepton 
channel at the $M_{G^*}=327$~GeV, but it was not confirmed in the more sensitive searches in the $\ell\ell jj$
and $\ell\ell\met$ final states\cite{PhysRevD.85.012008}. Figure~\ref{f8}(c) shows the results.

\begin{figure}[htb]
\begin{center}
\includegraphics[height=4.2cm]{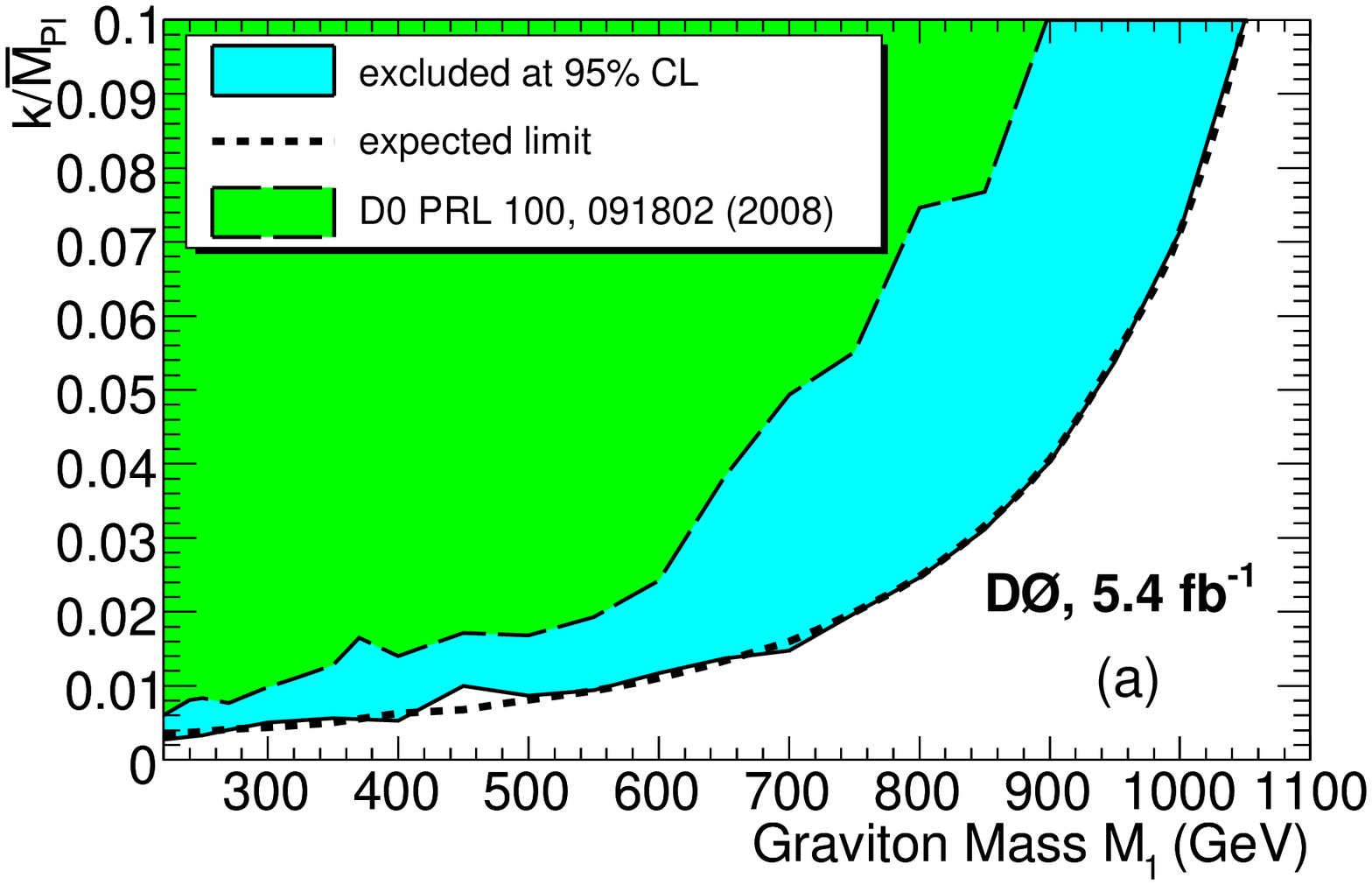}\hfill
\includegraphics[height=4.3cm]{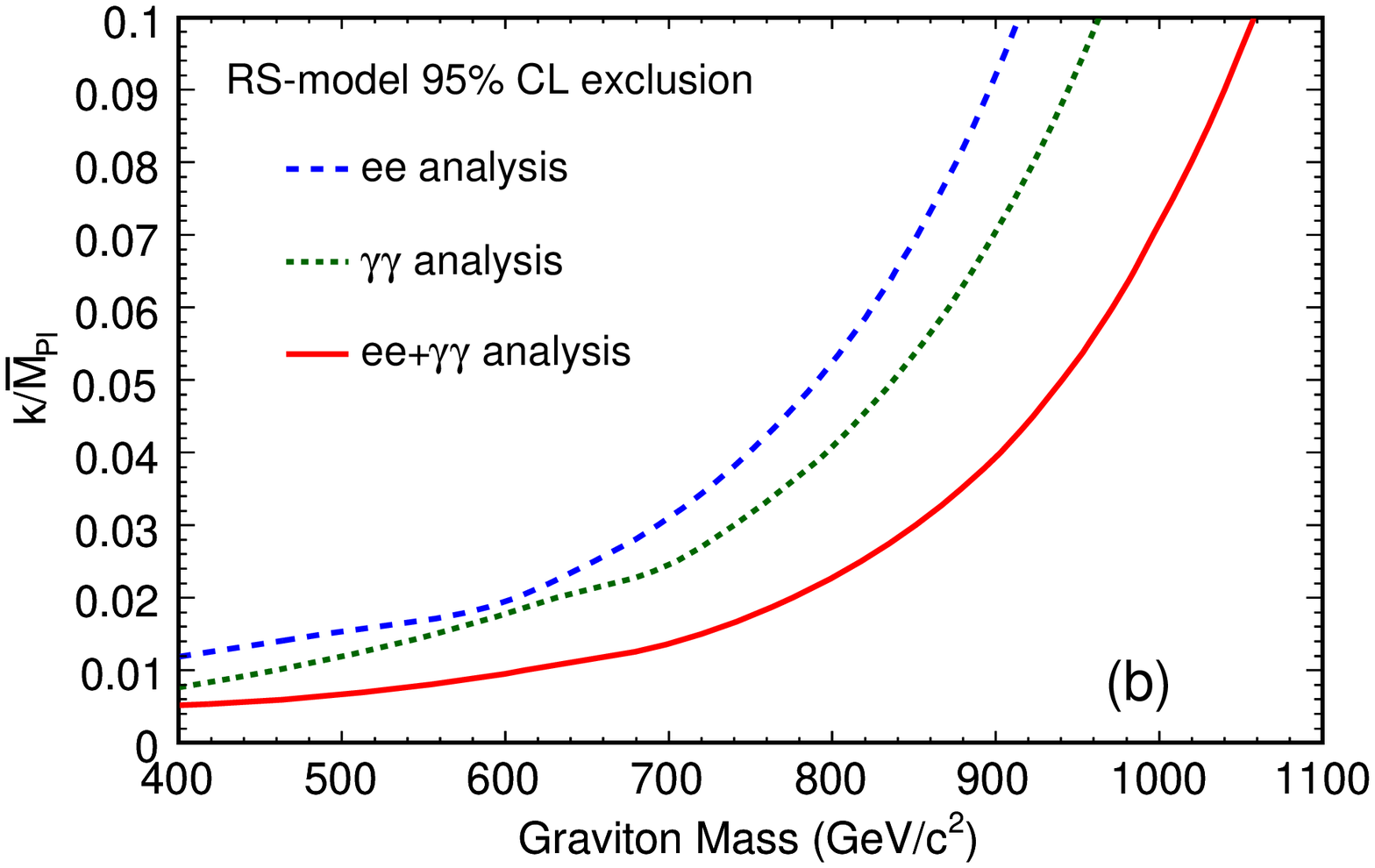}
\includegraphics[height=5.5cm]{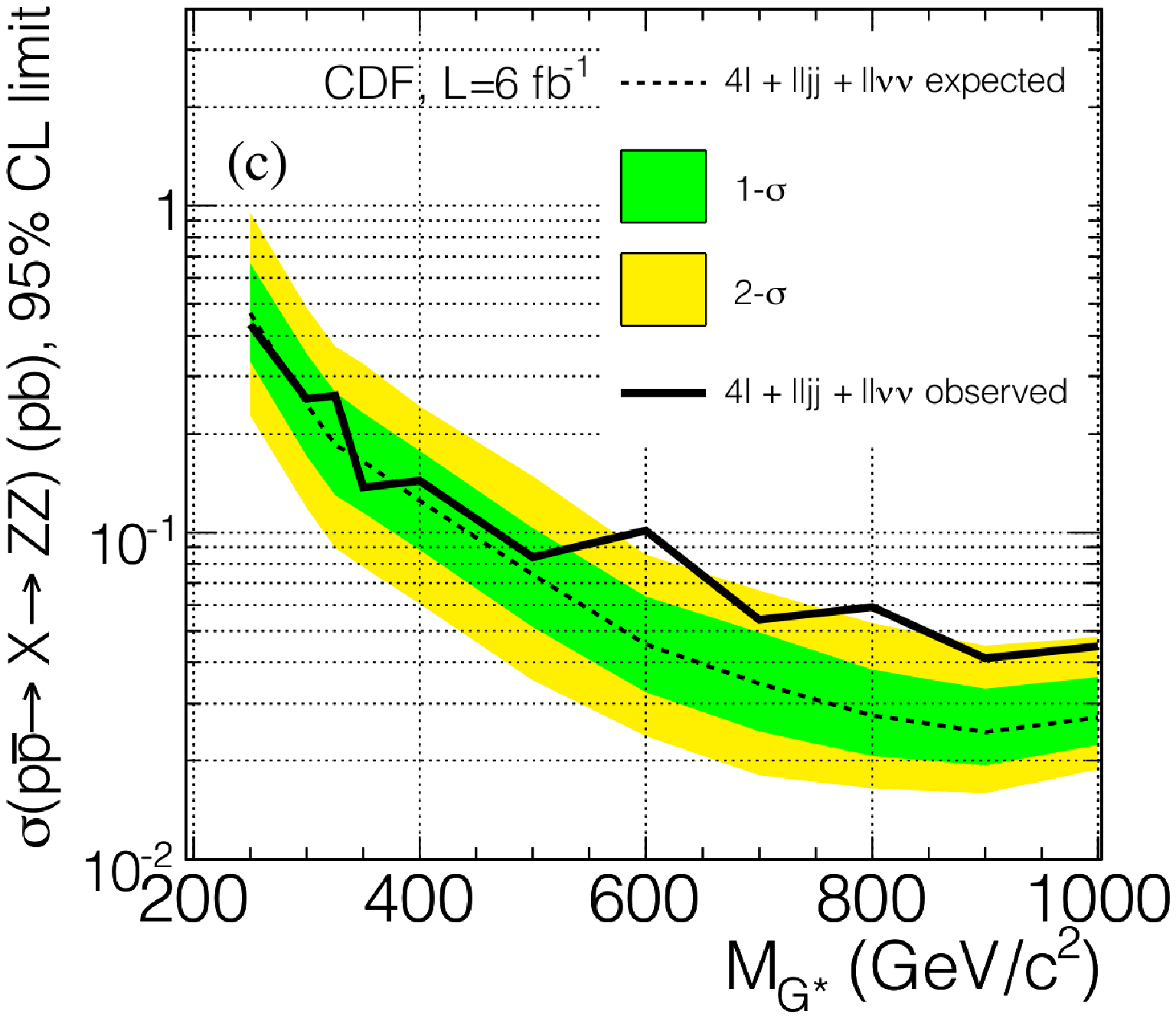}
\end{center}
\caption{(color online)  The 
 95\% C.L.\ excluded region in the $k/\overline{M}_{PL}$ vs. $M_G$ plane from  a search for RS gravitons  in the
$ee$ and $\gamma\gamma$ final states from (a)  the D0
 and (b)  the CDF experiments. 
(c)The 95\% C.L.\ upper limits on the cross section of the $G^*\to ZZ$
as a function of the $M_{G^*}$ from the CDF experiment. \label{f8}}
\end{figure}

There are many ways to search for dark matter (DM) in high energy collisions depending on the potential production model. 
 SUSY models, where the DM
is the LSP and is produced in the cascade decays of other sparticles, were
 described in section~\ref{sec_susy}.
However, direct production is possible and can be observed if the DM particles are produced in 
association with a high energy photon or jet produced via initial state radiation. 
The process $p\bar{p}\to DM+DM+\rm{jet}\to\rm{jet}+\met$
 was investigated at the CDF experiment\cite{PhysRevLett.108.211804}.
 No significant deviations are found and the 90\% C.L.\ cross section upper limits are set and converted
into constraints on the DM--nucleon cross
section. These results are shown together with several direct detection results in Fig.~\ref{fig-DM}. 

\begin{figure}[htb]
\begin{center}
\includegraphics[height=3.85cm]{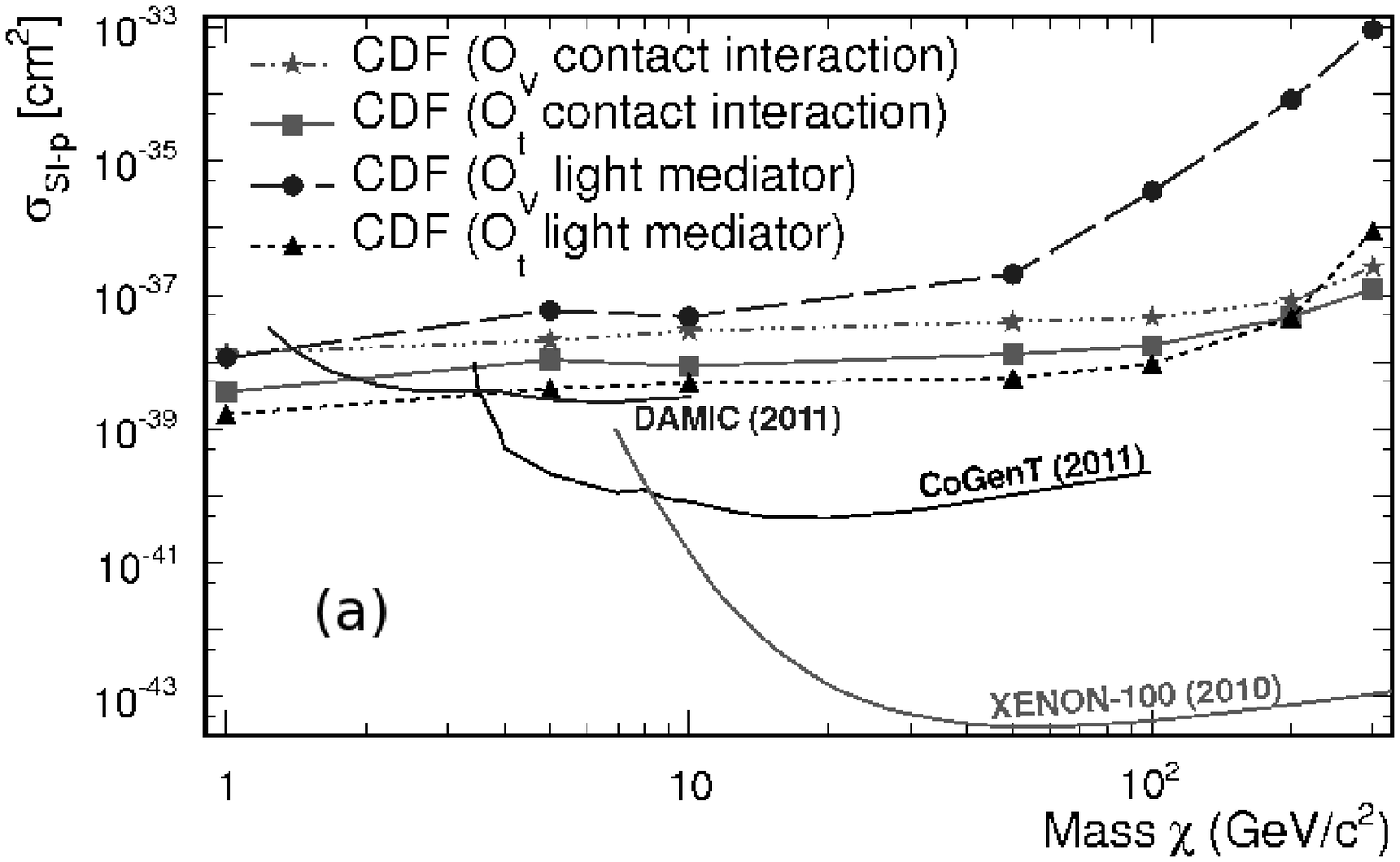}
\includegraphics[height=3.85cm]{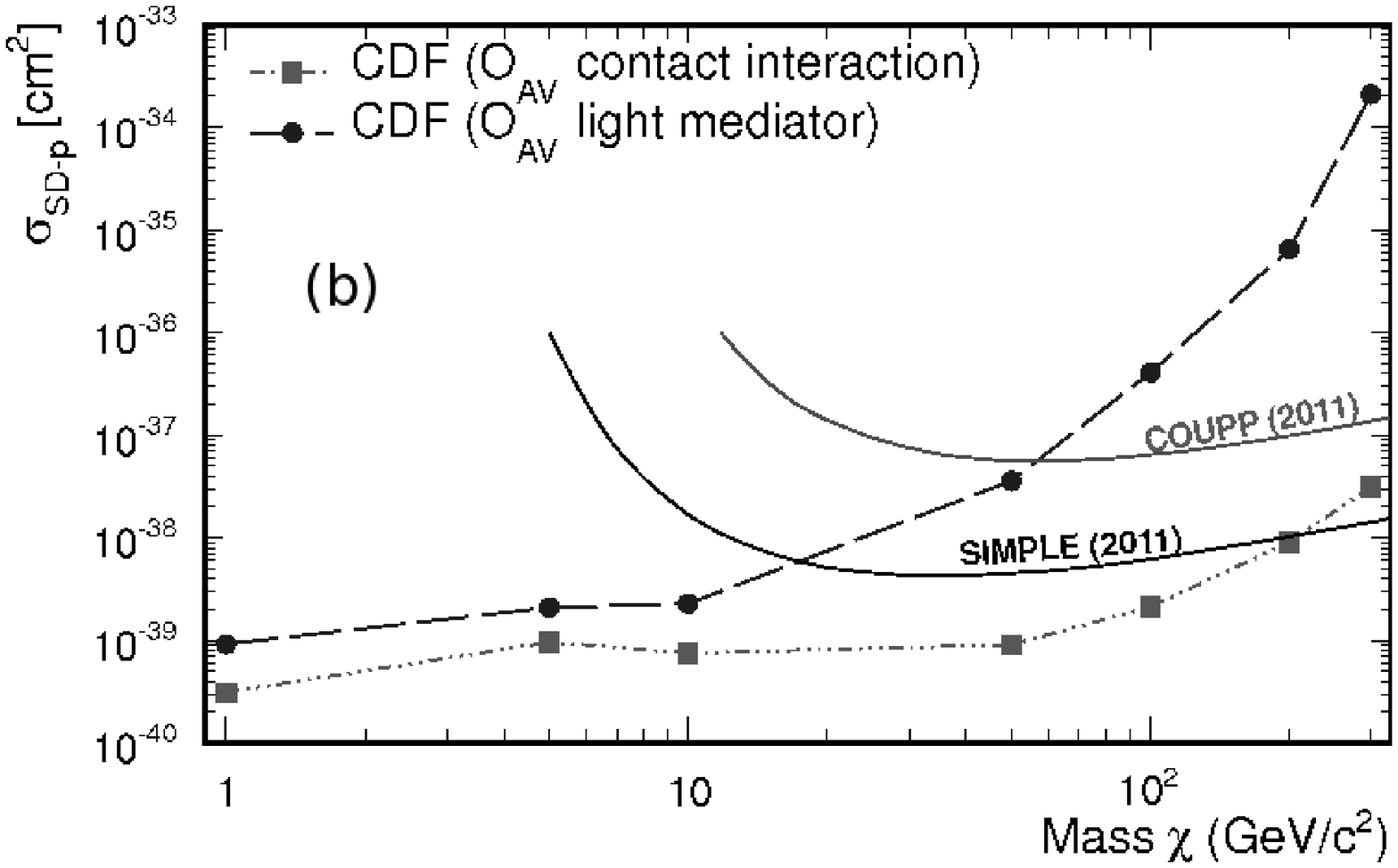}
\end{center}
\caption{(color online) The results on the DM--nucleon 
scattering from the CDF experiment, done at 90\% C.L.,  compared to the other direct dark matter 
experiments. For a detailed description, see Ref.~\citen{PhysRevLett.108.211804}. \label{fig-DM}}
\end{figure}

\subsection{\textbf{\textit{Signature--based searches and model--independent searches}}}\label{sec_sbmi}

Following the early development of signature--based searches in Run~I, as described
in section~\ref{sec_sleuth}, both the CDF
and the D0 
experiments did model--independent searches for new physics looking for discrepancies between data
and SM predictions in the events characterized with  high transverse momentum. These were done 
 using the {\sc sleuth}, {\sc bump hunter} and {\sc vista}
 programs~\cite{PhysRevD.78.012002, PhysRevD.79.011101} at the 
 CDF experiment and similar methods at the D0 experiment~\cite{Abazov:2011ma}. Despite the huge number of final states considered, 
({\sc sleuth} considered 399 final states, {\sc bump hunter} 5036 
final states and {\sc vista} considered 19650 final states), no true anomalies emerged (although 
the methods did serve to improve the MC simulation when discrepancies were noticed). 
The most discrepant final state contained $e\met+b$, but was found to be consistent when taking into account the trials factor.
In addition, the CDF experiment searched for new physics in a number of dedicated signature--based searches, specifically:
(i)~$\gamma\gamma$, $\ell\gamma+\met$ and $\ell\ell\gamma$ events\cite{PhysRevD.75.112001,PhysRevD.82.052005}, 
where the famous \eeggMet\ event from Run~I
would have been confirmed; 
(ii)~$\gamma+\rm{jet}+b+\met$ final state\cite{PhysRevD.80.052003};
(iii)~two jets and large \met\ events\cite{PhysRevLett.105.131801};
 (iv)~$ZZ+\met\to\ell\ell qq+\met$ events\cite{PhysRevD.85.011104};
 and 
 (v)~$p\bar{p} \to (3jets)(3jets)$ \cite{PhysRevLett.107.042001}.
In all of these searches data agreed with the SM prediction, and no new physics was found.

 \section{\textbf{Summary and Conclusions}}\label{sec_sum}
 The legacy of the Fermilab Tevatron collider experiments in searches for new particles and interactions is a powerful and glorious one. 
 The searches for new particles, such as SUSY, new fermions and bosons,
 excited fermions, leptoquarks, technicolor, hidden--valley model particles,
long--lived particles,
extra dimensions, dark matter particles,
 and a host of other interesting signatures was broad and deep,
  and produced interesting hints that changed the way we look at searches today. Indeed many new theoretical 
  models and experimental techniques 
  came into favor because of the CDF and D0 experiments, and are followed closely by the LHC which has taken over the high energy 
  frontier. In its time the answers from the Tevatron, both in terms of long--established models and new important 
  ones that cropped up, quickly responded to the best ideas in the field and 
and provided inspiration for new theoretical ideas.  

\section*{Acknowledgments}

We thank Ray Culbertson, Michael Eads, Paul Grannis, Oscar Gonzalez Lopez, and Stephen Mrenna for useful comments and discussions.

We thank the Fermilab staff and technical staffs of
the participating institutions for their vital contributions.
We acknowledge support from the DOE and NSF
(USA), ARC (Australia), CNPq, FAPERJ, FAPESP
and FUNDUNESP (Brazil), NSERC (Canada), NSC,
CAS and CNSF (China), Colciencias (Colombia), MSMT
and GACR (Czech Republic), the Academy of Finland,
CEA and CNRS/IN2P3 (France), BMBF and DFG (Germany),
DAE and DST (India), SFI (Ireland), INFN
(Italy), MEXT (Japan), the KoreanWorld Class University
Program and NRF (Korea), CONACyT (Mexico),
FOM (Netherlands), MON, NRC KI and RFBR (Russia),
the Slovak R\&D Agency, the Ministerio de Ciencia
e InnovaciÂ´on, and Programa Consolider--Ingenio 2010
(Spain), The Swedish Research Council (Sweden), SNSF
(Switzerland), STFC and the Royal Society (United
Kingdom), the A.P. Sloan Foundation (USA), and the
EU community Marie Curie Fellowship contract 302103.

One author (L. Z.) is supported by Serbian Ministry of Education, Science and  Technological development project 171004.
 One author (D.\ T.) is supported by the Mitchell Institute of Fundamental Physics and Astronomy.

\clearpage

\bibliographystyle{ws-ijmpa}
\bibliography{NP}

\end{document}